\documentclass[twocolumn,english,amsmath,amssymb,aps,prl,superscriptaddress]{revtex4-2}
\usepackage[T1]{fontenc}
\usepackage[latin9]{inputenc}
\usepackage{babel}
\usepackage{float}
\usepackage{amsmath}
\usepackage{graphicx}
\usepackage{colortbl}

\usepackage[unicode=true,
 bookmarks=false,
 breaklinks=true,pdfborder={0 0 1},backref=false,colorlinks=true]
 {hyperref}
\hypersetup{pdfcreator={},pdfproducer={LaTeX with hyperref},linkcolor=blue,anchorcolor=blue,citecolor=blue,filecolor=red,menucolor=red,urlcolor=blue,pdfstartview=FitV,pdfhighlight=/I,hypertexnames=true}

\newcommand{\new}[1]{{\color{black}#1}}

\makeatletter

\DeclareMathOperator {\Tr}{Tr}

\DeclareMathOperator {\sech}{sech}

\usepackage{natbib}
\setcitestyle{numbers}

\makeatother

\begin{document}
\title{Magnetic-field-tuned randomness in inhomogeneous altermagnets}
\author{Anzumaan R. Chakraborty}
\affiliation{Department of Physics, The Grainger College of Engineering, University of Illinois Urbana-Champaign, Urbana, IL 61801, USA}
\affiliation{Anthony J. Leggett Institute for Condensed Matter Theory, The Grainger College of Engineering, University of Illinois Urbana-Champaign, Urbana, 61801, IL, USA}

\author{J\"org Schmalian}
\affiliation{Institute for Theory of Condensed Matter, Karlsruhe
Institute of Technology, 76131 Karlsruhe, Germany}
\affiliation{Institute for Quantum Materials and Technologies,
Karlsruhe Institute of Technology, 76126 Karlsruhe, Germany}

\author{Rafael M. Fernandes}
\affiliation{Department of Physics, The Grainger College of Engineering, University of Illinois Urbana-Champaign, Urbana, IL 61801, USA}
\affiliation{Anthony J. Leggett Institute for Condensed Matter Theory, The Grainger College of Engineering, University of Illinois Urbana-Champaign, Urbana, 61801, IL, USA}

\date{\today}
\begin{abstract}
Altermagnetic (AM) states have compensated collinear magnetic configurations that are invariant under a combination of real-space rotation and time reversal. While these symmetries forbid a direct bilinear coupling of the AM order parameter with a magnetic field, they generally enable piezomagnetism, manifested as a trilinear coupling with magnetic field and strain. Here, we show that, because of this coupling, in an altermagnet subjected to random strain, the magnetic field triggers an effective random field conjugate to the AM order parameter, providing a rare realization of a tunable random-field Ising model. Specifically, we find two competing effects promoted by an external magnetic field: an increasing random-field disorder, which suppresses long-range AM order,
and an enhanced coupling to elastic fluctuations, which favors AM order. 
By solving the corresponding random-field transverse-field Ising model via a mean-field approach, we obtain the temperature-magnetic field phase diagram of an inhomogeneous AM state for different strengths of random-strain disorder, unveiling the emergence of a field-induced reentrant AM phase. We also discuss the fingerprints of this rich behavior on several experimentally-accessible quantities, such
as the shear modulus, the elasto-caloric effect coefficient, and the AM order parameter. Our results reveal an unusual but experimentally-feasible path to tune AM order with uniform magnetic fields.
\end{abstract}
\maketitle

\section{Introduction}

A recent classification of collinear magnetism revealed the existence of three different types of magnetic states: ferromagnetism, antiferromagnetism,
and altermagnetism \citep{Smejkal2020_AM,Smejkal2022_2_AM,Smejkal2022_AM}.
These different magnetic ground states, which are formally defined in the absence of spin-orbit coupling using the concept of spin groups \cite{Smejkal2022_AM,Liu2022,Fang2024,Liu2024,Song2024}, are distinguished
according to the type of space group operation which, when combined
with time-reversal, leaves the ground state invariant. In ferromagnets
(FM), there is no such operation and a uniform splitting between spin-up and spin-down bands emerges. Time-reversed
antiferromagnetic (AFM) ground states, on the other hand, are related by a translation
or inversion, resulting in a symmetry-protected Kramers spin degeneracy
throughout the entire Brillouin zone. Finally, time-reversed
altermagnetic (AM) ground states are related by any crystalline operation that is not translation or inversion, such as rotation, mirror reflection, or non-symmorphic operations like glides or screw rotations. The implication of this symmetry is a nodal $d$-wave, $g$-wave, or $i$-wave spin-split band structure, with spin degeneracy preserved only along certain high-symmetry momentum space planes (for a recent review, see \cite{Jungwirth2024review}). These unusual properties have motivated extensive theoretical works on the interplay between altermagnetism and other electronic phenomena such as topology \cite{Mazin2023_FeSe,Fernandes2024_AM,Cano2024,Antonenko2025,Schnyder2024,Agterberg2024,Knolle2024,Attias2024}, electronic correlations \cite{Yu_Agterberg2024,Leeb2024,Valenti2024,Kaushal2024,Sobral2024,Giuli2025,del2024dirac}, superconductivity \cite{Sudbo2023,Ouassou2023,Neupert2023,Beenakker2023,Sun2023,Papaj2023,Zhu2023,Wei2023,Li2023_Majorana,Chakraborty2024,Scheurer2024,Ghorashi2024,Chakraborty2024constraints,Heung2024,Carvalho2024}, non-trivial responses \cite{Smejkal2022chiral,Steward2023,Okamoto2023,Bhowal2024,Mcclarty2024,Vila2024,Radaelli2024,Schiff2024,Takahashi2025,Lin2025coulomb}, and multiferroics \cite{Smejkal2024_multiferroics,Gu2025,Duan2025}.

Many materials have been proposed to realize altermagnetism, from metals to Mott insulators \cite{Mazin2021,Facio2023,Mazin2023,Gao2023,Jaeschke2024,Haule2024,Sodequist2024,Li2024strain,Jiang2024monolayer,Ji2025}. Among those, experiments have directly demonstrated the altermagnetic character of materials such as MnTe \cite{Krempasky2024,Amin2024,Lee2023,Osumi2023}, CrSb \cite{Reimers2023,Yang2024,Li2024,Ding2024}, Co$_{1/4}$NbSe$_2$ \cite{Babu2024,Vita2025,Graham2025}, and $A$V$_2$$Ch_2$O (with alkali metal $A=\mathrm{Rb,K}$ and chalcogen $Ch=\mathrm{Se,Te}$) \cite{Jiang2024discovery}, while results for RuO$_2$ remain under debate \cite{Feng2022,Betancourt2023,Fedchenko2023,Bai2023,Lin2024observation,Smolyanyuk2023,Kessler2024,Hiraishi2024,Jeong2024}. More broadly, altermagnetism is connected to other problems of interest in condensed matter physics beyond spin-splitting in compensated magnets \cite{Naka2019,Hayami2019,Zunger2020,Kusunose2020}. For example, altermagetic order breaks the same symmetries as other states of interest in correlated electron systems, such as ferro-octupolar order \cite{Bhowal2024,Fernandes2024_AM,Jaeschke2025} in multipolar magnets \cite{Santini2009,Voleti2020,Fiore2022,winkler2023theory} and metals undergoing an even-parity spin-triplet Pomeranchuk instability \cite{Pomeranchuk1958,Wu2007,Kunes2019}. The microscopic mechanisms involved, however, are very different, as the crystalline potential plays an essential role in stabilizing altermagnetism \cite{Jungwirth2024supefluid}. Thus, given the rich landscape of materials and phenomena related to altermagnetism, it is important to establish which external perturbations can be used to control and probe these systems.  
\begin{figure}
\includegraphics[width=0.7\columnwidth]{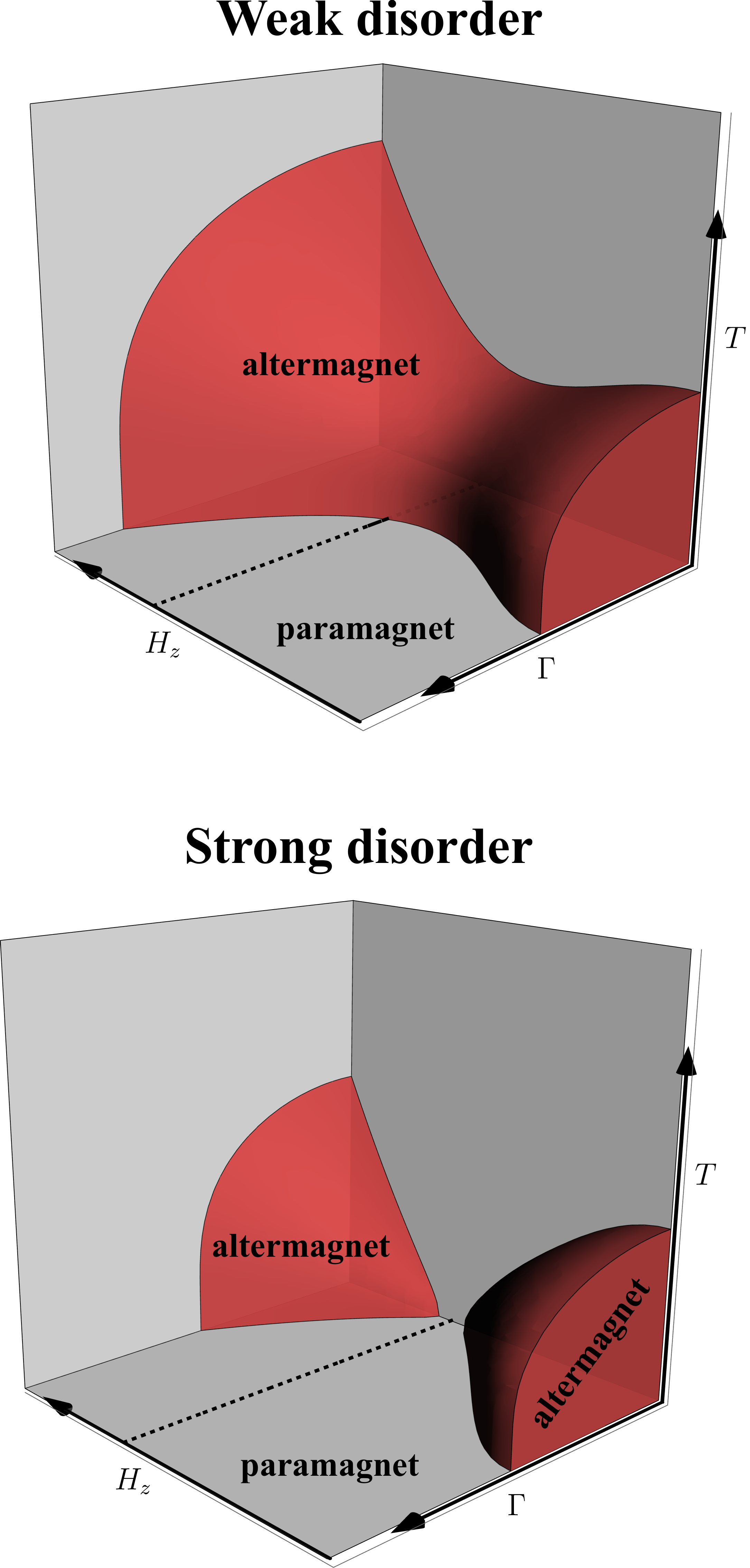}\caption{Schematic phase diagram based on the results of this paper for an Ising-like altermagnet
in the presence of weak (top) or strong (bottom) random strain disorder-strength as a function of magnetic
field $H_{z}$, quantum fluctuations tuning parameter $\Gamma$, and temperature $T$.
Above a critical disorder strength, the altermagnetic phase stabilizes
into separate low-field and high-field ordered phases.} \label{fig:schematic_phase_diagrams}
\end{figure}

At first sight, one may think that a magnetic field  $\mathbf{H}$ is not an ideal tuning parameter for altermagnetism. Indeed, since
$\mathbf{H}$ cannot couple directly (i.e., via a bilinear coupling) to the
AM order parameter $\boldsymbol{\Phi}(\mathbf{x})$, one would expect $\mathbf{H}$ to have a
minimal effect on the onset of AM order.  However, the symmetries that define altermagnetism also imply that nearly all altermagnets display piezomagnetism \cite{Steward2023,Fernandes2024_AM,Mcclarty2024,Aoyama2024_piezomagnetism,vandenBrink2024}.
Piezomagnetism is a phenomenon analogous to piezoelectricity,
in which the application of strain leads to a magnetic dipole moment,
and vice-versa \citep{Dzialoshinskii1958_piezomagnetism}.  It is
characterized by the linear response equation 
\begin{equation}
H_{i}=\Lambda_{ijk}\varepsilon_{jk}
\end{equation}
where $\Lambda_{ijk}$ is the piezomagnetic tensor, and $\varepsilon_{jk}$
is the strain tensor defined in terms of the crystal displacement
field $\mathbf{u}$ through $\varepsilon_{jk}=\frac{1}{2}(\partial_{j}u_{k}+\partial_{k}u_{j})$.
As discussed elsewhere, in altermagnets the components of $\Lambda_{ijk}$ are proportional to
odd powers of the AM order parameter $\boldsymbol{\Phi}$ \cite{Takahashi2025}. Moreover, 
$\varepsilon_{jk}$ is generally a shear strain, since a
pure symmetry-preserving dilatation $\sum_{i}\varepsilon_{ii}$ does not couple
to a product of magnetic field and an AM order parameter. We emphasize
that piezomagnetism is a qualitatively different response than magnetostriction,
which relates strain $\varepsilon_{ij}$ to a magnetic field bilinear $H_{j}H_{k}$ \citep{Lee1955_magnetostriction,Callen1968_magnetostriction}.
Because of piezomagnetism, the Landau free energy of an altermagnet must have a term of the form
\begin{equation}
\mathcal{F}_{\text{pzm}}=-\lambda_{ijkl}\Phi_{i}H_{j}\varepsilon_{kl}
\label{eq:F_pzm}
\end{equation}
where the coupling constants $\lambda_{ijkl}$  are related to the piezomagnetic tensor.
In this paper, we show that, because of piezomagnetism, a magnetic field can be used to tune the AM transition, provided that the system displays an inhomogeneous distribution of internal strain fields. This is generally expected to be the case in any crystal, since unavoidable lattice defects such as dislocations, vacancies, and dopants always generate random strain $\varepsilon_{ij}(\mathbf{x})$. The key point is that, from Eq. (\ref{eq:F_pzm}), the product of the appropriate component of  $\mathbf{H}$  with an inhomogeneous strain field $\varepsilon_{ij}(\mathbf{x})$ acts as an effective random longitudinal field that is conjugate to $\boldsymbol{\Phi}$. Thus, in the usual case in which spin-orbit coupling (SOC) lowers the symmetry of the vector AM order parameter to a single component $\Phi$ \cite{Fernandes2024_AM}, the inhomogeneous AM realizes a rare example of a
\textit{tunable random field Ising model} (RFIM). The RFIM is the prototypical model to elucidate the impact of disorder on critical phenomena \cite{Ma1996_RF,schneider1977_RF,Binder1983,Nattermann1988_RF,Toh1992_RF}. The tuning parameter of the model is the disorder strength, set by, e.g., the width of the longitudinal field distribution. While the RFIM is realized in certain magnetic \cite{Fishman1979,Christou2010} and nematic materials \cite{Carlson2006}, the disorder strength is usually (but not always \cite{Silevitch2007}) fixed for a given crystal. In contrast, in the case of this RFIM realization in altermagnets, the disorder strength is continuously tuned by the magnetic field even though the distribution of inhomogeneous strain is unchanged.

We show that the random-field effect generated by the combination of magnetic fields and residual strain, which tends to suppress AM order, competes with another effect that also arises from piezomagnetism but that tends to favor AM order. This latter effect arises because, in the presence of an external field, thermally excited elastic fluctuations mediate long-range correlations between the AM degrees of freedom. To investigate the interplay between these two piezomagnetic-generated effects, we employ a mean-field approach to calculate the temperature-magnetic field phase diagram of an inhomogeneous AM state. We find two qualitatively distinct behaviors, illustrated in Fig. \ref{fig:schematic_phase_diagrams}. For weak disorder strength, application of a magnetic field first suppresses the AM transition temperature but then enhances it. For large disorder strength, the random-field effect of the magnetic field is strong enough to completely suppress the AM phase for intermediate field values, leading to a guaranteed reentrance behavior for large enough fields. These behaviors are also reflected at $T=0$, where a non-thermal tuning parameter (denoted by $\Gamma$ in the phase diagrams) tunes a quantum AM transition. We also compute the behavior of several experimentally observable quantities across these phase diagrams, such as the AM order parameter, the shear modulus, and the elasto-caloric effect coefficient, thus providing concrete experimental predictions to verify this behavior in candidate AM materials.

This paper is organized as follows. In Sec. II we construct an effective
random-field transverse-field Ising model (RF-TFIM) for an AM system \citep{Senthil1998_RF}. This model consists of a uniform
transverse field $\Gamma$ encapsulating the role of quantum fluctuations, and
a longitudinal field $H_{z}$ that promotes the coupling between $\Phi$
and shear strain $\varepsilon$. We compute the mean-field phase diagram
of this model as a function of the parameters $T,$ $\Gamma$, $H_{z}$,
and intrinsic random strain. In Sec. III we compute the order parameter
self-consistently and obtain the AM susceptibility, shear modulus
renormalization, and elasto-caloric effect both as a function of $T$
and of $H_{z}$. Like in the case a nematic critical point, we find
a softening of the shear modulus at the AM critical point signaling
an accompanying structural transition. In Sec. IV, we discuss the effect of
fluctuations beyond mean-field, which are typical for random field
models without infinite-range interactions. Finally, we present our conclusions
in Sec. V, and in the Appendix, we provide additional details of the calculation for the elasto-caloric effect coefficient.

\section{Random-field Ising model with a uniform transverse field}

For concreteness, we consider a specific $d$-wave AM ordered state in the tetragonal lattice (with point group $D_{4h}$) that displays piezomagnetism. In the presence of SOC, the components of the vector AM order parameter $\boldsymbol{\Phi}$  transform as different irreducible representations (irreps) of the point group according to the direction of the magnetic moments. For out-of-plane moments, the system remains a ``pure'' altermagnet even in the presence of SOC \cite{Fernandes2024_AM}, and the AM order parameter $\Phi$ is Ising-like, transforming either as the  $B_{1g}^{-}$  irrep (in the case of a $d_{xy}$-wave AM, as relevant for rutile AM like MnF$_2$ \cite{Smejkal2022_AM,Bhowal2024}) or as the  $B_{2g}^{-}$  irrep (in the case of a $d_{x^2-y^2}$-wave AM, as relevant for La$_2$Mn$_2$Se$_2$O$_3$ \cite{Ji2025} and $A$V$_2$$Ch_2$O \cite{Jiang2024discovery,Jiang2024monolayer,Li2024strain,Zhang2024crystal}). Here, the minus superscript indicates that the irrep is odd under time-reversal symmetry. Using the group-theory result $B_{1g}^{\pm}\otimes B_{2g}^{\mp}\otimes A_{2g}^{-}=A_{1g}^+$, it is straightforward to derive Eq. (\ref{eq:F_pzm}) and obtain the Landau free-energy invariant \cite{Steward2023}:

\begin{equation}
\mathcal{F}_{\text{pzm}}=-\lambda\Phi H_{z}\varepsilon\label{eq:homogenous_PZMcoupling}
\end{equation}
where $\lambda$ is the piezomagnetic coupling, $\varepsilon \equiv \varepsilon_{xy}$ in the case of a $d_{xy}$-wave AM and $\varepsilon \equiv \varepsilon_{x^2-y^2}$ in the case of a $d_{x^2-y^2}$-wave AM.

To proceed, we write down an effective low-energy model for the coupled AM-strain degrees of freedom. \new{As explained above, for a tetragonal lattice, the $d$-wave AM order parameter transforms as either the $B_{1g}^{-}$  or the $B_{2g}^{-}$ irrep, which corresponds to an Ising-like scalar order parameter ($\Phi$ above). To account for the role of spatial and quantum fluctuations, we promote this homogeneous order parameter to a coarse-grained local Ising (i.e., pseudo-spin) variable $\tau^z_i$, such that $\langle \tau_i^z \rangle = \Phi $. Spatial fluctuations are encoded in the effective interaction between the pseudo-spins, whereas quantum fluctuations appear as a transverse field to the pseudo-spin. Thus, our phenomenological Hamiltonian for the AM phase is that of a transverse-field Ising model:}

\begin{equation}
\mathcal{H}_{\mathrm{AM}}=-J\sum_{\langle ij\rangle}\tau_{i}^z\tau_{j}^z -\Gamma\sum_{i}\tau_{i}^{x}
\end{equation}
where, as discussed above, $\tau_i^z$ and $\tau_i^x$ are Pauli matrices. $J$ is an effective AM interaction, \new{whose origin depends on the microscopic mechanism behind AM order}, that sets the scale of the thermal AM transition (and should not be confused with an exchange interaction). The transverse field $\Gamma$ is a non-thermal parameter (such as doping or pressure) that promotes quantum fluctuations, i.e., tunneling
between local pseudo-spin states. Thus, as $\Gamma$ is enhanced, the ordered state is suppressed and the system is driven to a quantum critical point (QCP). \new{Note that, being phenomenological, our approach is agnostic about the microscopic origin of the AM order parameter. For instance, in the Lieb lattice model of Ref. \cite{Antonenko2025}, $\tau_i^z$ could correspond to the staggered magnetization between the two sublattices related by a $90^\circ$ rotation, coarse-grained over a single square plaquette. It could also correspond to the average magnetic octupole moment on the third site of the Lieb lattice, or to some combination of them \cite{Mcclarty2024} determined by the energetics of the microscopic model \cite{Jaeschke2025}. The key point is that, regardless of the microscopic mechanism, the AM ordered state will occur due to the condensation of an Ising-like order parameter that transforms as the appropriate irrep.}

Strain fields in crystals are usually inhomogeneous,
$\varepsilon\to\varepsilon(\mathbf{x})$, with two contributions arising from distinct phenomena. The first is a non-singular contribution from thermally
excited elastic fluctuations $\varepsilon_{0}(\mathbf{q})$, associated with the shear modulus $C_0$ defined as $C_0 \equiv C_{66}$ for a $d_{xy}$-wave AM and $C_0 \equiv (C_{11}-C_{12})/2$ for a $d_{x^2-y^2}$-wave AM.
The second is a singular contribution from random strain arising
from crystal defects $\varepsilon_{s}(\mathbf{x})$ in the form of
a quenched random field. Therefore, from Eq. (\ref{eq:homogenous_PZMcoupling}),
we can cast the inhomogeneous piezomagnetic term as:
\begin{equation}
\mathcal{H}_{\text{pzm}}=-\lambda H_{z}\sum_{i}(\varepsilon_{0,i}+\varepsilon_{s,i})\tau_{i}^z
\end{equation}
Note that the elastic fluctuations are well-defined excitations; therefore $\varepsilon_{0,i}$ should be thought of as an
annealed\emph{ }dynamical field which one can integrate out. The effect
of the non-homogeneous part of $\varepsilon_{0,i}$, i.e., the $\mathbf{q}\neq0$
modes, is to drive the AM transition mean field and renormalize the
effective Hamiltonian \citep{Chou1996,Paul2017}. We therefore dispense with
$\varepsilon_{0}(\mathbf{q}\neq0$) modes and regard them as subleading
corrections to the full Hamiltonian. By contrast, $\varepsilon_{s,i}$
represents quenched disorder and remains as a fixed realization of
a random field drawn from a zero-mean probability distribution. Putting it all together, the effective Hamiltonian consisting of altermagnetic
and elastic degrees of freedom is 

\begin{align}
\mathcal{H} & =-J\sum_{\langle ij\rangle}\tau_{i}^{z}\tau_{j}^{z}-\frac{N}{2}C_{0}\varepsilon_{0}^{2}-\lambda H_{z}\sum_{i}(\varepsilon_{0}+\varepsilon_{s,i})\tau_{i}^{z}\label{eq:full_H}\\
 & -\Gamma\sum_{i}\tau_{i}^{x}\nonumber 
\end{align}
where $N$ denotes the number of sites.

\section{Field-tuned altermagnetic transition}

\subsection{Renormalized field-dependent Hamiltonian}

In this section, we analyze the phase diagram of Eq. (\ref{eq:full_H}) within mean-field
over a wide range of parameters, including the strength of quantum
fluctuations, temperature, disorder strength, and magnetic field. To formally perform a mean-field calculation, we extend the AM interaction $J$ to also be infinite-range rather than nearest-neighbors:
\begin{equation}
-J\sum_{\langle ij\rangle}\to-\frac{J}{N}\sum_{i<j}\label{eq:all-to-all_replacement}
\end{equation}

Next, we integrate out the uniform elastic field $\varepsilon_{0}$, which introduces an infinite-range interaction, effectively enhancing $J\to\tilde{J}$ for nonzero
$H_{z}$. This yields a renormalized Hamiltonian of the form
\begin{equation}
\tilde{\mathcal{H}}=-\frac{\tilde{J}}{N}\sum_{i<j}\tau_{i}^{z}\tau_{j}^{z}-\lambda H_{z}\sum_{i}\varepsilon_{s,i}\tau_{i}^{z}-\Gamma\sum_{i}\tau_{i}^{x}\label{eq:renormalized_H}
\end{equation}
where 
\begin{equation}
\tilde{J}\equiv J[1+(H_{z}/H_\lambda)^{2}]
\end{equation}
Here, we introduced a magnetic field scale $H_\lambda$
defined as
\begin{equation}
H_\lambda\equiv\frac{\sqrt{C_{0}J}}{\lambda}
\end{equation}
These results follow from a simple Gaussian
identity applied to the $\varepsilon_{0}$-dependent part of $\mathcal{H}$:
\begin{equation}
\begin{split}
-T\log\int& \mathrm{d}\varepsilon_{0}e^{-\frac{N}{2T}C_{0}\varepsilon_{0}^{2}+\frac{\lambda H_{z}}{T}\varepsilon_{0}\sum_{i}\tau_{i}^{z}}\\&=-\frac{1}{N}\Big(\frac{H_{z}}{H_\lambda}\Big)^{2}\sum_{i<j}\tau_{i}^{z}\tau_{j}^{z}\label{eq:integrate_out_eps0}
\end{split}
\end{equation}
and further justify replacing the original nearest-neighbor interaction
with an all-to-all interaction.

The second term of $\tilde{\mathcal{H}}$ describes the contribution to
the piezomagnetic coupling due to random strain, with the product $\lambda H_{z}\varepsilon_{s,i}$
taking the role of an effective random longitudinal field conjugate to the AM order
parameter. Given that an infinite-range interaction, a random field,
and a non-random transverse field are present, we call $\mathcal{\tilde{H}}$
an infinite-range \emph{random-field transverse-field} \emph{Ising
model }(RF-TFIM) \citep{Senthil1998_RF,Dutta1996_RF}. Including dynamical phonons changes the infinite-range coupling to a dipolar interaction, an effect that we ignore in our subsequent analysis.

We now make a few convenient substitutions. First, we express all energy scales
in units of $J$ by setting $J=1.$ Second, we express the Hamiltonian in a
dimensionless form by dividing through by the $H_{z}$-dependent renormalized
exchange $\tilde{J}=1+(H_{z}/H_\lambda)^{2}$ and arrive at
\begin{equation}
\tilde{\mathcal{H}}/\tilde{J}=-\frac{1}{N}\sum_{i<j}\tau_{i}^{z}\tau_{j}^{z}-\sum_{i}z_{i}\tau_{i}^{z}-\gamma\sum_{i}\tau_{i}^{x}\label{eq:renormalized_fullH}
\end{equation}
Here, $z_{i}$ and $\gamma$ are effective longitudinal (random) and
transverse (uniform) fields, which take into account the exchange
enhancement due to elastic fluctuations. By construction, they are
$H_{z}$-dependent. Upon defining a dimensionless magnetic field parameter
$h=H_{z}/H_\lambda$, the quantity $z_{i}$ takes the form

\begin{align}
z_{i}=z_{i}(h) & =\frac{2h}{1+h^{2}}\left(\frac{\lambda H_\lambda\varepsilon_{s,i}}{2}\right)
\end{align}
It is clear that the random field $z_{i}(h)$ reaches its maximum value at
$h=1$, i.e., at $H_{z}=H_\lambda$, when it is equal to the term inside the
brackets. For concreteness, we regard $\varepsilon_{s,i}$
as Gaussian-distributed strain, with standard deviation $\bar{\varepsilon}_{s}$
such that $z_{i}$ also follows a Gaussian distribution with $h$-dependent
standard deviation 
\begin{equation}
w(h)=\frac{2h}{1+h^{2}}W\label{eq:w(h)}
\end{equation}
where 
\begin{align}
W & =\frac{\lambda H_\lambda\bar{\varepsilon}_{s}}{2}\equiv\sqrt{\frac{E_{\text{el}}}{2}}
\end{align}
Here 
\begin{equation}
E_{\text{el}}=\frac{C_{0}\bar{\varepsilon}_{s}^{2}}{2}
\end{equation}
is a dimensionless quantity that gives the elastic energy of the random distribution of strain in units of the AM interaction $J$. In what follows,
the standard deviation of the random variables $W$
shall be regarded as a parameter of the problem. Similarly,
the temperature and transverse field in these dimensionless units
are 
\begin{align}
t(h) & =\frac{T}{1+h^{2}}\\
\gamma(h) & =\frac{\Gamma}{1+h^{2}}
\end{align}
where, we recall, $T$ and $\Gamma$ are given in units of $J$.

To set the stage, we provide a qualitative description of the phase
diagram in the four-dimensional $(h,W,\Gamma,T)$ parameter space.
In the classical limit $\Gamma=0$, Eq. (\ref{eq:renormalized_fullH})
describes the well-known infinite-range RFIM, for which there exists
a critical random field strength $w_{c}$ separating an ordered and a
disordered phase \cite{schneider1977_RF}. Unlike the prototypical RFIM, however, here the random field
strength $w(h)$ is tunable by magnetic field, suggesting the possibility
of field values $h_{1}$ and $h_{2}$ for which $w(h_{1})<w_{c}$
and $w(h_{2})>w_{c}$. This implies the existence of a critical field
$h_{c}\in[h_{1},h_{2}]$ marked by $w(h_{c})=w_{c}$ which separates
the ordered and disordered phases. Additionally, $w(h)$ is non-monotonic,
and reaches a maximum value of $W\propto\bar{\varepsilon}_{s}$ at $h=1$.
Therefore three possibilities exist when the system begins in the
ordered phase and a magnetic field is applied, which we delineate as
follows:

\begin{itemize}
\item In the regime of \emph{strong disorder}, $W$ exceeds $w_{c}$.
The non-monotonicity of $w(h)$ implies the existence of two critical
fields $h_{c}^{-}<1$ and $h_{c}^{+}>1$, both satisfying $w(h_{c}^{\pm})=w_{c}$,
for which the system is disordered over a finite window $[h_{c}^{-},h_{c}^{+}]$.
This implies the phenomenon of altermagnetic \emph{reentrance}, in
which the AM order parameter $\Phi$ is nonzero for $h<h_{c}^{-}$
and $h>h_{c}^{+}$ and is identically zero for $h\in[h_{c}^{-},h_{c}^{+}]$.

\item In the regime of \emph{weak disorder}, $W$ is smaller than $w_{c}$. As a result,
the system remains trapped in the ordered phase. Interestingly,
the order parameter $\Phi$ exhibits a vestige of the reentrance behavior characteristic of the strong disorder case, as it  attains a non-zero minimum value $\Phi_{\text{min}}$ at a field value $h_{\text{min}}<1$
for arbitrarily small $W\propto\bar{\varepsilon}_{s}$.

\item In the \textit{marginal disorder regime}, $W=w_{c}$. As a result,  $w(h)=w_{c}$
exhibits a double root $h=h_{*}$. Given that $h_{*}$ is a double
root, tuning the magnetic field across $h_{*}$ amounts to running
tangent to the co-dimension $1$ phase boundary at precisely one point
($h_{*},W_{*},\Gamma_{*},T_{*}$). The process of taking $W$ from $W<w_{c}$
to $W=w_{c}$ causes $\Phi(h)$ to smoothly deform with $\Phi_{\text{min}}$
going to zero at a single point $h_{*}$. This gives rise to
a ``V''-like behavior of the order parameter, $\Phi\propto|h-h_{*}|$.
This unusual phenomenon, which we explore in Section III, reflects the nontrivial
RF-TFIM parameter space enabled
by the piezomagnetic coupling.

\end{itemize}

The presence of thermal and quantum fluctuations reduces the critical
strength $w_{c}$, i.e., increases the effective disorder. Formally, this
means that the function $w_{c}(\Gamma,T)$ decreases with increasing
$\Gamma$ and $T$, and consequently, the reentrant behavior is enhanced, as
the paramagnetic window $\Delta h=h_{c}^{+}-h_{c}^{-}$ grows. Because
$W$ is an intrinsic property of the crystal, it is unlikely to be
tunable. Therefore, to assess the features of the phase diagram delineated
above one must tune $\Gamma$ and $T$ so that the function $w_{c}(\Gamma,T)$
becomes greater than, equal to, or less than the intrinsic random
strain scale $W$. 

\subsection{Zero-temperature phase diagram}

\begin{figure}
\includegraphics[width=0.9\columnwidth]{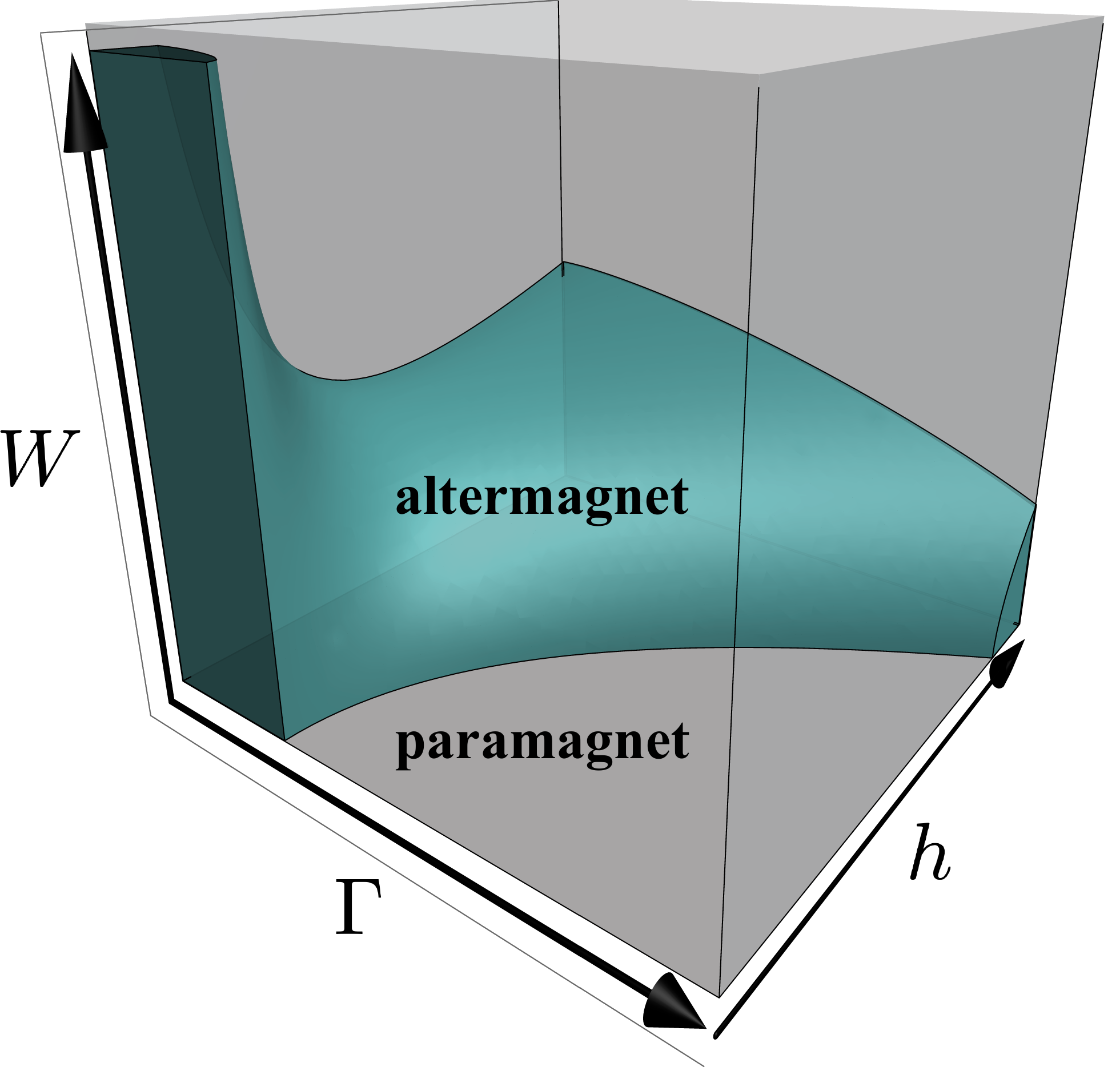}\caption{Zero-temperature mean-field phase diagram of the inhomogeneous altermagnet
as a function of relative disorder strength $W=\sqrt{E_{\text{el}}/2}$,
quantum-fluctuations promoting transverse-field $\Gamma$, and scaled magnetic field $h\equiv H_{z}/H_\lambda$.
For $W>W_{*}(\Gamma)$, a horizontal line intersects the AM-PM critical
surface twice, corresponding to AM reentrance. We show constant-$W$
cuts of this plot in Fig. \ref{fig:2D_T=00003D0_PDs}.} \label{fig:T0_phase_diagram}
\end{figure}

\begin{figure*}[t]
\centering{}\includegraphics[width=0.8\paperwidth]{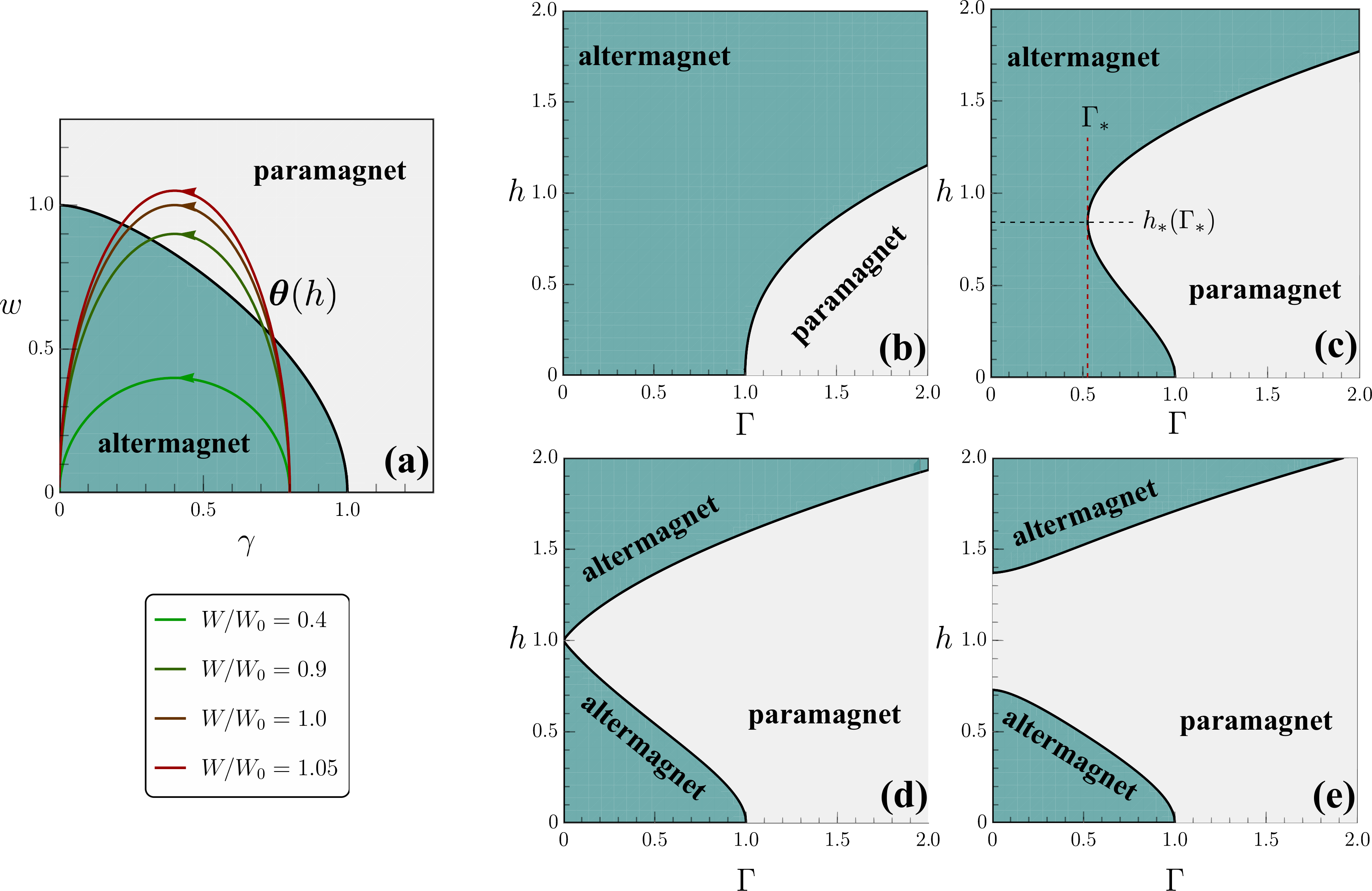}\caption{\label{fig:2D_T=00003D0_PDs} \textbf{(a)}: $T=0$ phase diagram with
magnetic-field scaled axes $\gamma\equiv\Gamma/(1+h^{2})$ and $w\equiv2Wh/(1+h^{2})$, where $\Gamma=0.8J$ sets the
starting point of the phase trajectory along the $\gamma$-axis and
$W$ (listed in the legend) sets the maximum extent of the trajectories along the $w$-axis. Increasing the magnetic field moves the system along the trajectory as indicated by the arrows.
\textbf{(b-e)}: $T=0$ phase diagram with unscaled axes $h=H_{z}/H_\lambda$
and $\Gamma$ corresponding to $W/W_{0}=0.4$, $0.9$, $1.0$, and
$1.05$, respectively. As the disorder strength $W$ increases, the
paramagnetic region grows and the AM-PM transition line forms a bulge at $(\Gamma_{*},h_{*})$
that annihilates at $(0,1)$.}
\end{figure*}

We now construct the mean-field $T=0$ phase diagram of Eq. (\ref{eq:renormalized_H})
as a function of $h$, $\Gamma$, and $W$, which is shown in Fig. \ref{fig:T0_phase_diagram}. In preparation for the
finite temperature case, for which there are four parameters ($h,W,\Gamma,T$)
rather than three, we describe two complementary approaches to visualize
the phase diagram. The first is to construct a single 2D phase
diagram with $h$-dependent axes $w(h)$ and $\gamma(h)$, where the
AM-PM phase boundary appears as a convex curve bounding the origin
(Fig. \ref{fig:2D_T=00003D0_PDs}a). Tuning $h$ amounts to following
an $h$-parameterized curved trajectory in this ($w,\gamma$)-plane,
which may or may not intersect with the phase boundary depending on
the values of $W$ and $\Gamma$. The second approach is to construct
multiple 2D phase diagrams, one for each $W$ -- corresponding to
different strengths of random strain -- with $h$ and $\Gamma$ as
dimensionless axes (Fig. \ref{fig:2D_T=00003D0_PDs}b-e). In this
case, tuning away from the $h=0$ limit amounts to following a line
parallel to the $h$-axis. The appearance of reentrance over some
parameter regime manifests as a ``bulge'' in the AM-PM phase boundary.
Finally, a third, which is relevant only for the $T=0$ case, is to
represent the full ($h,W,\Gamma$) phase diagram, as depicted in Fig. \ref{fig:T0_phase_diagram}.

We delineate the first approach now. The $h$-parameterized trajectory
$\boldsymbol{\theta}$ in this ($w,\gamma$)-plane is given by 
\begin{equation}
\boldsymbol{\theta}(W,\Gamma;h)=\langle w(h),\gamma(h)\rangle=\frac{\langle2hW,\Gamma\rangle}{1+h^{2}}
\end{equation}
Note the special values 
\begin{equation}
\begin{cases}
\boldsymbol{\theta}(0) & =\langle0,\Gamma\rangle\\
\boldsymbol{\theta}(1) & =\langle W,\Gamma/2\rangle\\
\boldsymbol{\theta}(\infty) & =\langle0,0\rangle
\end{cases}
\end{equation}
When $h=0$, the effective coupling between $\Phi$ and random strain
vanishes, leading to no renormalization of $\gamma$, so $\gamma(h)=\Gamma$.
When $h=1$, the effective random field width $w(h)$ reaches its
maximum value of $W$. Finally, as $h\to\infty$ the enhanced interaction
suppresses $\gamma$ and $w$ and the order parameter saturates to
unity in the fully ordered state corresponding to $w=\gamma=t=0$. 

Appropriate for the case of infinite-range interactions, we solve the RF-TFIM of Eq. (\ref{eq:renormalized_fullH})
at $T\ge0$, following the mean-field approach of \cite{yokota1988reentrant}. We set $\tau_{i}^{z}=\Phi+\delta\Phi_{i}$ where $\Phi=\langle\tau_{i}^{z}\rangle$
and $\delta\Phi_{i}=\tau_{i}^{z}-\langle\tau_{i}^{z}\rangle$, and
neglect the fluctuations, $\delta\Phi_{i}^{2}\approx0$. This
gives the mean field Hamiltonian
\begin{equation}
\mathcal{H}_{\text{MF}}/\tilde{J}=\frac{N}{2}\Phi^{2}-\sum_{i}[(\Phi+z_{i})\tau_{i}^{z}+\gamma\tau_{i}^{x}]
\end{equation}
We compute the partition function $\mathcal{Z}[z_{i}]$ by employing
the identity $\Tr e^{a\tau^{z}+b\tau^{x}}=2\cosh\sqrt{a^{2}+b^{2}}$
for real numbers $a$ and $b$, and we assume each disorder realization
$\{z_{i}\}$ to consist of independent and identically distributed
Gaussian random variables drawn from the distribution
\begin{equation}
p(z_{i})\equiv\frac{1}{\sqrt{2\pi}w}e^{-z_{i}^{2}/2w^{2}} \label{eq:pz}
\end{equation}
For each disorder realization, we compute the dimensionless free energy
density $f=-\frac{T}{\tilde{J}N}\log\mathcal{Z}[z_{i}]$
\begin{equation}
f(\Phi)=\frac{1}{2}\Phi^{2}-\frac{t}{N}\sum_{i}\log\Bigg[2\cosh\frac{\sqrt{(\Phi+z_{i})^{2}+\gamma^{2}}}{t}\Bigg]
\end{equation}
Assuming replica symmetry, we subsequently disorder-average $f\to\bar{f}$
over the joint distribution $\prod_{i}p(z_{i})$ and obtain the Landau
expansion
\begin{align}
\bar{f}(\Phi) & =\frac{1}{2}\Phi^{2}-\int_{-\infty}^{\infty}p(z-\Phi)\Lambda(z)\mathrm{d}z\label{eq:free_energy_exact}\\
 & =\bar{f}(0)+\frac{\mathcal{A}}{2}\Phi^{2}+\frac{\mathcal{U}}{4}\Phi^{4}+\frac{\mathcal{G}}{6}\Phi^{6}+...
\end{align}
In Eq. (\ref{eq:free_energy_exact}), we have defined the non-negative
function $\Lambda(z)$ which encapsulates all of the $t(h)$ and $\gamma(h)$
dependence of the problem
\begin{equation}
\Lambda(z)=t\log\Bigg[2\cosh\frac{\sqrt{z^{2}+\gamma^{2}}}{t}\Bigg]\label{eq:kernel}
\end{equation}
By expanding the integrand in powers of $\Phi$,
the quadratic and quartic Landau coefficients can be written compactly
as 
\begin{align}
\mathcal{A} & =1-\int_{-\infty}^{\infty}\Lambda(z)\partial_{z}^{2}p(z)\mathrm{d}z\label{eq:Landau_params1}\\
\mathcal{U} & =-\frac{1}{6}\int_{-\infty}^{\infty}\Lambda(z)\partial_{z}^{4}p(z)\mathrm{d}z \label{eq:Landau_params2}
\end{align}
Since $p(z)$ is Gaussian and since $\Lambda(z)$ grows algebraically,
total derivative terms vanish. Consequently, one can ``trade'' $z$
derivatives between $p(z)$ and $\Lambda(z)$ at will using integration-by-parts.
In the limit of zero temperature, $\Lambda$ simplifies dramatically:
\begin{equation}
\lim_{T\to0}\Lambda(z)=\lim_{t\to0}\Lambda(z)=\sqrt{z^{2}+\gamma^{2}}
\end{equation}
Analytical expressions for the quadratic and quartic coefficients
can be expressed using special functions: 
\begin{align}
\mathcal{A} & =1-\frac{U(1/2,0,2x)}{\sqrt{2}w}\\
\mathcal{U} & =\frac{xe^{x}}{3\sqrt{2\pi}w^{3}}[(1+4x)K_{1}(x)-(3+4x)K_{0}(x)]>0
\end{align}
where
\begin{equation}
x\equiv\frac{\gamma^{2}}{4w^{2}}=\frac{\Gamma^{2}}{4\lambda^{2}H_{z}^{2}\bar{\varepsilon}_{s}^{2}}
\end{equation}
and $U(a,b,z)$ is a confluent hypergeometric function of the second
kind, whereas $K_{n}(x)$ is a modified Bessel function of order $n$.
Since $\mathcal{U}$ is strictly positive whenever $\mathcal{A}\ge0$,
the transition
between AM and PM phases is second-order. Therefore, vanishing
of the quadratic Landau coefficient $\mathcal{A}=0$ gives the phase
boundary. 

\begin{figure}[H]

\includegraphics[width=0.8\columnwidth]{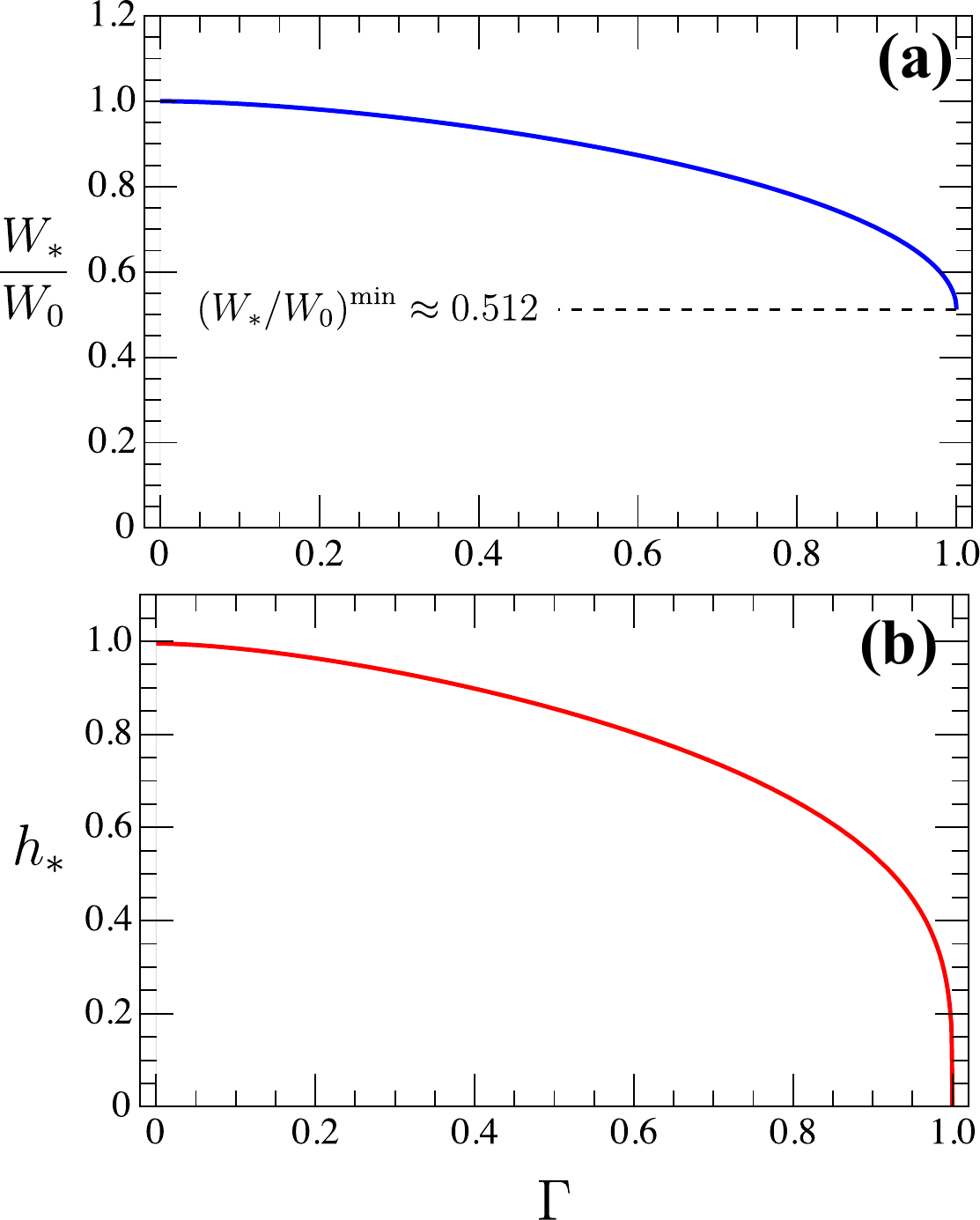}\caption{\label{fig:T=00003D0_reentrance_thresholdscales} \textbf{(a) } Threshold disorder strength $W_{*}/W_{0}$ and \textbf{(b) } corresponding magnetic field value $h_{*}$ as functions of $\Gamma$ at $T=0$. $W_{*}/W_{0}$
decreases closer to the QCP $\Gamma=1$ but remains of order one,
whereas $h_{*}$ vanishes rapidly. This means that AM reentrance and
order parameter non-monotonicity as a function of $h$ are favored
when starting close to the zero magnetic field critical point. }
\label{fig:Wstar_hstar}
\end{figure}

In the case of the classical RFIM with $\Gamma=0$ the quadratic Landau
coefficient takes the form
\begin{equation}
\mathcal{A}=1-\frac{U(1/2,0,0)}{\sqrt{2}w}=1-\frac{\sqrt{2/\pi}}{w}
\end{equation}
By solving $\mathcal{A}=0$ for $w$, we find the well-known result \cite{schneider1977_RF}
for the critical disorder strength
\begin{equation}
w_{c}=W_{0}\equiv\sqrt{2/\pi}\approx0.798
\end{equation}
When $W>W_{0}$, AM reentrance occurs via a paramagnetic phase bounded by two field values given by  $w(h_{c}^{\pm})=W_{0}$:
\begin{equation}
h_{c}^{\pm}=\frac{W_{0}}{W}\Big[1\pm\sqrt{1-(W_{0}/W)^{2}}\Big]\label{eq:reentrance_hc_T_Gamma=00003D0}
\end{equation}
For the case of nonzero $\Gamma$, the equation for the phase boundary
$\mathcal{A}=0$ yields a critical disorder strength that decreases monotonically with
$\Gamma$. We solve this equation and obtain the function $w_{c}(\gamma,t)$
restricted to the $t=0$ plane, i.e., $w_{c}(\gamma,0)$. We give the
asymptotic behaviors near the clean QCP and classical disorder-induced
transitions, respectively:
\begin{equation}
w_{c}(\gamma,0)\approx\begin{cases}
\sqrt{2/3}\sqrt{1-\gamma} & \text{for }\gamma\to1\\
W_{0}\left(1+\frac{\pi}{4}\gamma^{2}\log\gamma\right) & \text{for }\gamma\to0
\end{cases}
\end{equation}
For fixed $\Gamma$, we may also obtain the minimum $W$ required
to achieve reentrance. This minimum $W$, which we call $W_{*}$,
is given by the condition of tangency between the $h$-parametrized
trajectory $\boldsymbol{\theta}(W,\Gamma;h)$ and $\mathcal{A}=0$:
\begin{equation}
\begin{cases}
\mathcal{A}(\boldsymbol{\theta}(h))=0\\
\partial_{h}\mathcal{A}(\boldsymbol{\theta}(h))=0
\end{cases}
\end{equation}
Solving these two equations yields $W_{*}$ and the magnetic field
scale $h_{*}$ at which tangency occurs as explicit functions of $\Gamma$.
Note that $W_{*}(\Gamma)$ and $h_{*}(\Gamma)$ are related via $w_{c}$
as an identity through the functional equation
\begin{equation}
\frac{2h_{*}(\Gamma)}{1+h_{*}^{2}(\Gamma)}W_{*}(\Gamma)=w_{c}\Big(\frac{\Gamma}{1+h_{*}^{2}(\Gamma)},0\Big)
\end{equation}
We plot $W_{*}(\Gamma)$ and $h_{*}(\Gamma)$ in Fig. \ref{fig:T=00003D0_reentrance_thresholdscales}a-b,
showing that the threshold disorder strength $W_{*}(\Gamma)$ and threshold field $h_{*}(\Gamma)$
fall monotonically with increasing $\Gamma$. The values at the
classical limit $\Gamma=0$ and at the QCP $\Gamma=1$ are, respectively
\begin{equation}
\begin{cases}
W_{*}(0)=W_{0}=\sqrt{2/\pi}\approx0.798\\
h_{*}(0)=1\\
W_{*}(1)=1/\sqrt{6}\approx0.408\\
h_{*}(1)=0
\end{cases}
\end{equation}
Interestingly, at $\Gamma=1$, $W_{*}$ is not zero, and instead approaches
a minimum value of $W_{*}^{\mathrm{min}} = 1/\sqrt{6}\approx0.408$. This implies that from
the definition of $W$, reentrance requires the energy scale of the random strain
$E_{\text{el}}=\frac{1}{2}C_{0}\bar{\varepsilon}_{s}^{2}$  to be of order
$J$ even at the QCP, where quantum disordering effects are the strongest.
Conversely, the magnetic fields at which reentrance occurs are suppressed
rapidly close to the QCP because $h_{*}\to0$ as $\Gamma\to1$.

The key advantage of representing the zero temperature phase diagram
in the scaled coordinates $\gamma$ and $w$, as in Fig. \ref{fig:2D_T=00003D0_PDs}(a), is that it does not explicitly
depend on $W$ and $\Gamma$ and only does so implicitly through $w(h)$
and $\gamma(h)$. The drawback is that different $W$ correspond to
differently curved phase trajectories $\boldsymbol{\theta}$. Thus,
this representation may not be the most obvious physically motivated
one and could obscure global features of the AM-PM phase boundary
without a direct quantitative calculation. In some cases, it is more
revealing to construct the zero-temperature phase diagram in coordinates
for which the experimentally tunable quantity $h$ appears as an independent
axis. This ensures that tuning magnetic field amounts to following
a straight line, rather than a curve, for a fixed value of $\Gamma$
and $W$, with the drawback being that the phase boundary in these
new coordinates depends explicitly on $W$. 

We plot these $(\Gamma,h)$ phase diagrams for increasing values of fixed $W$ in Fig. \ref{fig:2D_T=00003D0_PDs}(b)-(e). In panels (b) and (c),  $W<W_0$, implying that the phenomenon of reentrant AM does not occur in the classical regime of $\Gamma=0$.  Nevertheless, there is an important distinction between these two cases: in panel (b), because $W<W_*^{\mathrm{min}}<W_0$, reentrant AM does not happen anywhere in the phase diagram -- recall that $W_{*}^{\mathrm{min}} = 1/\sqrt{6}\approx0.512 W_0$. In this case, all that the magnetic field can do is tune the system from the PM to the AM phase, but it cannot tune the AM transition to zero.  In contrast, in panel (c), $W_*^{\mathrm{min}}<W<W_0$, and the PM-AM phase boundary shows a ``bulge'' signaling a reentrant AM phase. This bulge first develops at $(\Gamma_*,h_*)=(1,0)$
when $W=W_*^{\mathrm{min}}$ and moves rapidly upward and to the left as $W$ increases further toward  $\Gamma_{*}\rightarrow0$
and $h_{*}\rightarrow1$. When $W=W_{0}=\sqrt{2/\pi}$, which is the case shown in panel (d), the tip of the
bulge annihilates at the $\Gamma=0$ axis, resulting in a non-analytic behavior of the free energy, as we discuss in Section IV. Finally, when $W>W_0$, illustrated in panel (e), reentrance occurs across the entire phase diagram between the critical fields
$h_{c}^{\pm}$ satisfying Eq. (\ref{eq:reentrance_hc_T_Gamma=00003D0}).  The combination of these panels gives the schematic three-dimensional phase diagram of Fig. \ref{fig:T0_phase_diagram}.

\begin{figure}
\centering{}\includegraphics[width=0.75\columnwidth]{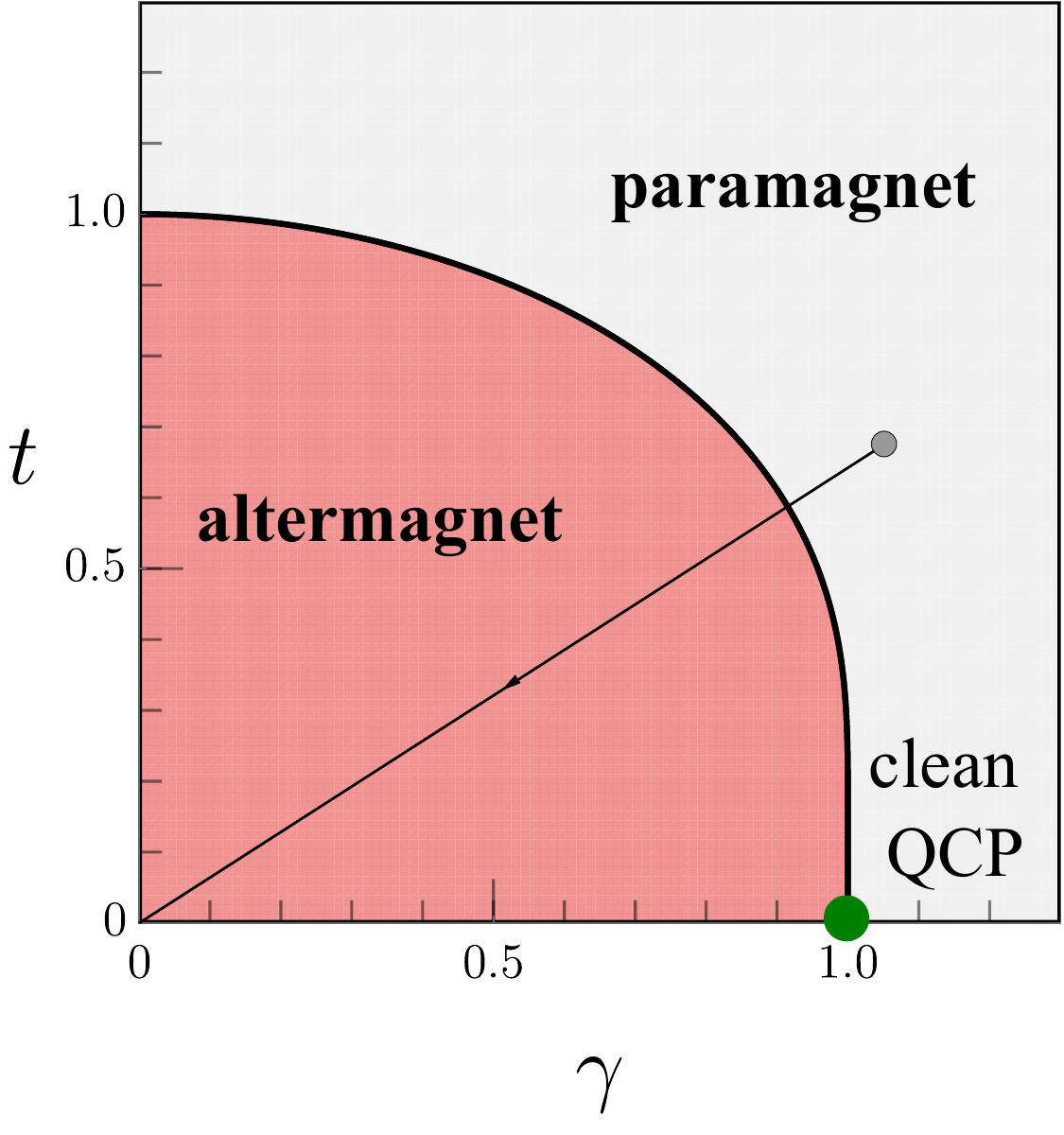}\caption{\label{fig:2D_scaled_phasediagram_W=00003D0}Mean-field TFIM
phase diagram, corresponding to $W=0$ in our case, plotted against scaled axes
$t\equiv T/(1+h^{2})$ and $\gamma\equiv\Gamma/(1+h^{2})$. The scaled
critical temperature $t_{c}=\gamma/\tanh^{-1}\gamma$ decreases with increasing $\gamma$, vanishing
at a quantum critical point at $\gamma=1$ (in green). Increasing
$h\equiv H_{z}/H_\lambda$ amounts to following a straight-line
trajectory from $(\gamma,t)=(\Gamma,T)$ to the origin $(\gamma,t)=(0,0)$.
The system orders monotonically due to the coupling to elastic fluctuations
but no coupling to random strain.}
\end{figure}

\subsection{Finite temperature phase diagram}

\begin{figure*}[t]
\centering{}\includegraphics[width=0.8\paperwidth]{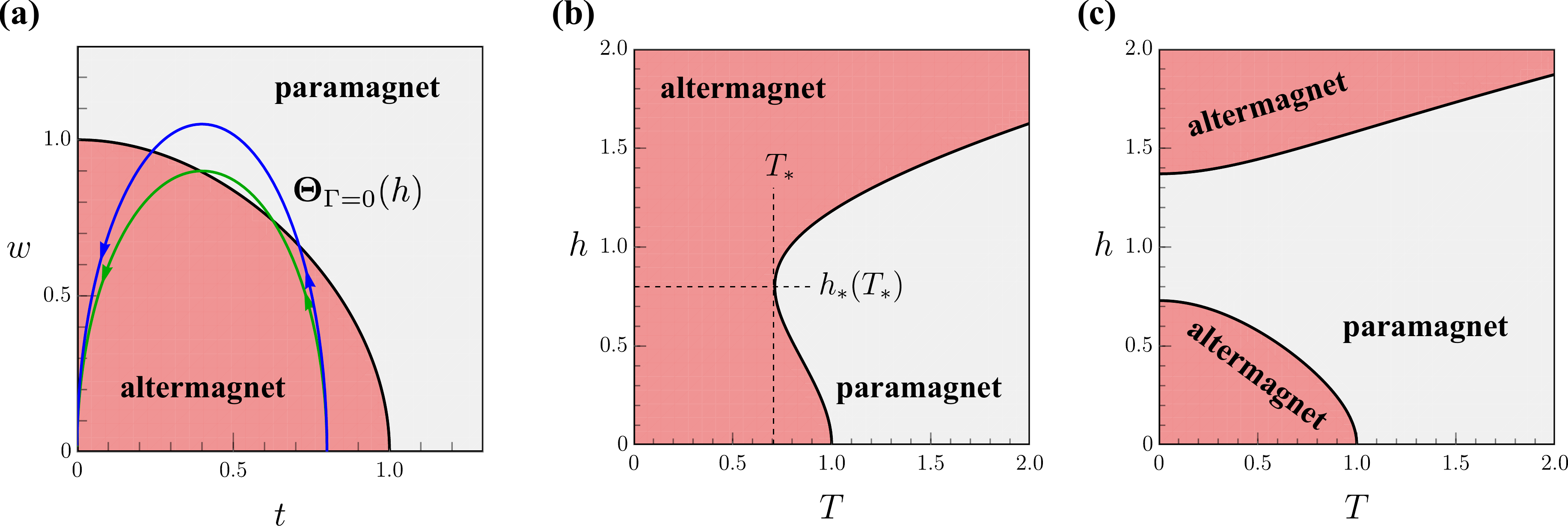}\caption{\textbf{(a)}: Classical (i.e., $\Gamma=0$) phase diagram with scaled axes $t\equiv T/(1+h^{2})$
and $w\equiv2Wh/(1+h^{2})$, where $T=0.8J$ sets the starting point of the phase
trajectory along the $t$-axis and $W$ sets the maximum extent
of the trajectory along the $w$-axis. Green and blue curves correspond to $W=0.9W_0$ and $W=1.05W_0$, respectively. Panels \textbf{(b) }and \textbf{(c) }show the phase diagrams with
unscaled axes $h=H_{z}/H_\lambda$ and $T$ corresponding
to $W=0.9W_0$ and $1.05W_0$, respectively. As
the relative disorder $W$ is increased, the paramagnetic region grows
via the formation of a bulge in the AM-PM transition line at $(T_{*},h_{*})$, which annihilates at $(0,1)$ when $W=W_0=\sqrt{2/\pi}\approx0.798$.} \label{fig:Gamma0_phase_diagram}
\end{figure*}

We now construct the finite temperature phase diagram in the
four dimensional parameter space spanned by ($h,W,\Gamma,T$). As
with the $T=0$ case, we represent the phase diagram in two different
coordinate systems. The first is a 3D phase diagram with $h$-dependent
coordinates ($w,\gamma,t$), where a changing magnetic field amounts
to following a trajectory $\boldsymbol{\Theta}(h)$ in this space:
\begin{equation}
\boldsymbol{\Theta}(W,\Gamma,T;h)=\langle w(h),\gamma(h),t(h)\rangle=\frac{\langle2hW,\Gamma,T\rangle}{1+h^{2}}
\end{equation}
Note the special values 
\begin{equation}
\begin{cases}
\boldsymbol{\Theta}(0) & =\langle0,\Gamma,T\rangle\\
\boldsymbol{\Theta}(1) & =\langle W,\Gamma/2,T/2\rangle\\
\boldsymbol{\Theta}(\infty) & =\langle0,0,0\rangle
\end{cases}
\end{equation}
As in the case of zero temperature, $W$ represents the maximum extent
of this trajectory along the $w$-axis.

Before going further, we consider two limiting cases for $T\ge0$:
the pure \emph{transverse field Ising model }TFIM with $W=0$ and the
classical RFIM with $\Gamma=0$. In both cases, $T_{c}$ decreases
monotonically with increasing $\Gamma$ and $W$, resulting in QCPs
at $\Gamma=1$ and $W=W_{0}$, respectively. These two QCPs are connected
by a line of QCPs (Fig. \ref{fig:2D_T=00003D0_PDs}a) whose 
shape on the ($\Gamma,W$)-plane depends on $h$. 

When treating the pure TFIM, $p(z_{i})$ in Eq. (\ref{eq:pz}) becomes a delta function,
and the equation $\mathcal{A}=0$ for the phase boundary simplifies
to 
\begin{equation}
1=\Lambda''(0)=\frac{\tanh(\gamma/t)}{\gamma}=(1+h^{2})\frac{\tanh(\Gamma/T)}{\Gamma}
\end{equation}
This yields the equation on $T_{c}$:
\begin{equation}
T_{c}(\Gamma,h)=\frac{\Gamma}{\tanh^{-1}\frac{\Gamma}{1+h^{2}}}\label{eq:Tc_vs_Gamma_W=00003D0}
\end{equation}
or, equivalently, on $t_{c}=\frac{T_{c}}{1+h^{2}}$: 
\begin{equation}
t_{c}(\gamma)=\frac{\gamma}{\tanh^{-1}\gamma}\label{eq:tc_vs_gamma_W=00003D0}
\end{equation}
In this case without disorder, $\Phi$ couples only to elastic fluctuations, and
increasing the magnetic field can only further order the system. This
agrees with Eq. (\ref{eq:Tc_vs_Gamma_W=00003D0}), from which we see
that increasing $h$ suppresses the denominator and enhances $T_{c}$.
The TFIM phase diagram represented in the ($t,\gamma$)-plane gives
a convex curve -- shown in Fig. \ref{fig:2D_scaled_phasediagram_W=00003D0}
-- with the $h$-parameterized trajectory $\boldsymbol{\Theta}(h)$
being a straight line connecting the initial point to the origin:
\begin{equation}
\boldsymbol{\Theta}(0,\Gamma,T;h)=\frac{\langle\Gamma,T\rangle}{1+h^{2}}
\end{equation}
Therefore, in the clean case, the magnetic field can be used to tune the system from the PM phase to the AM phase, but not the other way around.

We now consider the classical RFIM case, for which $\Gamma=0$. In
this limit, $\Lambda(z)$ simplifies, such that $\mathcal{A}=0$ gives
\begin{equation}
t=\int_{-\infty}^{\infty}\,p(z)\sech^{2}(z/t)\mathrm{d}z
\end{equation}
Solving for $w$ yields the function $w_{c}(\gamma,t)$ along the
$\gamma=0$ plane which has asymptotic behaviors given by
\begin{equation}
w_{c}(0,t)\approx\begin{cases}
\sqrt{1-t} & \text{for }t\to1\\
W_{0}\big(1-\frac{\pi^{2}}{24}t^{2}\big) & \text{for }t\to0
\end{cases}
\end{equation}
Temperature and transverse field take qualitatively the same role.
Both suppress $\Phi$ and both are renormalized in the same way due
to elastic fluctuations. Therefore, we can obtain $W_{*}(T)$ analogously
to how we obtained $W_{*}(\Gamma)$ in the previous section by determining the
point of tangency on the ($w,t$)-plane, as shown in Fig. \ref{fig:Gamma0_phase_diagram}(a): 
\begin{equation}
\begin{cases}
\mathcal{A}(\boldsymbol{\Theta}_{\Gamma=0}(h))=0\\
\partial_{h}\mathcal{A}(\boldsymbol{\Theta}_{\Gamma=0}(h))=0
\end{cases}
\end{equation}
Solving these two equations yields $W_{*}$ and the magnetic field
scale at which tangency occurs, $h_{*}$, as explicit functions of $T$.
Note that $W_{*}(T)$ and $h_{*}(T)$ are related via $w_{c}$ through
the functional equation
\begin{equation}
\frac{2h_{*}(T)}{1+h_{*}^{2}(T)}W_{*}(T)=w_{c}\Big(0,\frac{T}{1+h_{*}^{2}(T)}\Big)
\end{equation}
As expected, the threshold scale falls monotonically with increasing
$T$:
\begin{equation}
\begin{cases}
W_{*}(0)=W_{0}=\sqrt{2/\pi}\approx0.798\\
h_{*}(0)=1\\
W_{*}(1)=1/2\\
h_{*}(1)=0
\end{cases}
\end{equation}
Fig. \ref{fig:Gamma0_phase_diagram}(b)-(c) show two phase diagrams in unscaled coordinates $h$ and $T$ for representative values of disorder strength $W$.

The so far three convex phase boundaries that we have uncovered in
the ($w,\gamma$)-, ($\gamma,t$)-, and ($w,t$)- planes suggest
a convex critical surface in the ($w,\gamma,t$) parameter space.
As with the 2D phase diagrams represented with $h$-dependent axes,
the phase trajectory $\boldsymbol{\Theta}$ may avoid, intersect,
or run tangent to this convex surface depending on the starting parameters
$W$, $\Gamma$, and $T$. 

Similarly to the $T=0$ and $\Gamma=0$ cases, we apply the condition
for tangency on the whole ($w,\gamma,t$) parameter space, namely,
\begin{equation}
\begin{cases}
\mathcal{A}(\boldsymbol{\Theta}(h))=0\\
\partial_{h}\mathcal{A}(\boldsymbol{\Theta}(h))=0
\end{cases}
\end{equation}
This yields the threshold values $W_{*}$ and $h_{*}$ as functions
of $\Gamma$ and $T$, which, at finite temperatures, have a qualitatively similar behavior than the $T=0$ case shown in Fig. \ref{fig:Wstar_hstar}. We note that $W_{*}(\Gamma,T)$
remains of order one and is never suppressed to zero, suggesting that
$E_{\text{el}}$ must be of order $J$ for reentrance to occur,
like in the zero temperature case. However, we stress that $h_{*}$
can be made arbitrarily small depending on how close ($\Gamma,T$)
is to the transition when $H_{z}=0$. Generalizing from the two-dimensional
cases, $h_{*}(\Gamma,T)$ and $W_{*}(\Gamma,T)$ are related through
the functional equation 
\begin{equation}
\frac{2h_{*}}{1+h_{*}^{2}}W_{*}=w_{c}\Big(\frac{\Gamma}{1+h_{*}^{2}},\frac{T}{1+h_{*}^{2}}\Big)
\end{equation}

\begin{figure}[H]

\includegraphics[width=0.95\columnwidth]{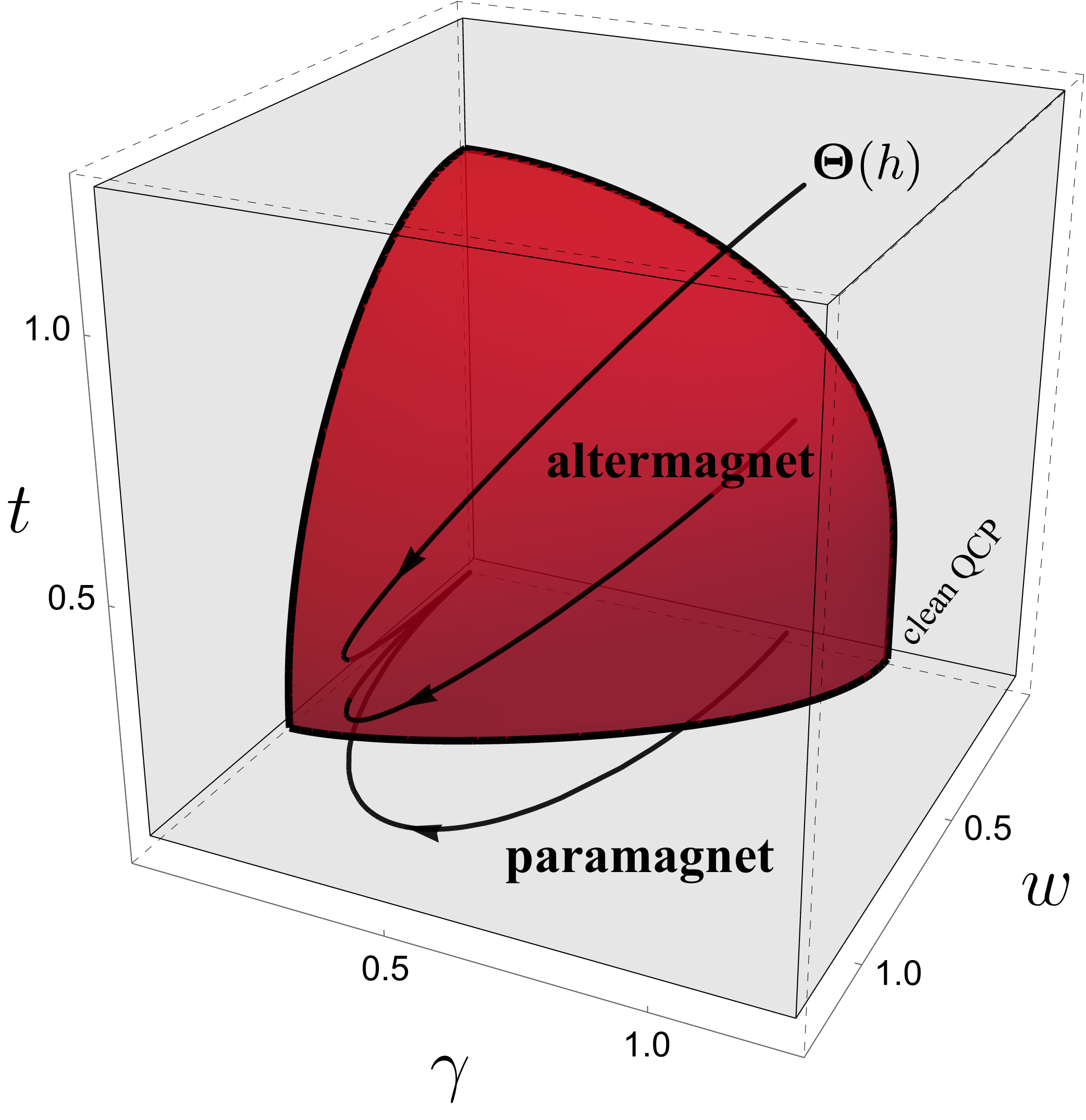}\caption{Mean-field RF-TFIM phase diagram of an inhomogeneous $d$-wave altermagnet in ($w,\gamma,t$)-space. Shown as gray directed curves are three $H_z$-parameterized trajectories $\bf{\Theta}$ starting at fixed ($W,\Gamma$)=($J,0.75J$) for $T=0$, $0.5J$, and $J$.} \label{fig:3D_scaled_phasediagram}
\end{figure}

\begin{figure*}
\centering{}\includegraphics[width=0.8\paperwidth]{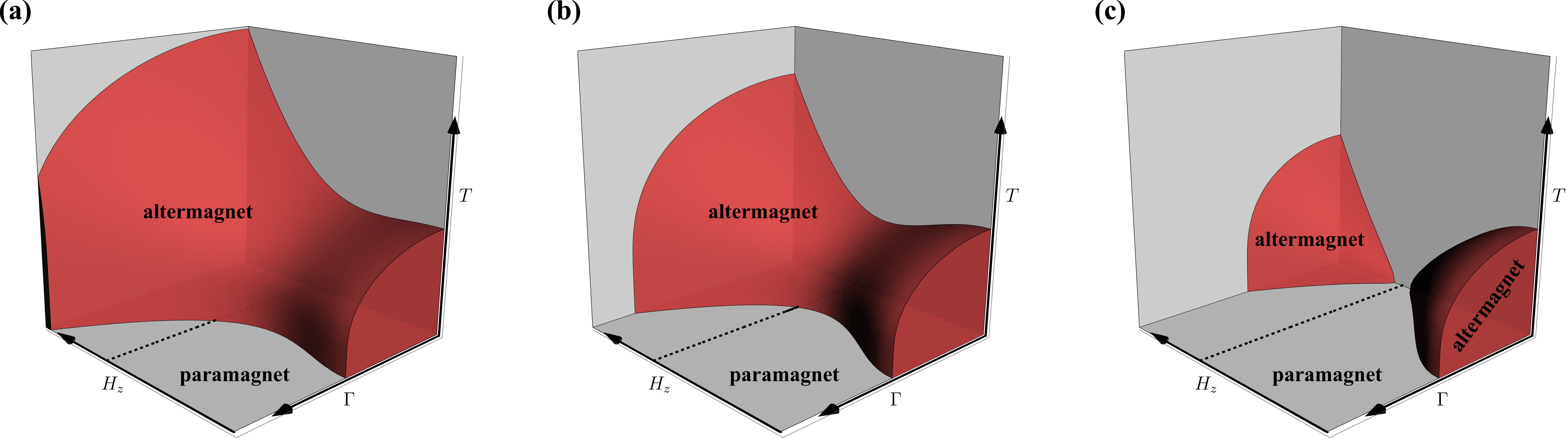}\caption{\label{fig:3D_unscaled}Mean-field phase diagram of an inhomogeneous $d$-wave altermagnet as a function of magnetic field $H_{z}$, transverse field $\Gamma$ (responsible for quantum fluctuations), and temperature $T$ for different values
of relative disorder strength $W=0.6J$ \textbf{(a)},
$W=0.7J$ \textbf{(b)}, and $W=0.8J>W_0$ \textbf{(c)}.
For $W>W_0$, the single AM phase region splits into two at $(H_{z},\Gamma,T)=(H_\lambda,0,0)$
(dashed line).} \label{fig:T_Gamma_H_phase_diagrams}
\end{figure*}

In Fig. \ref{fig:3D_scaled_phasediagram}, we plot the critical surface in these scaled coordinates
and show several phase trajectories. As with the previous section,
we also display the critical surface in the ($h,\Gamma,T$) parameter
space for different values of $W$. In Fig. \ref{fig:T_Gamma_H_phase_diagrams}, we see that when
\begin{equation}
W>\max_{\Gamma,T}W_{*}(\Gamma,T)=W_{0}
\end{equation}
the single altermagnetic phase region separates at $(h,\Gamma,T)=(1,0,0)$
into two disconnected AM phase regions, one at low fields $h<1$ and one
at high fields $h>1$. In this $W$ regime, altermagnetic reentrance
is guaranteed to occur even at $T=0$.

\section{Experimental signatures}

The phase diagrams shown in Fig. \ref{fig:T_Gamma_H_phase_diagrams} reveal that, in an inhomogeneous altermagnet, application of a magnetic field can induce both a PM-AM transition, when the disorder strength is weak, as well as an AM-PM-AM transition with reentrant altermagnetic order, when the disorder strength is strong. In this section, we go beyond the determination of the phase transition boundaries and discuss the behavior of three experimentally observable quantities as the phase boundaries are traversed by a magnetic field: the altermagnetic order parameter, the shear modulus, and the elasto-caloric effect coefficient. Note that, in principle, the AM order parameter could be measured in momentum space from the spin splitting of the band structure, or from extracting response functions that are proportional to the order parameter, such as the piezomagnetic and the elasto-Hall conductivity tensors \cite{Takahashi2025}. 

\subsection{Altermagnetic order parameter}

We first consider ($\Gamma,T$) values that place the system inside the AM ordered phase in the absence of a magnetic field. We also specialize to the case of $W<w_{c}(\Gamma,T)$
for which long-range AM order remains intact for all values of $h$. Although an AM reentrant behavior does not take place, we show that $\Phi(h)$ displays a non-monotonic behavior and a minimum value $\Phi_{\text{min}}$ for
$h=h_{\text{min}}$.

The mean field equation $\partial\bar{f}/\partial\Phi=0$ gives a
self-consistent condition for the order parameter:
\begin{align}
\Phi & =\int_{-\infty}^{\infty}p(z-\Phi)\partial_{z}\Lambda(z)\mathrm{d}z\label{eq:phi_selfconsistent}
\end{align}
where we recall that $p(z)$ is a normal distribution with width $w=w(h)$
and 
\begin{equation}
\partial_{z}\Lambda(z)=\frac{z}{\sqrt{z^{2}+\gamma^{2}}}\tanh\frac{\sqrt{z^{2}+\gamma^{2}}}{t}
\end{equation}
In the limit of $h=0$, for which $(w,\gamma,t)=(0,\Gamma,T)$, a nonzero
$\Phi$ satisfies the equation 
\begin{equation}
\sqrt{\Phi^{2}+\Gamma^{2}}=\tanh\frac{\sqrt{\Phi^{2}+\Gamma^{2}}}{T}
\end{equation}
for which we establish that the argument $\sqrt{\Phi^{2}+\Gamma^{2}}$
is less than unity due to the concavity of the hyperbolic tangent
function. Conversely, the limit $h\to\infty$ yields a maximally ordered
state for which $(w,\gamma,t)=(0,0,0)$. In this case, the self-consistent
equation simply yields $|\Phi|=1$, i.e., the order parameter saturates
to its maximum value. Finally, because $w(h)\propto h$ and $t(h),\gamma(h)\approx\text{const}$
for small $h$, increasing $h$ amounts to merely increasing the effective
random field width $w$, which can only suppress $\Phi$. These three
facts together, namely that $\Phi(h=0)\le1$, that $\Phi(h)$ decreases
for small $h$, and that $\Phi(h=\infty)=1$, establishes that $\Phi$
must have a minimum at some magnetic field value $h=h_{\text{min}}$. It is straightforward
to conclude that $h_{\text{min}}$ must be less than $1$, because
an upturn in $\Phi$ can only happen when a nontrivial competition
between the ordering and disordering tendencies occurs, i.e., when
the effective energy scales $w(h)$, $\gamma(h)$, and $t(h)$ are
not all decreasing, which can only happen if $h<1$. Generally, we
can then write
\begin{equation}
\Phi=\Phi_{\text{min}}+\frac{1}{2}\Upsilon(h-h_{\text{min}})^{2}+...\label{eq:phi_quadraticminimum}
\end{equation}
where the combination of the self-consistent equation and the condition $d\Phi/dh=0$
yields a system of equations that give $h_{\text{min}}=h_{\text{min}}(W,\Gamma,T)$
and $\Phi_{\text{min}}=\Phi_{\text{min}}(W,\Gamma,T)$. We stress,
however, that in the case of $\Phi=\mathcal{O}(1)$ deep inside the
ordered phase, the necessity of minimizing the infinite-order expression
for $\bar{f}(\Phi)$ yields a $\Phi_{\text{min}}=\Phi(h=h_{\text{min}})$
that is exponentially close to $\Phi(h=0)$, i.e.,
\begin{equation}
|\Phi(h=0)-\Phi(h=h_{\text{min}})|\sim e^{-1/W^{2}h_{\text{min}}^{2}}
\end{equation}

\begin{figure}[H]

\includegraphics[width=0.95\columnwidth]{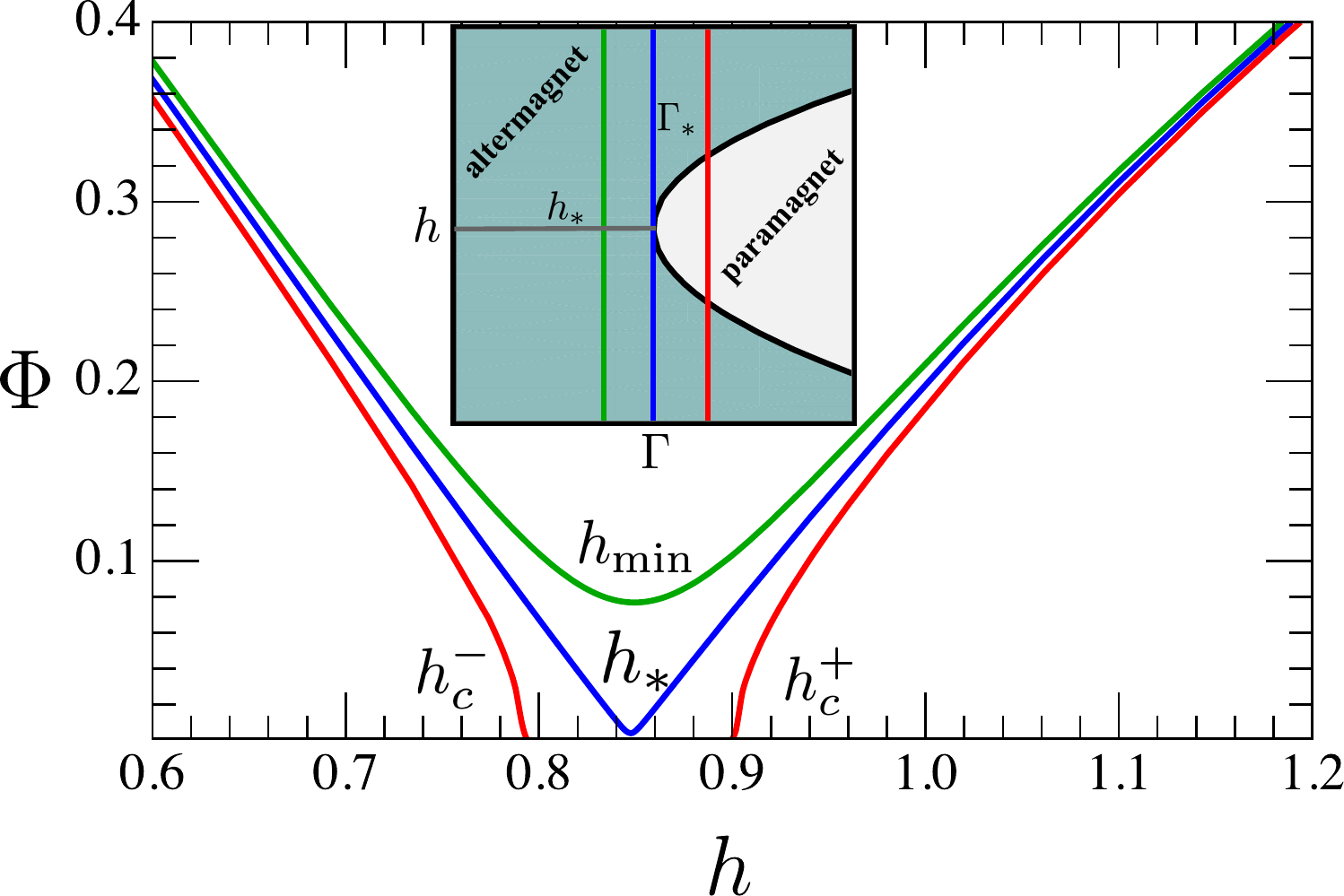}\caption{The AM order parameter $\Phi$ plotted as a function of the magnetic field $h$
at $T=0$ and $W=0.9W_0$ for $\Gamma=\Gamma_{*}(1-10^{-2})$ (green),
$\Gamma=\Gamma_{*}$ (blue), and $\Gamma=\Gamma_{*}(1+10^{-2})$ (red),
where $\Gamma_{*}\approx0.526$. The associated phase diagram (inset) shows the corresponding line cuts in the magnified $T=0$ plane. For $\Gamma<\Gamma_{*}$, the order
parameter displays an analytic local minimum at $h=h_{\text{min}}$,
crossing over between quadratic (near $h_{\text{min}}$) and linear
(away from $h_{\text{min}}$) regimes. For $\Gamma=\Gamma_{*}$, $h_{\text{min}}$
approaches $h_{*}\approx0.842$ at which point $\Phi$ vanishes, and
the quadratic regime is suppressed completely. For $\Gamma>\Gamma_{*}$,
$\Phi$ is singular at the two critical fields $h_{c}^{\pm}$, crossing over
between square-root (near $h_{c}^{\pm}$) and linear (away from $h_{c}^{\pm}$)
regimes.} \label{fig:OP_zoom}
\end{figure}

In light of this fact, it is convenient to focus on the case where the parameter values place the system close to the AM-PM phase boundary, where $\bar{f}$ may be truncated
to quartic order in $\Phi$. To proceed, we consider phase diagrams like those of Fig. \ref{fig:T_Gamma_H_phase_diagrams}(a)-(b), corresponding to  $W<W_0$ (weak disorder). In these cases, reentrant behavior as a function of $h$ can be obtained by tuning the value of $\Gamma$. For concreteness, we focus on the $T=0$ cut of the phase diagram in the vicinity of the point $(\Gamma_*, h_*) $ that marks the end of the bulge of the PM-AM phase boundary, as shown in the inset of Fig. \ref{fig:OP_zoom}.

To obtain the form of $\Phi(h)$ as the PM-AM phase boundary is traversed upon increasing $\Gamma$, we use the phenomenological Landau coefficients. Since $\Phi$
is small, we can approximate $\mathcal{U}$ to be a constant
and equal to $\mathcal{U}_{0}=\mathcal{U}(h=h_{\text{min}})$. Then, because $\mathcal{A}$ is analytic
in $h$ and closest to zero when $h=h_{\text{min}}$, we have the
approximation 
\begin{equation}
\mathcal{A}=-\mathcal{A}_{0}-\mathcal{A}_{2}(h-h_{\text{min}})^{2}
\end{equation}
where $\mathcal{A}_{0},\mathcal{A}_{2}>0$ are assumed to be $h$-independent
constants. This yields 
\begin{equation}
\Phi=\sqrt{\frac{|\mathcal{A}|}{\mathcal{U}}}\approx\sqrt{\frac{\mathcal{A}_{0}+\mathcal{A}_{2}(h-h_{\text{min}})^{2}}{\mathcal{U}_{0}}}
\end{equation}
which gives a crossover field scale
\begin{equation}
h_{t}\equiv\sqrt{\frac{\mathcal{A}_{0}}{\mathcal{A}_{2}}}
\end{equation}
that distinguishes between a quadratic and a linear dependence of the AM order parameter with respect to the distance to $h_{\mathrm{min}}$:
\begin{equation}
\Phi(h)\propto\begin{cases}
h_{t}\Big[1+\frac{(h-h_{\text{min}})^{2}}{2h_{t}^{2}}\Big] & \text{for }|h-h_{\text{min}}|\ll h_{t}\\
|h-h_{\text{min}}| & \text{for }h_{t}\ll|h-h_{\text{min}}|\ll1
\end{cases}
\end{equation}
If the tuning parameter $\Gamma$ is increased, we eventually encounter
the case in which the PM-AM phase boundary is touched tangentially. In this case, $\mathcal{A}_0 \rightarrow 0$ implying that $\Phi_{\min}\to0$ and
that the crossover scale $h_{t}$ vanishes, leaving a singular cusp-like
behavior in the $h$-dependence of the order parameter for which long-range
order is lost at precisely one value of magnetic field $h_{*}$:
\begin{equation}
\Phi\propto|h-h_{*}|\label{eq:phi_cusp}
\end{equation}
Note that, in the last step, we renamed $h_{\text{min}}$ to $h_{*}$ in accordance with the
notation of Sec. III. These two behaviors for $\Phi(h)$, corresponding to a quadratic minimum and to a cusp are illustrated by the green and blue lines in Fig. \ref{fig:OP_zoom} . 

\begin{figure*}
\begin{raggedright}
\includegraphics[width=0.85\paperwidth]{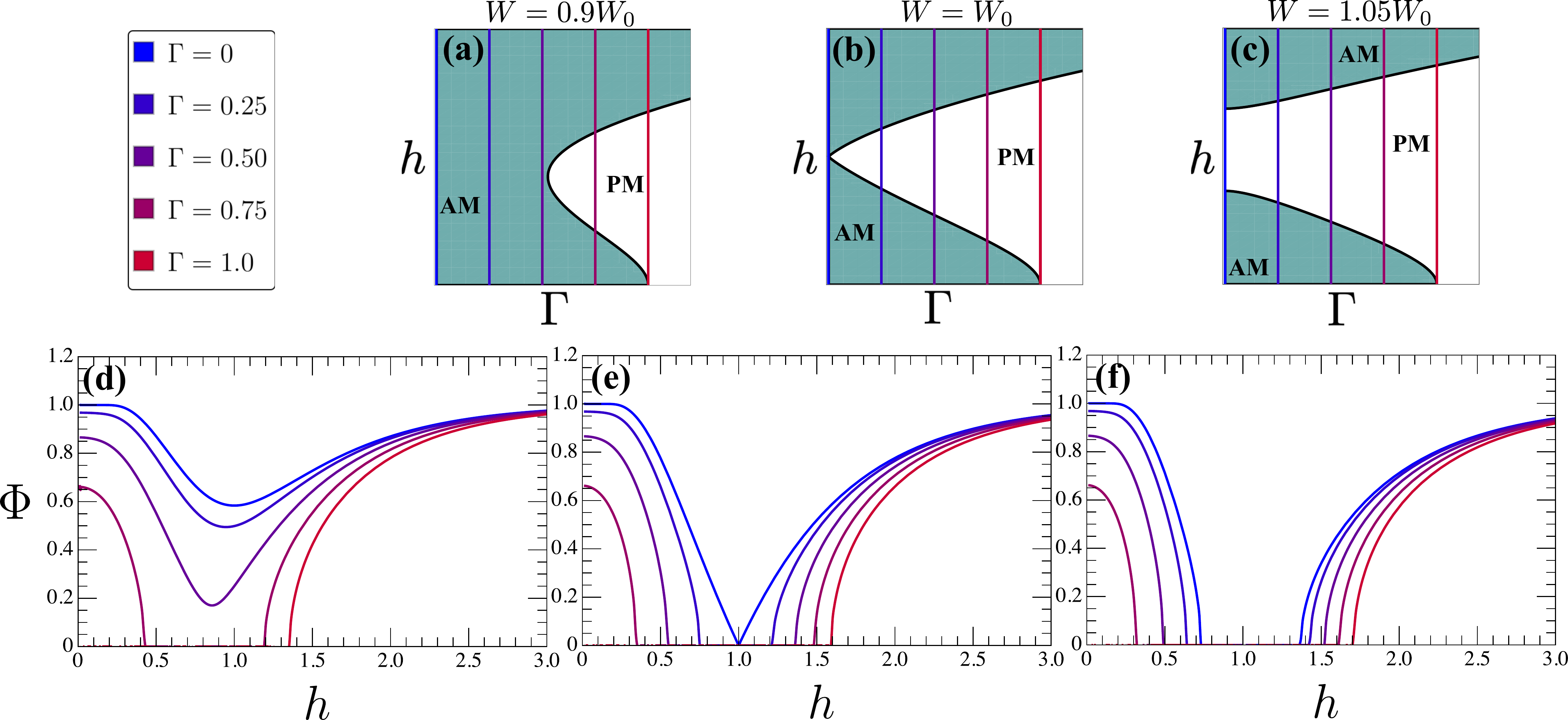}\caption{The zero-temperature mean-field order parameter as a function of $h=H_{z}/H_\lambda$
for different values of $\Gamma$ given in the legend, and $W=0.9W_0$
\textbf{(d)}, $W=W_0$ \textbf{(e)}, and $W=1.05W_0$ \textbf{(f)}. The corresponding cuts in the $(h,\Gamma)$-plane are given in panels \textbf{(a)}, \textbf{(b)}, and \textbf{(c)}, respectively. For
weak disorder, the order parameter behaves non-monotonically, whereas
for strong disorder it exhibits reentrance.} \label{fig:OP_all}
\par\end{raggedright}
\end{figure*}

Upon increasing $\Gamma$ further, the PM-AM phase boundary is intersected twice at the two critical fields $h_{c}^{\pm}$, signaling
the loss of long-range order over a finite range $h\in[h_{c}^{-},h_{c}^{+}]$.
In the limit of being near the intersection point, i.e., for $h\to h_{c}^{-}$
or $h\to h_{c}^{+}$, the phase trajectory $\boldsymbol{\Theta}(h)$
has a nonzero component normal to the phase boundary, and hence the
distance to the critical surface scales linearly with $|h-h_{c}^{\pm}|$.
For $\Phi(h)$, this translates to a ``splitting'' of the cusp at
$h_{*}$ into two successive square-root singularities at $h_{c}^{-}$
and $h_{c}^{+}$. 

We use our Landau theory to investigate what happens within the ordered
phase when $h_{c}^{-}$ and $h_{c}^{+}$ are sufficiently close together.
Within the ordered phase, but near $h_{c}^{\pm}$, we obtain for the
quadratic coefficient 
\begin{equation}
\mathcal{A}=-\mathcal{A}_{0}'(h-h_{c}^{-})(h-h_{c}^{+})
\end{equation}
where $\mathcal{A}_{0}'>0$ such that $h\in[h_{c}^{-},h_{c}^{+}]$
yields $\mathcal{A}>0$ and $h\notin[h_{c}^{-},h_{c}^{+}]$ yields
$\mathcal{A}<0$. Using $\Phi\sim\sqrt{-\mathcal{A}/\mathcal{U}}$
we obtain a crossover regime governed by 
\begin{equation}
h_{t}'\equiv h_{c}^{+}-h_{c}^{-}
\end{equation}
that distinguishes between square-root and linear behaviors:
\begin{equation}
\Phi(h)\propto\begin{cases}
\sqrt{h_{t}'}\sqrt{h_{c}^{-}-h} & \text{for }0<h_{c}^{-}-h\ll h_{t}'\\
\sqrt{h_{t}'}\sqrt{h-h_{c}^{+}} & \text{for }0<h-h_{c}^{+}\ll h_{t}'\\
|h-(h_{c}^{+}+h_{c}^{-})/2| & \text{for }h_{t}'\ll|h-h_{c}^{\pm}|\ll1
\end{cases}\label{eq:reentrant_crossover}
\end{equation}
This behavior is illustrated by the red line in Fig. \ref{fig:OP_zoom}. We can also decrease $\Gamma$ to
move back from the reentrant to the tangent case. The corresponding
limit is $h_{c}^{\pm}\to h_{*}$ and $h_{t}'\to0$, for which the
square-root behavior vanishes and only the cusp-like behavior of Eq.
(\ref{eq:phi_cusp}) remains. We note that Eq. (\ref{eq:reentrant_crossover})
is an approximation, and higher-order corrections in $h_{t}'$ will
lead to square-root prefactors that are different for the two critical
fields. 

While the analysis here focused on the $T=0$ phase diagram with weak disorder ($W<W_0$), similar behaviors for $\Phi(h)$ are carried over to the phase diagrams of the marginal disorder ($W=W_0$) and strong disorder ($W>W_0$) cases. As shown in Fig. \ref{fig:OP_all}, in the former, $\Phi(h)$ displays the cusp-behavior for $\Gamma=0$ whereas, in the latter, $\Phi(h)$ always shows the square-root-like behavior for any $\Gamma$ value.

\subsection{Shear modulus softening}

While in the previous section we focused on the behavior of the AM order parameter, in this and in the next subsection we focus on the behavior of the AM susceptibility. The key point is that piezomagnetism enables a coupling between altermagnetic and elastic
degrees of freedom such that their fluctuations also become correlated.
Consequently, order parameter fluctuations, which become pronounced
near the AM transition, enhance elastic fluctuations and lead to a
structural transition which accompanies the AM critical point (see also \cite{Steward2025}). The situation is analogous to a nematic transition softening the shear modulus \cite{Fernandes2010}. A straightforward
mean field calculation of the renormalized shear modulus $\tilde{C}_{0}$ via Eq. (\ref{eq:full_H}) gives:
\begin{equation}
\tilde{C}_{0}^{-1}=C_{0}^{-1}+\frac{\lambda^{2}H_{z}^{2}}{C_{0}^{2}}\chi\label{eq:unscaled_shear_modulus_renorm}
\end{equation}
where $\chi$ is the AM static susceptibility. As expected, the $H_{z}=0$
limit results in no renormalization since the effective bilinear strain-AM
coupling is zero, whereas $H_{z}\neq0$ results in the vanishing of
the shear modulus at the AM transition where $\chi\to\infty$. Using
the definition of $H_\lambda$, Eq. (\ref{eq:unscaled_shear_modulus_renorm})
can be written in a dimensionless fashion:
\begin{equation}
\frac{\tilde{C}_{0}}{C_{0}}=\frac{1}{1+h^{2}\chi}
\end{equation}

This analysis reveals a clear analogy between nematicity and altermagnetism \cite{Steward2025}.
In nematicity, a direct bilinear coupling between the order parameter
and shear strain is symmetry-allowed, resulting in a structural distortion
at the nematic transition. This coupling of an electronic order parameter
to the lattice, and subsequent structural distortion, is a key ingredient in the elucidation of nematicity in correlated
systems \cite{Fradkin2010,Fernandes2014}. By contrast, in the altermagnetic case, this effective coupling
and the resulting structural distortion magnitude is \emph{tunable
}by a magnetic field. 

To calculate the renormalized shear modulus, we use our Landau theory
to ascertain $\chi$ in both the PM and AM phases. In mean field theory,
$\chi$ exhibits Curie-Weiss-like universal behavior proximate to
the phase boundary in which
\begin{equation}
\chi\propto|x-x_{c}|^{-1}
\end{equation}

\begin{figure}[H]

\includegraphics[width=0.95\columnwidth]{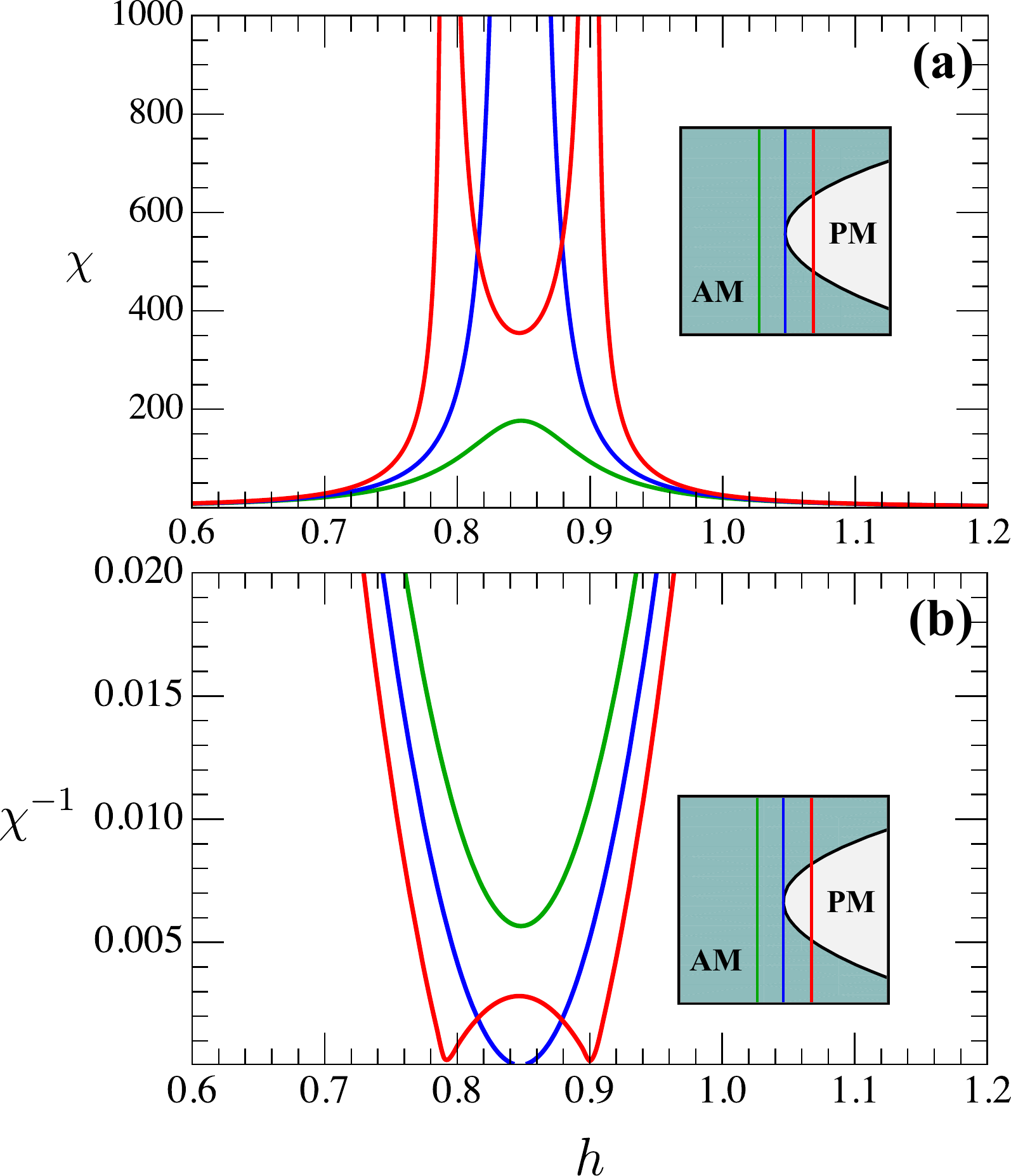}\caption{The altermagnetic susceptibility $\chi\equiv[\partial\Phi/\partial b]_{b\to0}$
\textbf{(a) }and its inverse (\textbf{b)} plotted as a function of
$h$ at $T=0$ and $W=0.9W_0$ for different values of $\Gamma$, as shown in the inset phase diagram. Here, $\Gamma_{*}\approx0.526J$ and
$h_{*}\approx0.842$. The green curve is the susceptibility for the
non-reentrant case $\Gamma=\Gamma_{*}(1-10^{-2})$, for which $\chi$
achieves a quadratic maximum at $h=h_{\text{min}}$. In blue, we plot
$\chi$ in the tangent case $\Gamma=\Gamma_{*}$ for which $\chi$
diverges as $|h-h_{*}|^{-2}$. In red, we plot $\chi$ in the reentrant
case $\Gamma=\Gamma_{*}(1+10^{-2})$, for which $\chi$ diverges as
$|h-h_{c}^{\pm}|^{-1}$ at the two successive critical fields. } \label{fig:shear_all}

\end{figure}

Here, $x$ is a generic tuning parameter parameterizing an axis with
a component that is normal to the phase boundary marked by $x_{c}$.
As with $\Phi$, the various trajectories followed on the phase diagram
with different choices of $\Gamma$, $W$, and $T$ leads to interesting
consequences on the behavior of $\chi$ as a function of $h$, such
as non-monotonicity or a suppression of the universal regime arbitrarily
close to the critical field. To calculate the AM susceptibility, we
introduce an infinitesimal conjugate field $b$ which couples to $\sum_{i}\tau_{i}^{z}$
in the Hamiltonian. To take into account the fact that we have
divided the Hamiltonian through by the renormalized exchange $\tilde{J}=J[1+h^{2}]$,
we define 
\begin{equation}
B(h)\equiv\frac{b}{1+h^{2}}
\end{equation}
In the Landau theory, $B$ simply shifts the mean of the effective
random field distribution $p(z)$. This yields the new free energy
\begin{align}
\bar{f}_{B}(\Phi) & =\frac{1}{2}\Phi^{2}-\int_{-\infty}^{\infty}p(z-\Phi-B)\Lambda(z)dz
\end{align}
To lowest order in $B$, the saddle-point equation $\partial_{\Phi}\bar{f}_{B}=0$
can be written as 
\begin{equation}
\partial_{\Phi}\bar{f}=B(1-\partial_{\Phi}^{2}\bar{f})\label{eq:EOM_smallB}
\end{equation}
where only the $B=0$ free energy $\bar{f}$, defined in Eq. (\ref{eq:free_energy_exact}), is invoked. Differentiating
Eq. (\ref{eq:EOM_smallB}) with respect to $b$ and taking $b\to0$ yields
the susceptibility in terms of $\bar{f}$ or, equivalently, its Landau
coefficients $ \mathcal{A}$ and  $ \mathcal{U}$: 
\begin{align}
\chi & =\frac{[\partial_{\Phi}^{2}\bar{f}]^{-1}-1}{1+h^{2}}\\
 & =\frac{1}{1+h^{2}}\times\begin{cases}
\mathcal{A}^{-1}-1 & \text{for }\Phi=0\\{}
[\mathcal{A}+3\mathcal{U}\Phi^{2}+...]^{-1}-1 & \text{for }\Phi\neq0
\end{cases}
\end{align}
Using this expression for $\chi$, the shear modulus renormalization
can be written as 
\begin{equation}
\frac{\tilde{C}_{0}}{C_{0}}=\frac{1+h^{2}}{\partial_{\Phi}^{2}\bar{f}+h^{2}}(\partial_{\Phi}^{2}\bar{f})
\end{equation}

In Fig. \ref{fig:shear_all}, we plot the behavior of the field-dependent AM susceptibility $\chi(h)$ and of its inverse along different cuts in the zero-temperature $(\Gamma,h)$ phase diagram shown in the inset. We focus on the regime of weak disorder $W_* < W<W_0$, for which reentrant AM order emerges to the right of the $(\Gamma_*,h_*)$  point that marks the end of the bulge of the PM-AM phase boundary. The non-monotonic behavior of $\chi$  is evident, as is the unusual quadratic behavior of $\chi^{-1}$ when the PM-AM phase boundary is tangentially touched (blue curve).

\begin{figure}[H]
\includegraphics[width=0.95\columnwidth]{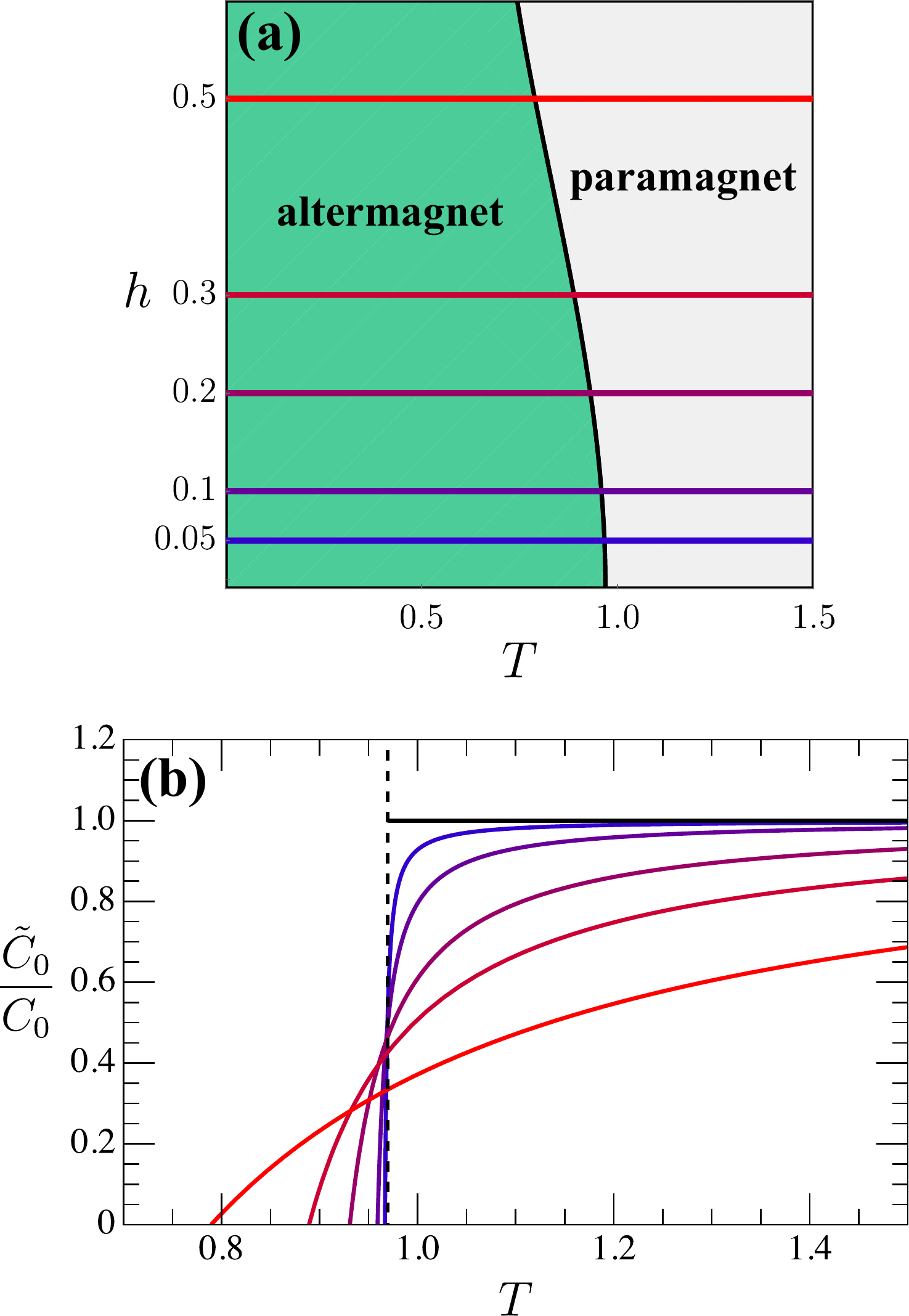}\caption{Shear modulus softening for different magnetic-field values $h=H_{z}/H_\lambda$
as a function of $T$ for $W=0.7J$, $\Gamma=0.3J$ \textbf{(b)}. The corresponding $(T,h)$ phase diagram with constant-$h$ line cuts is shown in panel \textbf{(a)}. When $h=0$, the order parameter does not couple to strain, and the shear modulus is unrenormalized, as shown in the solid black line. For  $h\neq0$, $\tilde{C}_{0}$
vanishes at $T_{c}$, signaling a tetragonal-to-orthorhombic
structural transition. Increasing $h$ results in an enhanced nemato-elastic coupling $\lambda H_z$, thereby suppressing $\tilde{C}_0$ for high temperatures. The black dashed line is the AM transition temperature at $h=0$.} \label{fig:shear_zoom}
\end{figure}

The temperature dependence of the renormalized shear modulus $\tilde{C}_{0}/C_{0}$ is shown in Fig. \ref{fig:shear_zoom} for different values of the magnetic field. Here, we consider the $(T,h)$ phase diagram cross section show in the inset, with fixed $W=0.7J$ and $\Gamma=0.3J$.  Without a magnetic field, as expected, there is no change in the shear modulus. However, as the field increases, the shear modulus vanishes as the transition is approached, signaling a tetragonal-to-orthorhombic transition coincident with the PM-AM transition. 

\subsection{The elasto-caloric effect}

Another experimental quantity that depends on the AM susceptibility is the elasto-caloric effect (ECE) coefficient.
The ECE can be thought of heat generation in response to
an applied ac strain \cite{ikeda2019ac}. The precise measurement of this effect, namely
the amount of heat arising from a given amount of strain, provides
a direct probe of the entropy landscape and is therefore sensitive
to phase transitions. Indeed, the ECE has been exploited to detect
strain-tunable phase transitions in nematic \cite{Ikeda2021_ECE, rosenberg2024nematic}, superconducting \citep{palle2023constraints,ghosh2024elastocaloric,li2022elastocaloric}, and
magnetic systems \cite{ye2023elastocaloric,Ye2024}. Due to the nontrivial role of strain
in AM systems, ECE provides a powerful tool for the
experimental detection and characterization of AM phases and phase
transitions. 

The ECE coefficient $\eta$ -- having units of energy -- is defined
as the strain derivative of temperature taken at constant entropy.
Recall that in our units, all energy scales including $\eta$ are
in units of the bare interaction $J$. The ECE coefficient is
\begin{equation}
\eta=\Big(\frac{\partial T}{\partial\varepsilon}\Big)_{S}=-\frac{T}{C_{\varepsilon}}\Big(\frac{\partial S}{\partial\varepsilon}\Big)_{T}\label{eq:ECE1}
\end{equation}
Here, $\varepsilon$ is a small applied shear strain. Note that $\eta$
can be thought of as a version of the Gr\"uneisen parameter
which connects $T$ to $\varepsilon$. The Maxwell relation $\Big(\frac{\partial S}{\partial\varepsilon}\Big)_{T}=-\Big(\frac{\partial S}{\partial T}\Big)_{\varepsilon}\Big(\frac{\partial T}{\partial\varepsilon}\Big)_{S}$
yields the second equality, where $C_{\varepsilon}=T\Big(\frac{\partial S}{\partial T}\Big)_{\varepsilon}$
is the heat-capacity at constant $\varepsilon$. We can equivalently
work with specific entropy $s\equiv S/N$ and specific heat $c\equiv C_{\varepsilon}/N$
since a scaling of either does not change the result for $\eta$.
Using our Landau theory, it is straightforward to compute the specific entropy $s=-\frac{\partial\bar{f}}{\partial T}$
via a partial derivative with respect to temperature rather than a total one, since we do not need to consider the temperature
dependence in the order parameter as $\partial\bar{f}/\partial\Phi=0$.

To set the stage, and in order to properly connect the ECE coefficient to other physical observables like AM susceptibility and the shear modulus, we take account of the quantities held fixed in the definition of each observable. Recall that the order parameter susceptibility and the shear modulus are defined at fixed zero \textit{stress}. This allowed us to treat the elastic degrees of freedom, i.e., $\varepsilon_0$ in Eq. (\ref{eq:full_H}), dynamically, thereby renormalizing $J\to \tilde{J}$. At nonzero $H_z$, dynamical elastic modes contributed to the ordering tendency that opposed the disordering tendency caused by random strain, resulting in the nontrivial phase diagrams found in previous sections.

\begin{figure}
\includegraphics[width=0.95\columnwidth]{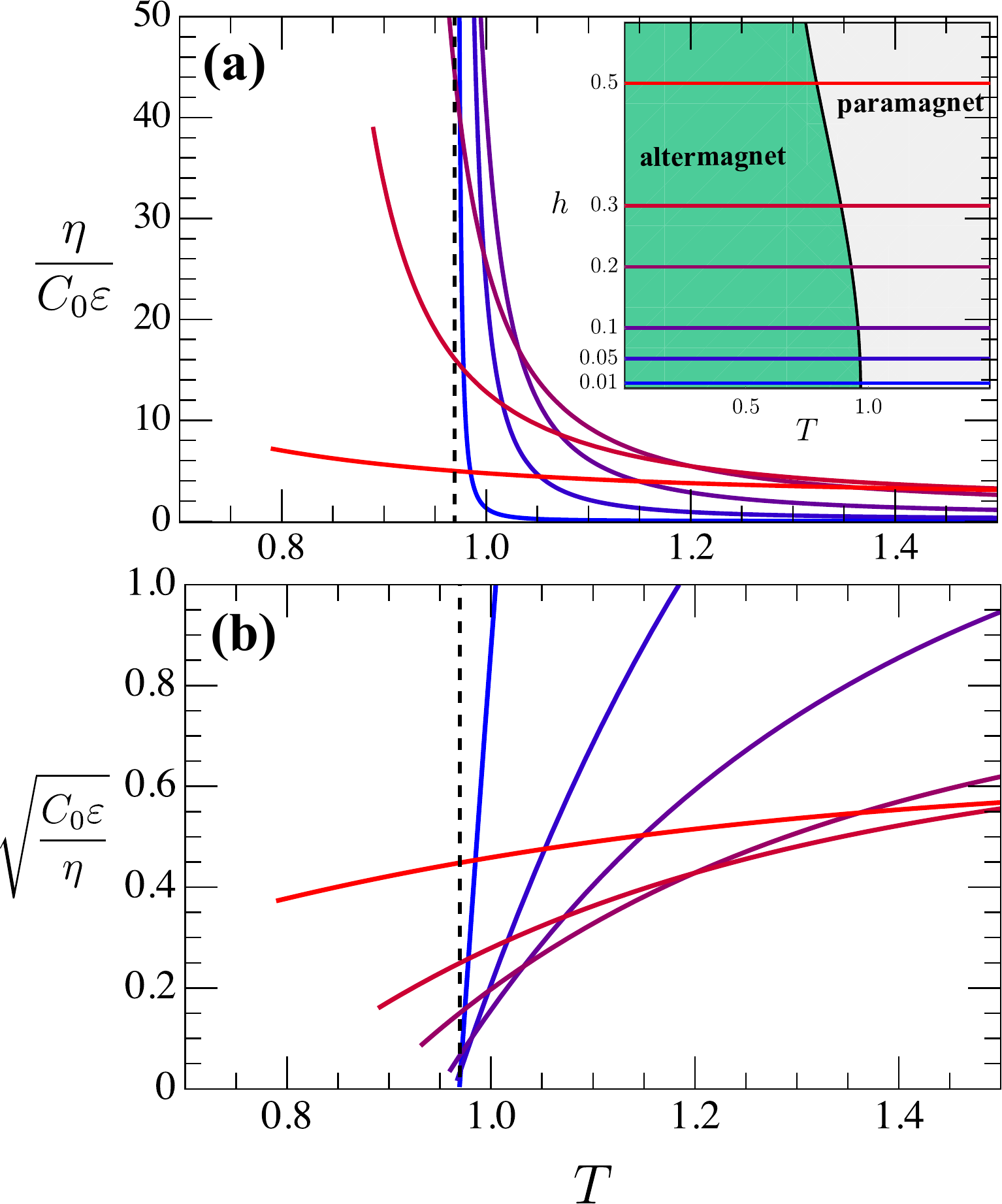}\caption{ECE coefficient $\eta/C_{0}\varepsilon$ \textbf{(a) }and its inverse square root $\sqrt{C_{0}\varepsilon/\eta}$
\textbf{(b) }for different field values $h=H_{z}/H_\lambda$
as a function of $T$ for $W=0.7J$, $\Gamma=0.3J$ (the same as in Fig. \ref{fig:shear_zoom}, as shown by the phase diagram in the inset). All curves terminate at the AM phase transition. The ECE is enhanced near the AM transition, but remains finite, and exhibits a $|T-T_c|^{-2}$ scaling regime at modest fields for temperatures satisfying $Jh^2 \ll T-T_c \ll J$ [Eq.(\ref{eq:h-regime})]. The dashed line is the AM transition temperature without a field.} \label{fig:ECE}

\end{figure}

On the other hand, the ECE is measured at fixed \textit{strain}, with the ac strain period being longer than any inherent relaxation timescale. We therefore cannot treat elastic modes dynamically in this scenario and, in our model, no renormalization of $J$ occurs. As a result, any divergence in the expression for $\eta$ at nonzero $H_z$ occurs at a temperature $T^*$ which is strictly \textit{lower} than $T_c$. In other words, the ECE coefficient depends on the bare AM susceptibility rather than the actual AM susceptibility. Consequently, $\eta$ does not diverge at the true AM transition and is only enhanced to a finite value, as we will show in the ECE coefficient curves of Fig. \ref{fig:ECE}. Note that, while only the curves associated with $H_z = 0.5H_\lambda$ and $H_z = 0.3H_\lambda$ in panel (a) seem to terminate at a finite value, a larger vertical axis window would show that the other curves do as well.

We compute $\eta$ in the disordered phase, and expand our Landau free energy to quartic order to obtain its behavior near the transition $T_c$ (assuming that  $h\lesssim 1$ so that $T^*$ and $T_c$ are not too far apart). An infinite-order analysis is done in the Appendix to obtain $\eta$ for all parameter values. As in the previous section, we introduce a conjugate field $b$, which we now assume to be arising from an applied infinitesimal shear strain $b=\lambda H_{z}\varepsilon$ and expand the free energy density to $\mathcal{O}(\Phi^4)$:
\begin{equation}
\bar{f}(\Phi) = \frac{\mathcal{A}}{2}\Phi^2 + \frac{\mathcal{U}}{4}\Phi^4 -b \Phi
\end{equation}
Crucially, the Landau parameters are given in Eq. (\ref{eq:Landau_params1}) and Eq. (\ref{eq:Landau_params2}) but with the important condition that $\tilde{J}=J$ due to the frozen out elastic modes. We can re-express $\mathcal{A}$ in terms of the temperature scale $T^*$ as 
\begin{equation}
    \mathcal{A}=\mathfrak{a}(T-T^*)
\end{equation}
where $\mathfrak{a}$ and $T^*$ generically depend on all non-thermal parameters in the problem. Taking $\mathcal{A}=0$ in the case of no random strain disorder $W=0$ yields
\begin{equation}
    T^*(\Gamma,h)=\frac{\Gamma}{\tanh^{-1}\Gamma}
\end{equation}
Using the expression for $T_c$ in Eq. (\ref{eq:Tc_vs_Gamma_W=00003D0}), we find to leading order in $h$
\begin{equation}
    T_c-T^*\approx \frac{\Gamma^2h^2}{(1-\Gamma^2)(\tanh^{-1}\Gamma)^2} 
\end{equation}
which in the classical limit of $\Gamma\to 0$ becomes $T_c-T^*\approx \lambda^2H_z^2/C_0$ and recovers the well-known case of coupled nematic and structural degrees of freedom \cite{chu2012divergent} (where we interpret $\lambda H_z$ as the effective nemato-elastic coupling).

From here, the specific entropy as a function of $\Phi$ follows as $s=-\partial \bar{f}/\partial{T} = -\mathfrak{a}\Phi^2/2$ where $\Phi$ solves $\partial \bar{f}/\partial \Phi=0$, yielding the expression 
\begin{equation}
    s=-\frac{b^2}{2\mathfrak{a}(T-T^*)^2}=-\frac{\lambda^2 H_z^2}{2\mathfrak{a}(T-T^*)^2}\varepsilon^2
\end{equation}
Hence, the ECE coefficient $\eta$ follows from Eq. (\ref{eq:ECE1}) as 
\begin{align}
    \eta &= \frac{T}{c_+}\frac{\lambda^2 H_z^2}{\mathfrak{a}(T-T^*)^2}\varepsilon \\ &=  \frac{T}{c_+}\frac{\lambda^2 H_z^2}{\mathfrak{a}(T-T_c + \varsigma\lambda^2 H_z^2/C_0)^2}\varepsilon
\end{align}
where $c_\text{+}$ refers to the specific heat immediately above $T^*$ and $\varsigma$ is an $H_z$-independent positive dimensionless $\mathcal{O}(1)$ parameter. In accordance with our mean field approach, the specific heat undergoes a finite jump $\Delta c=c_+ - c_-$ at $T=T^*$. In the limit of $H_z \ll H_\lambda$, the greatest contribution to the variation in $\eta$ arises from the temperature denominator. Accordingly, the maximum enhancement of $\eta$ occurs close to the AM transition, with a maximum value $\eta_\text{max}\propto \lambda^2 H_z^2(T_c - T^*)^{-2}$ that diverges in the limit $H_z \to 0$. This suggests a crossover temperature at small fields for which $\eta \sim (T-T_c)^{-2}$:
\begin{align}
    h^2 \ll \frac{T-T_c}{J} \ll 1 \implies \eta \propto \frac{\lambda ^2 H_z^2}{(T-T_c)^2}\varepsilon
    \label{eq:h-regime}
\end{align}
Note that the $(T-T_c)^{-2}$ divergence is cutoff when $T-T_c \sim Jh^2$.

For $h\sim 1$, $T_c$ and $T^*$ are generally quite different, and the quartic truncation for $\bar{f}$ is insufficient. An infinite-order expression for $\eta$ is derived in the Appendix. Using the equations derived in the Appendix, we obtain $\eta/C_0 \varepsilon$ as a function of $T$ for fixed $\Gamma, W$ and several values of $h$, as plotted in Fig. \ref{fig:ECE}. We clearly see that, as $h$ increases, the enhancement of the ECE coefficient $\eta$ at the AM transition becomes weaker. The behavior of $1/ \sqrt{\eta}$ is to be contrasted with that of the renormalized shear modulus $\tilde{C}_0$ in Fig. \ref{fig:shear_zoom} for the same magnetic field values in the same $(T,h)$ phase diagram. Specifically, $\tilde{C}_0$ always vanishes at the transition, as it is proportional to the actual AM susceptibility, whereas $1/\sqrt{\eta}$  always reaches a non-zero value at the AM transition, since it is proportional to the bare AM susceptibility.

\section{Beyond Mean field}

In realistic systems, true infinite-range interactions do not exist
and one must explicitly consider the effect of finite wavelength elastic fluctuations, which mediate long-range but not infinite-range interactions
between AM degrees of freedom. In this case, the effect of fluctuations
above mean-field become relevant for both bulk-phase quantities and
in the universal behavior near the transition. The additional complication
of random fields makes the RF-TFIM extremely difficult to solve beyond mean-field. Indeed, the effect of rare regions promotes
activated universal scaling of dynamic observables, causing super-exponential
growth of timescales near the phase transition \cite{Senthil1998_RF,neto2005quantum,vojta2006nonequilibrium}. Additionally, the
generally slow-dynamics close to the transition makes Monte Carlo
methods also quite resource-intensive. However, for some dynamical processes on not too long time scales and, in particular, for a qualitative understanding of some of the equilibrium properties, the mean field analysis discussed here still gives a sensible qualitative behavior.

Since we do not consider dynamics in this work, we assert that the
qualitative behavior of observables obtained from the mean field theory
should carry over to the general case. Moreover, the fact that the
phase diagram is a deformation of the RF-TFIM phase diagram must hold
in general, and not just in mean field. 

\section{Discussion and Conclusions}

Piezomagnetism is a generic phenomenon of altermagnetic systems that enables the tunability of the effective AM-strain coupling
by external magnetic fields. This trilinear coupling leads to competition
between two opposing phenomena: random strain and long-wavelength
elastic fluctuations. The former arises from crystalline defects,
and, when multiplied by a magnetic field, acts as an effective random
longitudinal field conjugate to the AM order parameter $\Phi$. Applying
strong magnetic fields continuously enhances this random field strength linearly in
$H_{z}$ and thereby pushes the system across the critical disorder strength $w_{c}$
of an effective RF-TFIM. On the other hand, long-wavelength elastic
fluctuations enable long-range interactions between AM degrees of freedom,
driving the system mean field, with the effective long-range coupling
strengthening as $H_{z}^{2}$ and thus favoring an ordered state. Thus, at weak magnetic fields, the random strain effect
dominates, whereas at strong magnetic fields, the elastic fluctuations effect
dominates, with a crossover field set by $H_\lambda$. At this field
strength, the effective strain-AM coupling $\lambda H_\lambda$
equals the geometric mean of the AM energy scale $J$ and the bare shear
modulus $C_{0}$.

By minimizing the mean-field free energy of the RF-TFIM, we uncovered
a nontrivial phase diagram as a function of effective random
strain strength $W$ (proportional to $\sqrt{E_{\text{el}}/J}$), on-site
tunneling amplitude $\Gamma$, temperature $T$, and magnetic field
$H_{z}$. This phase diagram is simply a $W$-dependent deformation
of the original convex critical surface of the RF-TFIM, where the
axes are given by magnetic-field dependent effective temperature $t$, effective random field strength
$w$, and effective transverse field $\gamma$. 

For fixed $\Gamma$, $T$, there is a threshold disorder strength value $W_{*}(\Gamma,T)$ that distinguishes three types of behaviors once the system starts in the AM phase. For $W>W_{*}$, an increasing magnetic field first drives the AM phase towards a PM phase at $H_{z,c}^{-}<H_\lambda$  and then back to the AM phase at $H_{z,c}^{+}>H_\lambda$, establishing a reentrant AM order. Exactly at $W=W_{*}$, the critical fields $H_{z,c}^{\pm}$
merge onto a single point $H_{z,*}$ and the corresponding critical point has critical exponents that are twice the values of the critical exponents at the separate points $H_{z,c}^{\pm}$. For  $W<W_{*}$, the system remains in the AM phase. In all cases, however, if the system starts in the PM phase, a large enough field may drive a transition to the AM phase. 

We found that each of these different regimes leave clear fingerprints on the AM order parameter and the AM susceptibility. For instance, even in the case $W<W_{*}$, where the AM phase cannot be completely suppressed by an external magnetic field, the AM order parameter and the AM susceptibility show a non-monotonic behavior. Importantly, we showed that the AM susceptibility $\chi$ is directly manifested in the renormalized shear modulus $\tilde{C}_{0}$, which vanishes as $1/\chi$, 
and in the elasto-caloric effect coefficient $\eta$ , which enhances strongly near the transition in the presence of a magnetic field, thus opening new avenues to probe AM fluctuations in the PM phase.

We emphasize that the tuning properties of the external magnetic field become much more pronounced if the system is closer to the AM-PM phase boundary, in which case the
critical fields $H_{z,c}^{\pm}$ become small. Nevertheless, it is instructive to estimate the typical field magnitude $H_\lambda$  that sets the scale for many of the phenomena discussed here to occur.  Minimizing
the free energy $F(\Phi,\varepsilon)$ of piezomagnetically coupled order parameter and strain with respect to uniform strain $\varepsilon$
yields the linear response relation $\varepsilon=\lambda\Phi H_{z}/C_0$, which allows us to 
directly identify the piezomagnetic tensor element $\Lambda^{-1}=\lambda\Phi/C_{0}$.
Note that $\Lambda$ has dimensions of magnetic field, since $\Phi$ is dimensionless, $C_0$ has dimensions of energy, and $\lambda$ has dimensions of energy times inverse field. Assuming $\Phi=\mathcal{O}(1)$ deep inside the AM phase, this enables us to relate the magnetic field scale $H_\lambda$ with other quantities:
\begin{equation}
H_\lambda=\frac{\sqrt{C_{0}J}}{\lambda}\sim\Lambda\sqrt{\frac{J}{C_{0}}}
\end{equation}
Now, the energy scale associated with the shear modulus $C_{0}$ is typically of the order of $30$ eV, corresponding to an elastic constant of the order of $\mathcal{C}_0 \sim 50$ GPa multiplied by a unit cell volume of the order of $v\sim 100$ \AA$^3$, i.e., $C_0 = \mathcal{C}_0 v$. Conversely, the energy scale $J$ can be estimated from the typical AM transition temperature, giving $10$ meV, and thus $H_\lambda\sim  0.02\Lambda$ . As for the piezomagnetic coupling $\Lambda$, we consider rutiles like CoF$_2$, FeF$_2$, and MnF$_2$, which are $d_{xy}$-wave altermagnets. The quantity reported in the literature for  CoF$_2$ \cite{Disa2020,Borovik1960}, $\bar{\Lambda} = 6  \times 10^{-3} \mu_B/\mathrm{GPa}$ , is actually the magnetization induced by applied stress, $m_z = \bar{\Lambda} \sigma_{xy}$. Since $m_z=\chi_0 H_z$ and $\sigma_{xy}=\mathcal{C}_0 \varepsilon $, we thus have $\Lambda= \bar{\Lambda} \mathcal{C}_0 /\chi_0 \sim 6 T$. In the last step, we approximated the uniform magnetic susceptibility to a characteristic value for these rutiles \cite{Schleck2010,Bakshi1982}, $\chi_0 \sim 0.05 \mu_B/T$.   Putting everything together, we find $H_\lambda\sim  0.1 T$, which is an accessible value. We can also use these estimates to express the disorder strength in terms of the root mean square strain $\bar{\varepsilon}_{s}$, obtaining $W=\bar{\varepsilon}_{s} \sqrt{C_0/4J}\sim 30 \bar{\varepsilon}_{s} $. This implies that $W \sim 1 $, which ensures the existence of reentrant AM behavior, requires $\bar{\varepsilon}_{s} \sim 3 \%$, which is a reasonable value. This estimate assumes that the most significant contribution to the FM fluctuations comes from the same degrees of freedom responsible for AM order. In contrast, the presence of separate FM degrees of freedom would suppress our estimate for $\lambda$. Whether this approximation applies to the rutiles requires further investigation.

\section*{acknowledgments}
We thank T. Birol, A. V. Chubukov, I. Jang, W. J. Meese, L.  \v{S}mejkal,
and C. Steward for fruitful discussions. A.R.C. and R.M.F. were supported by the Air Force Office of Scientific Research under Award No. FA9550-21-1-0423. J.S. was supported by the German Research Foundation TRR 288-422213477 ELASTO-Q-MAT, Project A07. R.M.F. also acknowledges a Mercator Fellowship from the German Research Foundation (DFG) through TRR 288, 422213477 Elasto-Q-Mat.

\bibliography{AMpaper_references}

\begin{thebibliography}{134}%
\makeatletter
\providecommand \@ifxundefined [1]{%
 \@ifx{#1\undefined}
}%
\providecommand \@ifnum [1]{%
 \ifnum #1\expandafter \@firstoftwo
 \else \expandafter \@secondoftwo
 \fi
}%
\providecommand \@ifx [1]{%
 \ifx #1\expandafter \@firstoftwo
 \else \expandafter \@secondoftwo
 \fi
}%
\providecommand \natexlab [1]{#1}%
\providecommand \enquote  [1]{``#1''}%
\providecommand \bibnamefont  [1]{#1}%
\providecommand \bibfnamefont [1]{#1}%
\providecommand \citenamefont [1]{#1}%
\providecommand \href@noop [0]{\@secondoftwo}%
\providecommand \href [0]{\begingroup \@sanitize@url \@href}%
\providecommand \@href[1]{\@@startlink{#1}\@@href}%
\providecommand \@@href[1]{\endgroup#1\@@endlink}%
\providecommand \@sanitize@url [0]{\catcode `\\12\catcode `\$12\catcode `\&12\catcode `\#12\catcode `\^12\catcode `\_12\catcode `\%12\relax}%
\providecommand \@@startlink[1]{}%
\providecommand \@@endlink[0]{}%
\providecommand \url  [0]{\begingroup\@sanitize@url \@url }%
\providecommand \@url [1]{\endgroup\@href {#1}{\urlprefix }}%
\providecommand \urlprefix  [0]{URL }%
\providecommand \Eprint [0]{\href }%
\providecommand \doibase [0]{https://doi.org/}%
\providecommand \selectlanguage [0]{\@gobble}%
\providecommand \bibinfo  [0]{\@secondoftwo}%
\providecommand \bibfield  [0]{\@secondoftwo}%
\providecommand \translation [1]{[#1]}%
\providecommand \BibitemOpen [0]{}%
\providecommand \bibitemStop [0]{}%
\providecommand \bibitemNoStop [0]{.\EOS\space}%
\providecommand \EOS [0]{\spacefactor3000\relax}%
\providecommand \BibitemShut  [1]{\csname bibitem#1\endcsname}%
\let\auto@bib@innerbib\@empty
\bibitem [{\citenamefont {{\v{S}}mejkal}\ \emph {et~al.}(2020)\citenamefont {{\v{S}}mejkal}, \citenamefont {Gonz{\'a}lez-Hern{\'a}ndez}, \citenamefont {Jungwirth},\ and\ \citenamefont {Sinova}}]{Smejkal2020_AM}%
  \BibitemOpen
  \bibfield  {author} {\bibinfo {author} {\bibfnamefont {L.}~\bibnamefont {{\v{S}}mejkal}}, \bibinfo {author} {\bibfnamefont {R.}~\bibnamefont {Gonz{\'a}lez-Hern{\'a}ndez}}, \bibinfo {author} {\bibfnamefont {T.}~\bibnamefont {Jungwirth}},\ and\ \bibinfo {author} {\bibfnamefont {J.}~\bibnamefont {Sinova}},\ }\bibfield  {title} {\bibinfo {title} {Crystal time-reversal symmetry breaking and spontaneous hall effect in collinear antiferromagnets},\ }\href {https://www.science.org/doi/abs/10.1126/sciadv.aaz8809} {\bibfield  {journal} {\bibinfo  {journal} {Science Advances}\ }\textbf {\bibinfo {volume} {6}},\ \bibinfo {pages} {eaaz8809} (\bibinfo {year} {2020})}\BibitemShut {NoStop}%
\bibitem [{\citenamefont {{\v{S}}mejkal}\ \emph {et~al.}(2022{\natexlab{a}})\citenamefont {{\v{S}}mejkal}, \citenamefont {Sinova},\ and\ \citenamefont {Jungwirth}}]{Smejkal2022_2_AM}%
  \BibitemOpen
  \bibfield  {author} {\bibinfo {author} {\bibfnamefont {L.}~\bibnamefont {{\v{S}}mejkal}}, \bibinfo {author} {\bibfnamefont {J.}~\bibnamefont {Sinova}},\ and\ \bibinfo {author} {\bibfnamefont {T.}~\bibnamefont {Jungwirth}},\ }\bibfield  {title} {\bibinfo {title} {Emerging research landscape of altermagnetism},\ }\href {https://link.aps.org/doi/10.1103/PhysRevX.12.040501} {\bibfield  {journal} {\bibinfo  {journal} {Physical Review X}\ }\textbf {\bibinfo {volume} {12}},\ \bibinfo {pages} {040501} (\bibinfo {year} {2022}{\natexlab{a}})}\BibitemShut {NoStop}%
\bibitem [{\citenamefont {{\v{S}}mejkal}\ \emph {et~al.}(2022{\natexlab{b}})\citenamefont {{\v{S}}mejkal}, \citenamefont {Sinova},\ and\ \citenamefont {Jungwirth}}]{Smejkal2022_AM}%
  \BibitemOpen
  \bibfield  {author} {\bibinfo {author} {\bibfnamefont {L.}~\bibnamefont {{\v{S}}mejkal}}, \bibinfo {author} {\bibfnamefont {J.}~\bibnamefont {Sinova}},\ and\ \bibinfo {author} {\bibfnamefont {T.}~\bibnamefont {Jungwirth}},\ }\bibfield  {title} {\bibinfo {title} {Beyond conventional ferromagnetism and antiferromagnetism: A phase with nonrelativistic spin and crystal rotation symmetry},\ }\href {https://link.aps.org/doi/10.1103/PhysRevX.12.031042} {\bibfield  {journal} {\bibinfo  {journal} {Physical Review X}\ }\textbf {\bibinfo {volume} {12}},\ \bibinfo {pages} {031042} (\bibinfo {year} {2022}{\natexlab{b}})}\BibitemShut {NoStop}%
\bibitem [{\citenamefont {Liu}\ \emph {et~al.}(2022)\citenamefont {Liu}, \citenamefont {Li}, \citenamefont {Han}, \citenamefont {Wan},\ and\ \citenamefont {Liu}}]{Liu2022}%
  \BibitemOpen
  \bibfield  {author} {\bibinfo {author} {\bibfnamefont {P.}~\bibnamefont {Liu}}, \bibinfo {author} {\bibfnamefont {J.}~\bibnamefont {Li}}, \bibinfo {author} {\bibfnamefont {J.}~\bibnamefont {Han}}, \bibinfo {author} {\bibfnamefont {X.}~\bibnamefont {Wan}},\ and\ \bibinfo {author} {\bibfnamefont {Q.}~\bibnamefont {Liu}},\ }\bibfield  {title} {\bibinfo {title} {{Spin-Group Symmetry in Magnetic Materials with Negligible Spin-Orbit Coupling}},\ }\href {https://doi.org/https://doi.org/10.1103/PhysRevX.12.021016} {\bibfield  {journal} {\bibinfo  {journal} {Physical Review X}\ }\textbf {\bibinfo {volume} {12}},\ \bibinfo {pages} {021016} (\bibinfo {year} {2022})}\BibitemShut {NoStop}%
\bibitem [{\citenamefont {Jiang}\ \emph {et~al.}(2024{\natexlab{a}})\citenamefont {Jiang}, \citenamefont {Song}, \citenamefont {Zhu}, \citenamefont {Fang}, \citenamefont {Weng}, \citenamefont {Liu}, \citenamefont {Yang},\ and\ \citenamefont {Fang}}]{Fang2024}%
  \BibitemOpen
  \bibfield  {author} {\bibinfo {author} {\bibfnamefont {Y.}~\bibnamefont {Jiang}}, \bibinfo {author} {\bibfnamefont {Z.}~\bibnamefont {Song}}, \bibinfo {author} {\bibfnamefont {T.}~\bibnamefont {Zhu}}, \bibinfo {author} {\bibfnamefont {Z.}~\bibnamefont {Fang}}, \bibinfo {author} {\bibfnamefont {H.}~\bibnamefont {Weng}}, \bibinfo {author} {\bibfnamefont {Z.-X.}\ \bibnamefont {Liu}}, \bibinfo {author} {\bibfnamefont {J.}~\bibnamefont {Yang}},\ and\ \bibinfo {author} {\bibfnamefont {C.}~\bibnamefont {Fang}},\ }\bibfield  {title} {\bibinfo {title} {{Enumeration of Spin-Space Groups: Toward a Complete Description of Symmetries of Magnetic Orders}},\ }\href {https://doi.org/10.1103/PhysRevX.14.031039} {\bibfield  {journal} {\bibinfo  {journal} {Physical Review X}\ }\textbf {\bibinfo {volume} {14}},\ \bibinfo {pages} {031039} (\bibinfo {year} {2024}{\natexlab{a}})}\BibitemShut {NoStop}%
\bibitem [{\citenamefont {Chen}\ \emph {et~al.}(2024)\citenamefont {Chen}, \citenamefont {Ren}, \citenamefont {Zhu}, \citenamefont {Yu}, \citenamefont {Zhang}, \citenamefont {Liu}, \citenamefont {Li}, \citenamefont {Liu}, \citenamefont {Li},\ and\ \citenamefont {Liu}}]{Liu2024}%
  \BibitemOpen
  \bibfield  {author} {\bibinfo {author} {\bibfnamefont {X.}~\bibnamefont {Chen}}, \bibinfo {author} {\bibfnamefont {J.}~\bibnamefont {Ren}}, \bibinfo {author} {\bibfnamefont {Y.}~\bibnamefont {Zhu}}, \bibinfo {author} {\bibfnamefont {Y.}~\bibnamefont {Yu}}, \bibinfo {author} {\bibfnamefont {A.}~\bibnamefont {Zhang}}, \bibinfo {author} {\bibfnamefont {P.}~\bibnamefont {Liu}}, \bibinfo {author} {\bibfnamefont {J.}~\bibnamefont {Li}}, \bibinfo {author} {\bibfnamefont {Y.}~\bibnamefont {Liu}}, \bibinfo {author} {\bibfnamefont {C.}~\bibnamefont {Li}},\ and\ \bibinfo {author} {\bibfnamefont {Q.}~\bibnamefont {Liu}},\ }\bibfield  {title} {\bibinfo {title} {{Enumeration and Representation Theory of Spin Space Groups}},\ }\href {https://doi.org/10.1103/PhysRevX.14.031038} {\bibfield  {journal} {\bibinfo  {journal} {Physical Review X}\ }\textbf {\bibinfo {volume} {14}},\ \bibinfo {pages} {031038} (\bibinfo {year} {2024})}\BibitemShut {NoStop}%
\bibitem [{\citenamefont {Xiao}\ \emph {et~al.}(2024)\citenamefont {Xiao}, \citenamefont {Zhao}, \citenamefont {Li}, \citenamefont {Shindou},\ and\ \citenamefont {Song}}]{Song2024}%
  \BibitemOpen
  \bibfield  {author} {\bibinfo {author} {\bibfnamefont {Z.}~\bibnamefont {Xiao}}, \bibinfo {author} {\bibfnamefont {J.}~\bibnamefont {Zhao}}, \bibinfo {author} {\bibfnamefont {Y.}~\bibnamefont {Li}}, \bibinfo {author} {\bibfnamefont {R.}~\bibnamefont {Shindou}},\ and\ \bibinfo {author} {\bibfnamefont {Z.-D.}\ \bibnamefont {Song}},\ }\bibfield  {title} {\bibinfo {title} {{Spin Space Groups: Full Classification and Applications}},\ }\href {https://doi.org/10.1103/PhysRevX.14.031037} {\bibfield  {journal} {\bibinfo  {journal} {Physical Review X}\ }\textbf {\bibinfo {volume} {14}},\ \bibinfo {pages} {031037} (\bibinfo {year} {2024})}\BibitemShut {NoStop}%
\bibitem [{\citenamefont {Jungwirth}\ \emph {et~al.}(2024{\natexlab{a}})\citenamefont {Jungwirth}, \citenamefont {Fernandes}, \citenamefont {Sinova},\ and\ \citenamefont {{\v{S}}mejkal}}]{Jungwirth2024review}%
  \BibitemOpen
  \bibfield  {author} {\bibinfo {author} {\bibfnamefont {T.}~\bibnamefont {Jungwirth}}, \bibinfo {author} {\bibfnamefont {R.~M.}\ \bibnamefont {Fernandes}}, \bibinfo {author} {\bibfnamefont {J.}~\bibnamefont {Sinova}},\ and\ \bibinfo {author} {\bibfnamefont {L.}~\bibnamefont {{\v{S}}mejkal}},\ }\bibfield  {title} {\bibinfo {title} {{Altermagnets and beyond: Nodal magnetically-ordered phases}},\ }\href {https://arxiv.org/pdf/2409.10034} {\bibfield  {journal} {\bibinfo  {journal} {arXiv:2409.10034}\ } (\bibinfo {year} {2024}{\natexlab{a}})}\BibitemShut {NoStop}%
\bibitem [{\citenamefont {Mazin}\ \emph {et~al.}(2023)\citenamefont {Mazin}, \citenamefont {Gonz{\'a}lez-Hern{\'a}ndez},\ and\ \citenamefont {{\v{S}}mejkal}}]{Mazin2023_FeSe}%
  \BibitemOpen
  \bibfield  {author} {\bibinfo {author} {\bibfnamefont {I.}~\bibnamefont {Mazin}}, \bibinfo {author} {\bibfnamefont {R.}~\bibnamefont {Gonz{\'a}lez-Hern{\'a}ndez}},\ and\ \bibinfo {author} {\bibfnamefont {L.}~\bibnamefont {{\v{S}}mejkal}},\ }\bibfield  {title} {\bibinfo {title} {{Induced Monolayer Altermagnetism in {MnP(S,Se)$_3$ and FeSe}}},\ }\href {https://arxiv.org/pdf/2309.02355} {\bibfield  {journal} {\bibinfo  {journal} {arXiv:2309.02355}\ } (\bibinfo {year} {2023})}\BibitemShut {NoStop}%
\bibitem [{\citenamefont {Fernandes}\ \emph {et~al.}(2024)\citenamefont {Fernandes}, \citenamefont {de~Carvalho}, \citenamefont {Birol},\ and\ \citenamefont {Pereira}}]{Fernandes2024_AM}%
  \BibitemOpen
  \bibfield  {author} {\bibinfo {author} {\bibfnamefont {R.~M.}\ \bibnamefont {Fernandes}}, \bibinfo {author} {\bibfnamefont {V.~S.}\ \bibnamefont {de~Carvalho}}, \bibinfo {author} {\bibfnamefont {T.}~\bibnamefont {Birol}},\ and\ \bibinfo {author} {\bibfnamefont {R.~G.}\ \bibnamefont {Pereira}},\ }\bibfield  {title} {\bibinfo {title} {{Topological transition from nodal to nodeless Zeeman splitting in altermagnets}},\ }\href {https://doi.org/10.1103/PhysRevB.109.024404} {\bibfield  {journal} {\bibinfo  {journal} {Physical Review B}\ }\textbf {\bibinfo {volume} {109}},\ \bibinfo {pages} {024404} (\bibinfo {year} {2024})}\BibitemShut {NoStop}%
\bibitem [{\citenamefont {Fang}\ \emph {et~al.}(2024)\citenamefont {Fang}, \citenamefont {Cano},\ and\ \citenamefont {Ghorashi}}]{Cano2024}%
  \BibitemOpen
  \bibfield  {author} {\bibinfo {author} {\bibfnamefont {Y.}~\bibnamefont {Fang}}, \bibinfo {author} {\bibfnamefont {J.}~\bibnamefont {Cano}},\ and\ \bibinfo {author} {\bibfnamefont {S.~A.~A.}\ \bibnamefont {Ghorashi}},\ }\bibfield  {title} {\bibinfo {title} {{Quantum Geometry Induced Nonlinear Transport in Altermagnets}},\ }\href {https://doi.org/10.1103/PhysRevLett.133.106701} {\bibfield  {journal} {\bibinfo  {journal} {Physical Review Letters}\ }\textbf {\bibinfo {volume} {133}},\ \bibinfo {pages} {106701} (\bibinfo {year} {2024})}\BibitemShut {NoStop}%
\bibitem [{\citenamefont {Antonenko}\ \emph {et~al.}(2025)\citenamefont {Antonenko}, \citenamefont {Fernandes},\ and\ \citenamefont {Venderbos}}]{Antonenko2025}%
  \BibitemOpen
  \bibfield  {author} {\bibinfo {author} {\bibfnamefont {D.~S.}\ \bibnamefont {Antonenko}}, \bibinfo {author} {\bibfnamefont {R.~M.}\ \bibnamefont {Fernandes}},\ and\ \bibinfo {author} {\bibfnamefont {J.~W.~F.}\ \bibnamefont {Venderbos}},\ }\bibfield  {title} {\bibinfo {title} {{Mirror Chern Bands and Weyl Nodal Loops in Altermagnets}},\ }\href {https://doi.org/10.1103/PhysRevLett.134.096703} {\bibfield  {journal} {\bibinfo  {journal} {Physical Review Letters}\ }\textbf {\bibinfo {volume} {134}},\ \bibinfo {pages} {096703} (\bibinfo {year} {2025})}\BibitemShut {NoStop}%
\bibitem [{\citenamefont {Parshukov}\ \emph {et~al.}(2024)\citenamefont {Parshukov}, \citenamefont {Wiedmann},\ and\ \citenamefont {Schnyder}}]{Schnyder2024}%
  \BibitemOpen
  \bibfield  {author} {\bibinfo {author} {\bibfnamefont {K.}~\bibnamefont {Parshukov}}, \bibinfo {author} {\bibfnamefont {R.}~\bibnamefont {Wiedmann}},\ and\ \bibinfo {author} {\bibfnamefont {A.~P.}\ \bibnamefont {Schnyder}},\ }\bibfield  {title} {\bibinfo {title} {{Topological responses from gapped {Weyl} points in {2D} altermagnets}},\ }\href {https://arxiv.org/pdf/2403.09520} {\bibfield  {journal} {\bibinfo  {journal} {arXiv:2403.09520}\ } (\bibinfo {year} {2024})}\BibitemShut {NoStop}%
\bibitem [{\citenamefont {Roig}\ \emph {et~al.}(2024)\citenamefont {Roig}, \citenamefont {Kreisel}, \citenamefont {Yu}, \citenamefont {Andersen},\ and\ \citenamefont {Agterberg}}]{Agterberg2024}%
  \BibitemOpen
  \bibfield  {author} {\bibinfo {author} {\bibfnamefont {M.}~\bibnamefont {Roig}}, \bibinfo {author} {\bibfnamefont {A.}~\bibnamefont {Kreisel}}, \bibinfo {author} {\bibfnamefont {Y.}~\bibnamefont {Yu}}, \bibinfo {author} {\bibfnamefont {B.~M.}\ \bibnamefont {Andersen}},\ and\ \bibinfo {author} {\bibfnamefont {D.~F.}\ \bibnamefont {Agterberg}},\ }\bibfield  {title} {\bibinfo {title} {{Minimal models for altermagnetism}},\ }\href {https://doi.org/10.1103/PhysRevB.110.144412} {\bibfield  {journal} {\bibinfo  {journal} {Physical Review B}\ }\textbf {\bibinfo {volume} {110}},\ \bibinfo {pages} {144412} (\bibinfo {year} {2024})}\BibitemShut {NoStop}%
\bibitem [{\citenamefont {Rao}\ \emph {et~al.}(2024)\citenamefont {Rao}, \citenamefont {Mook},\ and\ \citenamefont {Knolle}}]{Knolle2024}%
  \BibitemOpen
  \bibfield  {author} {\bibinfo {author} {\bibfnamefont {P.}~\bibnamefont {Rao}}, \bibinfo {author} {\bibfnamefont {A.}~\bibnamefont {Mook}},\ and\ \bibinfo {author} {\bibfnamefont {J.}~\bibnamefont {Knolle}},\ }\bibfield  {title} {\bibinfo {title} {{Tunable band topology and optical conductivity in altermagnets}},\ }\href {https://doi.org/10.1103/PhysRevB.110.024425} {\bibfield  {journal} {\bibinfo  {journal} {Physical Review B}\ }\textbf {\bibinfo {volume} {110}},\ \bibinfo {pages} {024425} (\bibinfo {year} {2024})}\BibitemShut {NoStop}%
\bibitem [{\citenamefont {Attias}\ \emph {et~al.}(2024)\citenamefont {Attias}, \citenamefont {Levchenko},\ and\ \citenamefont {Khodas}}]{Attias2024}%
  \BibitemOpen
  \bibfield  {author} {\bibinfo {author} {\bibfnamefont {L.}~\bibnamefont {Attias}}, \bibinfo {author} {\bibfnamefont {A.}~\bibnamefont {Levchenko}},\ and\ \bibinfo {author} {\bibfnamefont {M.}~\bibnamefont {Khodas}},\ }\bibfield  {title} {\bibinfo {title} {{Intrinsic anomalous Hall effect in altermagnets}},\ }\href {https://doi.org/10.1103/PhysRevB.110.094425} {\bibfield  {journal} {\bibinfo  {journal} {Physical Review B}\ }\textbf {\bibinfo {volume} {110}},\ \bibinfo {pages} {094425} (\bibinfo {year} {2024})}\BibitemShut {NoStop}%
\bibitem [{\citenamefont {Yu}\ \emph {et~al.}(2024)\citenamefont {Yu}, \citenamefont {Shishidou}, \citenamefont {Sumita}, \citenamefont {Weinert},\ and\ \citenamefont {Agterberg}}]{Yu_Agterberg2024}%
  \BibitemOpen
  \bibfield  {author} {\bibinfo {author} {\bibfnamefont {Y.}~\bibnamefont {Yu}}, \bibinfo {author} {\bibfnamefont {T.}~\bibnamefont {Shishidou}}, \bibinfo {author} {\bibfnamefont {S.}~\bibnamefont {Sumita}}, \bibinfo {author} {\bibfnamefont {M.}~\bibnamefont {Weinert}},\ and\ \bibinfo {author} {\bibfnamefont {D.~F.}\ \bibnamefont {Agterberg}},\ }\bibfield  {title} {\bibinfo {title} {{Spin--orbit enabled unconventional Stoner magnetism}},\ }\href {https://www.pnas.org/doi/10.1073/pnas.2411038121} {\bibfield  {journal} {\bibinfo  {journal} {Proceedings of the National Academy of Sciences}\ }\textbf {\bibinfo {volume} {121}},\ \bibinfo {pages} {e2411038121} (\bibinfo {year} {2024})}\BibitemShut {NoStop}%
\bibitem [{\citenamefont {Leeb}\ \emph {et~al.}(2024)\citenamefont {Leeb}, \citenamefont {Mook}, \citenamefont {\ifmmode~\check{S}\else \v{S}\fi{}mejkal},\ and\ \citenamefont {Knolle}}]{Leeb2024}%
  \BibitemOpen
  \bibfield  {author} {\bibinfo {author} {\bibfnamefont {V.}~\bibnamefont {Leeb}}, \bibinfo {author} {\bibfnamefont {A.}~\bibnamefont {Mook}}, \bibinfo {author} {\bibfnamefont {L.}~\bibnamefont {\ifmmode~\check{S}\else \v{S}\fi{}mejkal}},\ and\ \bibinfo {author} {\bibfnamefont {J.}~\bibnamefont {Knolle}},\ }\bibfield  {title} {\bibinfo {title} {{Spontaneous Formation of Altermagnetism from Orbital Ordering}},\ }\href {https://doi.org/10.1103/PhysRevLett.132.236701} {\bibfield  {journal} {\bibinfo  {journal} {Physical Review Letters}\ }\textbf {\bibinfo {volume} {132}},\ \bibinfo {pages} {236701} (\bibinfo {year} {2024})}\BibitemShut {NoStop}%
\bibitem [{\citenamefont {Ferrari}\ and\ \citenamefont {Valent\'{\i}}(2024)}]{Valenti2024}%
  \BibitemOpen
  \bibfield  {author} {\bibinfo {author} {\bibfnamefont {F.}~\bibnamefont {Ferrari}}\ and\ \bibinfo {author} {\bibfnamefont {R.}~\bibnamefont {Valent\'{\i}}},\ }\bibfield  {title} {\bibinfo {title} {{Altermagnetism on the Shastry-Sutherland lattice}},\ }\href {https://doi.org/10.1103/PhysRevB.110.205140} {\bibfield  {journal} {\bibinfo  {journal} {Physical Review B}\ }\textbf {\bibinfo {volume} {110}},\ \bibinfo {pages} {205140} (\bibinfo {year} {2024})}\BibitemShut {NoStop}%
\bibitem [{\citenamefont {Kaushal}\ and\ \citenamefont {Franz}(2024)}]{Kaushal2024}%
  \BibitemOpen
  \bibfield  {author} {\bibinfo {author} {\bibfnamefont {N.}~\bibnamefont {Kaushal}}\ and\ \bibinfo {author} {\bibfnamefont {M.}~\bibnamefont {Franz}},\ }\bibfield  {title} {\bibinfo {title} {{Altermagnetism in modified Lieb lattice Hubbard model}},\ }\href {https://arxiv.org/pdf/2412.16421} {\bibfield  {journal} {\bibinfo  {journal} {arXiv:2412.16421}\ } (\bibinfo {year} {2024})}\BibitemShut {NoStop}%
\bibitem [{\citenamefont {Sobral}\ \emph {et~al.}(2024)\citenamefont {Sobral}, \citenamefont {Mandal},\ and\ \citenamefont {Scheurer}}]{Sobral2024}%
  \BibitemOpen
  \bibfield  {author} {\bibinfo {author} {\bibfnamefont {J.~A.}\ \bibnamefont {Sobral}}, \bibinfo {author} {\bibfnamefont {S.}~\bibnamefont {Mandal}},\ and\ \bibinfo {author} {\bibfnamefont {M.~S.}\ \bibnamefont {Scheurer}},\ }\bibfield  {title} {\bibinfo {title} {{Fractionalized Altermagnets: from neighboring and altermagnetic spin-liquids to fractionalized spin-orbit coupling}},\ }\href {https://arxiv.org/pdf/2410.10949} {\bibfield  {journal} {\bibinfo  {journal} {arXiv:2410.10949}\ } (\bibinfo {year} {2024})}\BibitemShut {NoStop}%
\bibitem [{\citenamefont {Giuli}\ \emph {et~al.}(2025)\citenamefont {Giuli}, \citenamefont {Mejuto-Zaera},\ and\ \citenamefont {Capone}}]{Giuli2025}%
  \BibitemOpen
  \bibfield  {author} {\bibinfo {author} {\bibfnamefont {S.}~\bibnamefont {Giuli}}, \bibinfo {author} {\bibfnamefont {C.}~\bibnamefont {Mejuto-Zaera}},\ and\ \bibinfo {author} {\bibfnamefont {M.}~\bibnamefont {Capone}},\ }\bibfield  {title} {\bibinfo {title} {{Altermagnetism from interaction-driven itinerant magnetism}},\ }\href {https://doi.org/10.1103/PhysRevB.111.L020401} {\bibfield  {journal} {\bibinfo  {journal} {Physical Review B}\ }\textbf {\bibinfo {volume} {111}},\ \bibinfo {pages} {L020401} (\bibinfo {year} {2025})}\BibitemShut {NoStop}%
\bibitem [{\citenamefont {Del~Re}(2024)}]{del2024dirac}%
  \BibitemOpen
  \bibfield  {author} {\bibinfo {author} {\bibfnamefont {L.}~\bibnamefont {Del~Re}},\ }\bibfield  {title} {\bibinfo {title} {{Dirac points and topological phases in correlated altermagnets}},\ }\href {https://arxiv.org/abs/2408.14288} {\bibfield  {journal} {\bibinfo  {journal} {arXiv:2408.14288}\ } (\bibinfo {year} {2024})}\BibitemShut {NoStop}%
\bibitem [{\citenamefont {Brekke}\ \emph {et~al.}(2023)\citenamefont {Brekke}, \citenamefont {Brataas},\ and\ \citenamefont {Sudb\o{}}}]{Sudbo2023}%
  \BibitemOpen
  \bibfield  {author} {\bibinfo {author} {\bibfnamefont {B.}~\bibnamefont {Brekke}}, \bibinfo {author} {\bibfnamefont {A.}~\bibnamefont {Brataas}},\ and\ \bibinfo {author} {\bibfnamefont {A.}~\bibnamefont {Sudb\o{}}},\ }\bibfield  {title} {\bibinfo {title} {{Two-dimensional altermagnets: Superconductivity in a minimal microscopic model}},\ }\href {https://doi.org/10.1103/PhysRevB.108.224421} {\bibfield  {journal} {\bibinfo  {journal} {Physical Review B}\ }\textbf {\bibinfo {volume} {108}},\ \bibinfo {pages} {224421} (\bibinfo {year} {2023})}\BibitemShut {NoStop}%
\bibitem [{\citenamefont {Ouassou}\ \emph {et~al.}(2023)\citenamefont {Ouassou}, \citenamefont {Brataas},\ and\ \citenamefont {Linder}}]{Ouassou2023}%
  \BibitemOpen
  \bibfield  {author} {\bibinfo {author} {\bibfnamefont {J.~A.}\ \bibnamefont {Ouassou}}, \bibinfo {author} {\bibfnamefont {A.}~\bibnamefont {Brataas}},\ and\ \bibinfo {author} {\bibfnamefont {J.}~\bibnamefont {Linder}},\ }\bibfield  {title} {\bibinfo {title} {{dc {Josephson} Effect in Altermagnets}},\ }\href {https://doi.org/10.1103/PhysRevLett.131.076003} {\bibfield  {journal} {\bibinfo  {journal} {Physical Review Letters}\ }\textbf {\bibinfo {volume} {131}},\ \bibinfo {pages} {076003} (\bibinfo {year} {2023})}\BibitemShut {NoStop}%
\bibitem [{\citenamefont {Zhang}\ \emph {et~al.}(2024{\natexlab{a}})\citenamefont {Zhang}, \citenamefont {Hu},\ and\ \citenamefont {Neupert}}]{Neupert2023}%
  \BibitemOpen
  \bibfield  {author} {\bibinfo {author} {\bibfnamefont {S.-B.}\ \bibnamefont {Zhang}}, \bibinfo {author} {\bibfnamefont {L.-H.}\ \bibnamefont {Hu}},\ and\ \bibinfo {author} {\bibfnamefont {T.}~\bibnamefont {Neupert}},\ }\bibfield  {title} {\bibinfo {title} {{Finite-momentum {Cooper} pairing in proximitized altermagnets}},\ }\href {https://doi.org/10.1038/s41467-024-45951-3} {\bibfield  {journal} {\bibinfo  {journal} {Nat. Commun}\ }\textbf {\bibinfo {volume} {15}},\ \bibinfo {pages} {1801} (\bibinfo {year} {2024}{\natexlab{a}})}\BibitemShut {NoStop}%
\bibitem [{\citenamefont {Beenakker}\ and\ \citenamefont {Vakhtel}(2023)}]{Beenakker2023}%
  \BibitemOpen
  \bibfield  {author} {\bibinfo {author} {\bibfnamefont {C.~W.~J.}\ \bibnamefont {Beenakker}}\ and\ \bibinfo {author} {\bibfnamefont {T.}~\bibnamefont {Vakhtel}},\ }\bibfield  {title} {\bibinfo {title} {{Phase-shifted {Andreev} levels in an altermagnet {Josephson} junction}},\ }\href {https://doi.org/10.1103/PhysRevB.108.075425} {\bibfield  {journal} {\bibinfo  {journal} {Physical Review B}\ }\textbf {\bibinfo {volume} {108}},\ \bibinfo {pages} {075425} (\bibinfo {year} {2023})}\BibitemShut {NoStop}%
\bibitem [{\citenamefont {Sun}\ \emph {et~al.}(2023)\citenamefont {Sun}, \citenamefont {Brataas},\ and\ \citenamefont {Linder}}]{Sun2023}%
  \BibitemOpen
  \bibfield  {author} {\bibinfo {author} {\bibfnamefont {C.}~\bibnamefont {Sun}}, \bibinfo {author} {\bibfnamefont {A.}~\bibnamefont {Brataas}},\ and\ \bibinfo {author} {\bibfnamefont {J.}~\bibnamefont {Linder}},\ }\bibfield  {title} {\bibinfo {title} {{{Andreev} reflection in altermagnets}},\ }\href {https://doi.org/10.1103/PhysRevB.108.054511} {\bibfield  {journal} {\bibinfo  {journal} {Physical Review B}\ }\textbf {\bibinfo {volume} {108}},\ \bibinfo {pages} {054511} (\bibinfo {year} {2023})}\BibitemShut {NoStop}%
\bibitem [{\citenamefont {Papaj}(2023)}]{Papaj2023}%
  \BibitemOpen
  \bibfield  {author} {\bibinfo {author} {\bibfnamefont {M.}~\bibnamefont {Papaj}},\ }\bibfield  {title} {\bibinfo {title} {{{Andreev} reflection at altermagnet/superconductor interface}},\ }\href {https://doi.org/10.1103/PhysRevB.108.L060508} {\bibfield  {journal} {\bibinfo  {journal} {Physical Review B}\ }\textbf {\bibinfo {volume} {108}},\ \bibinfo {pages} {L060508} (\bibinfo {year} {2023})}\BibitemShut {NoStop}%
\bibitem [{\citenamefont {Zhu}\ \emph {et~al.}(2023)\citenamefont {Zhu}, \citenamefont {Zhuang}, \citenamefont {Wu},\ and\ \citenamefont {Yan}}]{Zhu2023}%
  \BibitemOpen
  \bibfield  {author} {\bibinfo {author} {\bibfnamefont {D.}~\bibnamefont {Zhu}}, \bibinfo {author} {\bibfnamefont {Z.-Y.}\ \bibnamefont {Zhuang}}, \bibinfo {author} {\bibfnamefont {Z.}~\bibnamefont {Wu}},\ and\ \bibinfo {author} {\bibfnamefont {Z.}~\bibnamefont {Yan}},\ }\bibfield  {title} {\bibinfo {title} {{Topological superconductivity in two-dimensional altermagnetic metals}},\ }\href {https://doi.org/10.1103/PhysRevB.108.184505} {\bibfield  {journal} {\bibinfo  {journal} {Physical Review B}\ }\textbf {\bibinfo {volume} {108}},\ \bibinfo {pages} {184505} (\bibinfo {year} {2023})}\BibitemShut {NoStop}%
\bibitem [{\citenamefont {Wei}\ \emph {et~al.}(2024)\citenamefont {Wei}, \citenamefont {Xiang}, \citenamefont {Xu}, \citenamefont {Zhang}, \citenamefont {Tang},\ and\ \citenamefont {Wang}}]{Wei2023}%
  \BibitemOpen
  \bibfield  {author} {\bibinfo {author} {\bibfnamefont {M.}~\bibnamefont {Wei}}, \bibinfo {author} {\bibfnamefont {L.}~\bibnamefont {Xiang}}, \bibinfo {author} {\bibfnamefont {F.}~\bibnamefont {Xu}}, \bibinfo {author} {\bibfnamefont {L.}~\bibnamefont {Zhang}}, \bibinfo {author} {\bibfnamefont {G.}~\bibnamefont {Tang}},\ and\ \bibinfo {author} {\bibfnamefont {J.}~\bibnamefont {Wang}},\ }\bibfield  {title} {\bibinfo {title} {{Gapless superconducting state and mirage gap in altermagnets}},\ }\href {https://doi.org/10.1103/PhysRevB.109.L201404} {\bibfield  {journal} {\bibinfo  {journal} {Physical Review B}\ }\textbf {\bibinfo {volume} {109}},\ \bibinfo {pages} {L201404} (\bibinfo {year} {2024})}\BibitemShut {NoStop}%
\bibitem [{\citenamefont {Li}\ and\ \citenamefont {Liu}(2023)}]{Li2023_Majorana}%
  \BibitemOpen
  \bibfield  {author} {\bibinfo {author} {\bibfnamefont {Y.-X.}\ \bibnamefont {Li}}\ and\ \bibinfo {author} {\bibfnamefont {C.-C.}\ \bibnamefont {Liu}},\ }\bibfield  {title} {\bibinfo {title} {{{Majorana} corner modes and tunable patterns in an altermagnet heterostructure}},\ }\href {https://doi.org/10.1103/PhysRevB.108.205410} {\bibfield  {journal} {\bibinfo  {journal} {Physical Review B}\ }\textbf {\bibinfo {volume} {108}},\ \bibinfo {pages} {205410} (\bibinfo {year} {2023})}\BibitemShut {NoStop}%
\bibitem [{\citenamefont {Chakraborty}\ and\ \citenamefont {Black-Schaffer}(2024{\natexlab{a}})}]{Chakraborty2024}%
  \BibitemOpen
  \bibfield  {author} {\bibinfo {author} {\bibfnamefont {D.}~\bibnamefont {Chakraborty}}\ and\ \bibinfo {author} {\bibfnamefont {A.~M.}\ \bibnamefont {Black-Schaffer}},\ }\bibfield  {title} {\bibinfo {title} {{Zero-field finite-momentum and field-induced superconductivity in altermagnets}},\ }\href {https://doi.org/10.1103/PhysRevB.110.L060508} {\bibfield  {journal} {\bibinfo  {journal} {Physical Review B}\ }\textbf {\bibinfo {volume} {110}},\ \bibinfo {pages} {L060508} (\bibinfo {year} {2024}{\natexlab{a}})}\BibitemShut {NoStop}%
\bibitem [{\citenamefont {Banerjee}\ and\ \citenamefont {Scheurer}(2024)}]{Scheurer2024}%
  \BibitemOpen
  \bibfield  {author} {\bibinfo {author} {\bibfnamefont {S.}~\bibnamefont {Banerjee}}\ and\ \bibinfo {author} {\bibfnamefont {M.~S.}\ \bibnamefont {Scheurer}},\ }\bibfield  {title} {\bibinfo {title} {{Altermagnetic superconducting diode effect}},\ }\href {https://doi.org/10.1103/PhysRevB.110.024503} {\bibfield  {journal} {\bibinfo  {journal} {Physical Review B}\ }\textbf {\bibinfo {volume} {110}},\ \bibinfo {pages} {024503} (\bibinfo {year} {2024})}\BibitemShut {NoStop}%
\bibitem [{\citenamefont {Ghorashi}\ \emph {et~al.}(2024)\citenamefont {Ghorashi}, \citenamefont {Hughes},\ and\ \citenamefont {Cano}}]{Ghorashi2024}%
  \BibitemOpen
  \bibfield  {author} {\bibinfo {author} {\bibfnamefont {S.~A.~A.}\ \bibnamefont {Ghorashi}}, \bibinfo {author} {\bibfnamefont {T.~L.}\ \bibnamefont {Hughes}},\ and\ \bibinfo {author} {\bibfnamefont {J.}~\bibnamefont {Cano}},\ }\bibfield  {title} {\bibinfo {title} {{Altermagnetic Routes to Majorana Modes in Zero Net Magnetization}},\ }\href {https://doi.org/10.1103/PhysRevLett.133.106601} {\bibfield  {journal} {\bibinfo  {journal} {Physical Review Letters}\ }\textbf {\bibinfo {volume} {133}},\ \bibinfo {pages} {106601} (\bibinfo {year} {2024})}\BibitemShut {NoStop}%
\bibitem [{\citenamefont {Chakraborty}\ and\ \citenamefont {Black-Schaffer}(2024{\natexlab{b}})}]{Chakraborty2024constraints}%
  \BibitemOpen
  \bibfield  {author} {\bibinfo {author} {\bibfnamefont {D.}~\bibnamefont {Chakraborty}}\ and\ \bibinfo {author} {\bibfnamefont {A.~M.}\ \bibnamefont {Black-Schaffer}},\ }\bibfield  {title} {\bibinfo {title} {{Constraints on superconducting pairing in altermagnets}},\ }\href {https://arxiv.org/pdf/2408.03999} {\bibfield  {journal} {\bibinfo  {journal} {arXiv:2408.03999}\ } (\bibinfo {year} {2024}{\natexlab{b}})}\BibitemShut {NoStop}%
\bibitem [{\citenamefont {Heung}\ and\ \citenamefont {Franz}(2024)}]{Heung2024}%
  \BibitemOpen
  \bibfield  {author} {\bibinfo {author} {\bibfnamefont {T.~F.}\ \bibnamefont {Heung}}\ and\ \bibinfo {author} {\bibfnamefont {M.}~\bibnamefont {Franz}},\ }\bibfield  {title} {\bibinfo {title} {{Probing topological degeneracy on a torus using superconducting altermagnets}},\ }\href {https://arxiv.org/pdf/2411.17964} {\bibfield  {journal} {\bibinfo  {journal} {arXiv:2411.17964}\ } (\bibinfo {year} {2024})}\BibitemShut {NoStop}%
\bibitem [{\citenamefont {de~Carvalho}\ and\ \citenamefont {Freire}(2024)}]{Carvalho2024}%
  \BibitemOpen
  \bibfield  {author} {\bibinfo {author} {\bibfnamefont {V.~S.}\ \bibnamefont {de~Carvalho}}\ and\ \bibinfo {author} {\bibfnamefont {H.}~\bibnamefont {Freire}},\ }\bibfield  {title} {\bibinfo {title} {{Unconventional superconductivity in altermagnets with spin-orbit coupling}},\ }\href {https://doi.org/10.1103/PhysRevB.110.L220503} {\bibfield  {journal} {\bibinfo  {journal} {Physical Review B}\ }\textbf {\bibinfo {volume} {110}},\ \bibinfo {pages} {L220503} (\bibinfo {year} {2024})}\BibitemShut {NoStop}%
\bibitem [{\citenamefont {{\v{S}}mejkal}\ \emph {et~al.}(2023)\citenamefont {{\v{S}}mejkal}, \citenamefont {Marmodoro}, \citenamefont {Ahn}, \citenamefont {Gonzalez-Hernandez}, \citenamefont {Turek}, \citenamefont {Mankovsky}, \citenamefont {Ebert}, \citenamefont {D'Souza}, \citenamefont {{\v{S}}ipr}, \citenamefont {Sinova},\ and\ \citenamefont {Jungwirth}}]{Smejkal2022chiral}%
  \BibitemOpen
  \bibfield  {author} {\bibinfo {author} {\bibfnamefont {L.}~\bibnamefont {{\v{S}}mejkal}}, \bibinfo {author} {\bibfnamefont {A.}~\bibnamefont {Marmodoro}}, \bibinfo {author} {\bibfnamefont {K.-H.}\ \bibnamefont {Ahn}}, \bibinfo {author} {\bibfnamefont {R.}~\bibnamefont {Gonzalez-Hernandez}}, \bibinfo {author} {\bibfnamefont {I.}~\bibnamefont {Turek}}, \bibinfo {author} {\bibfnamefont {S.}~\bibnamefont {Mankovsky}}, \bibinfo {author} {\bibfnamefont {H.}~\bibnamefont {Ebert}}, \bibinfo {author} {\bibfnamefont {S.~W.}\ \bibnamefont {D'Souza}}, \bibinfo {author} {\bibfnamefont {O.}~\bibnamefont {{\v{S}}ipr}}, \bibinfo {author} {\bibfnamefont {J.}~\bibnamefont {Sinova}},\ and\ \bibinfo {author} {\bibfnamefont {T.}~\bibnamefont {Jungwirth}},\ }\bibfield  {title} {\bibinfo {title} {{Chiral magnons in altermagnetic {RuO}$_2$}},\ }\href {https://doi.org/10.1103/PhysRevLett.131.256703} {\bibfield  {journal} {\bibinfo  {journal} {Physical Review Letters}\ }\textbf {\bibinfo {volume} {131}},\ \bibinfo {pages} {256703}
  (\bibinfo {year} {2023})}\BibitemShut {NoStop}%
\bibitem [{\citenamefont {Steward}\ \emph {et~al.}(2023)\citenamefont {Steward}, \citenamefont {Fernandes},\ and\ \citenamefont {Schmalian}}]{Steward2023}%
  \BibitemOpen
  \bibfield  {author} {\bibinfo {author} {\bibfnamefont {C.~R.~W.}\ \bibnamefont {Steward}}, \bibinfo {author} {\bibfnamefont {R.~M.}\ \bibnamefont {Fernandes}},\ and\ \bibinfo {author} {\bibfnamefont {J.}~\bibnamefont {Schmalian}},\ }\bibfield  {title} {\bibinfo {title} {Dynamic paramagnon-polarons in altermagnets},\ }\href {https://doi.org/10.1103/PhysRevB.108.144418} {\bibfield  {journal} {\bibinfo  {journal} {Physical Review B}\ }\textbf {\bibinfo {volume} {108}},\ \bibinfo {pages} {144418} (\bibinfo {year} {2023})}\BibitemShut {NoStop}%
\bibitem [{\citenamefont {Maier}\ and\ \citenamefont {Okamoto}(2023)}]{Okamoto2023}%
  \BibitemOpen
  \bibfield  {author} {\bibinfo {author} {\bibfnamefont {T.~A.}\ \bibnamefont {Maier}}\ and\ \bibinfo {author} {\bibfnamefont {S.}~\bibnamefont {Okamoto}},\ }\bibfield  {title} {\bibinfo {title} {{Weak-Coupling Theory of Neutron Scattering as a Probe of Altermagnetism}},\ }\href {https://doi.org/10.1103/PhysRevB.108.L100402} {\bibfield  {journal} {\bibinfo  {journal} {Physical Review B}\ }\textbf {\bibinfo {volume} {108}},\ \bibinfo {pages} {L100402} (\bibinfo {year} {2023})}\BibitemShut {NoStop}%
\bibitem [{\citenamefont {Bhowal}\ and\ \citenamefont {Spaldin}(2024)}]{Bhowal2024}%
  \BibitemOpen
  \bibfield  {author} {\bibinfo {author} {\bibfnamefont {S.}~\bibnamefont {Bhowal}}\ and\ \bibinfo {author} {\bibfnamefont {N.~A.}\ \bibnamefont {Spaldin}},\ }\bibfield  {title} {\bibinfo {title} {Ferroically ordered magnetic octupoles in d-wave altermagnets},\ }\href {https://link.aps.org/doi/10.1103/PhysRevX.14.011019} {\bibfield  {journal} {\bibinfo  {journal} {Physical Review X}\ }\textbf {\bibinfo {volume} {14}},\ \bibinfo {pages} {011019} (\bibinfo {year} {2024})}\BibitemShut {NoStop}%
\bibitem [{\citenamefont {McClarty}\ and\ \citenamefont {Rau}(2024)}]{Mcclarty2024}%
  \BibitemOpen
  \bibfield  {author} {\bibinfo {author} {\bibfnamefont {P.~A.}\ \bibnamefont {McClarty}}\ and\ \bibinfo {author} {\bibfnamefont {J.~G.}\ \bibnamefont {Rau}},\ }\bibfield  {title} {\bibinfo {title} {{Landau Theory of Altermagnetism}},\ }\href {https://doi.org/https://doi.org/10.1103/PhysRevLett.132.176702} {\bibfield  {journal} {\bibinfo  {journal} {Physical Review Letters}\ }\textbf {\bibinfo {volume} {132}},\ \bibinfo {pages} {176702} (\bibinfo {year} {2024})}\BibitemShut {NoStop}%
\bibitem [{\citenamefont {Vila}\ \emph {et~al.}(2024)\citenamefont {Vila}, \citenamefont {Sunko},\ and\ \citenamefont {Moore}}]{Vila2024}%
  \BibitemOpen
  \bibfield  {author} {\bibinfo {author} {\bibfnamefont {M.}~\bibnamefont {Vila}}, \bibinfo {author} {\bibfnamefont {V.}~\bibnamefont {Sunko}},\ and\ \bibinfo {author} {\bibfnamefont {J.~E.}\ \bibnamefont {Moore}},\ }\bibfield  {title} {\bibinfo {title} {Orbital-spin locking and its optical signatures in altermagnets},\ }\href {https://arxiv.org/pdf/2410.23513} {\bibfield  {journal} {\bibinfo  {journal} {arXiv:2410.23513}\ } (\bibinfo {year} {2024})}\BibitemShut {NoStop}%
\bibitem [{\citenamefont {Radaelli}(2024)}]{Radaelli2024}%
  \BibitemOpen
  \bibfield  {author} {\bibinfo {author} {\bibfnamefont {P.~G.}\ \bibnamefont {Radaelli}},\ }\bibfield  {title} {\bibinfo {title} {Tensorial approach to altermagnetism},\ }\href {https://doi.org/10.1103/PhysRevB.110.214428} {\bibfield  {journal} {\bibinfo  {journal} {Physical Review B}\ }\textbf {\bibinfo {volume} {110}},\ \bibinfo {pages} {214428} (\bibinfo {year} {2024})}\BibitemShut {NoStop}%
\bibitem [{\citenamefont {Schiff}\ \emph {et~al.}(2024)\citenamefont {Schiff}, \citenamefont {McClarty}, \citenamefont {Rau},\ and\ \citenamefont {Romhanyi}}]{Schiff2024}%
  \BibitemOpen
  \bibfield  {author} {\bibinfo {author} {\bibfnamefont {H.}~\bibnamefont {Schiff}}, \bibinfo {author} {\bibfnamefont {P.}~\bibnamefont {McClarty}}, \bibinfo {author} {\bibfnamefont {J.~G.}\ \bibnamefont {Rau}},\ and\ \bibinfo {author} {\bibfnamefont {J.}~\bibnamefont {Romhanyi}},\ }\bibfield  {title} {\bibinfo {title} {{Collinear Altermagnets and their Landau Theories}},\ }\href {https://arxiv.org/pdf/2412.18025} {\bibfield  {journal} {\bibinfo  {journal} {arXiv:2412.18025}\ } (\bibinfo {year} {2024})}\BibitemShut {NoStop}%
\bibitem [{\citenamefont {Takahashi}\ \emph {et~al.}(2025)\citenamefont {Takahashi}, \citenamefont {Steward}, \citenamefont {Ogata}, \citenamefont {Fernandes},\ and\ \citenamefont {Schmalian}}]{Takahashi2025}%
  \BibitemOpen
  \bibfield  {author} {\bibinfo {author} {\bibfnamefont {K.}~\bibnamefont {Takahashi}}, \bibinfo {author} {\bibfnamefont {C.~R.}\ \bibnamefont {Steward}}, \bibinfo {author} {\bibfnamefont {M.}~\bibnamefont {Ogata}}, \bibinfo {author} {\bibfnamefont {R.~M.}\ \bibnamefont {Fernandes}},\ and\ \bibinfo {author} {\bibfnamefont {J.}~\bibnamefont {Schmalian}},\ }\bibfield  {title} {\bibinfo {title} {Elasto-hall conductivity and the anomalous hall effect in altermagnets},\ }\href {https://arxiv.org/pdf/2502.03517} {\bibfield  {journal} {\bibinfo  {journal} {arXiv:2502.03517}\ } (\bibinfo {year} {2025})}\BibitemShut {NoStop}%
\bibitem [{\citenamefont {Lin}\ \emph {et~al.}(2025)\citenamefont {Lin}, \citenamefont {Zhang}, \citenamefont {Lu},\ and\ \citenamefont {Xie}}]{Lin2025coulomb}%
  \BibitemOpen
  \bibfield  {author} {\bibinfo {author} {\bibfnamefont {H.-J.}\ \bibnamefont {Lin}}, \bibinfo {author} {\bibfnamefont {S.-B.}\ \bibnamefont {Zhang}}, \bibinfo {author} {\bibfnamefont {H.-Z.}\ \bibnamefont {Lu}},\ and\ \bibinfo {author} {\bibfnamefont {X.}~\bibnamefont {Xie}},\ }\bibfield  {title} {\bibinfo {title} {Coulomb drag in altermagnets},\ }\href {https://link.aps.org/doi/10.1103/PhysRevLett.134.136301} {\bibfield  {journal} {\bibinfo  {journal} {Physical Review Letters}\ }\textbf {\bibinfo {volume} {134}},\ \bibinfo {pages} {136301} (\bibinfo {year} {2025})}\BibitemShut {NoStop}%
\bibitem [{\citenamefont {{\v{S}}mejkal}(2024)}]{Smejkal2024_multiferroics}%
  \BibitemOpen
  \bibfield  {author} {\bibinfo {author} {\bibfnamefont {L.}~\bibnamefont {{\v{S}}mejkal}},\ }\bibfield  {title} {\bibinfo {title} {{Altermagnetic multiferroics and altermagnetoelectric effect}},\ }\href {https://arxiv.org/pdf/2411.19928} {\bibfield  {journal} {\bibinfo  {journal} {arXiv:2411.19928}\ } (\bibinfo {year} {2024})}\BibitemShut {NoStop}%
\bibitem [{\citenamefont {Gu}\ \emph {et~al.}(2025)\citenamefont {Gu}, \citenamefont {Liu}, \citenamefont {Zhu}, \citenamefont {Yananose}, \citenamefont {Chen}, \citenamefont {Hu}, \citenamefont {Stroppa},\ and\ \citenamefont {Liu}}]{Gu2025}%
  \BibitemOpen
  \bibfield  {author} {\bibinfo {author} {\bibfnamefont {M.}~\bibnamefont {Gu}}, \bibinfo {author} {\bibfnamefont {Y.}~\bibnamefont {Liu}}, \bibinfo {author} {\bibfnamefont {H.}~\bibnamefont {Zhu}}, \bibinfo {author} {\bibfnamefont {K.}~\bibnamefont {Yananose}}, \bibinfo {author} {\bibfnamefont {X.}~\bibnamefont {Chen}}, \bibinfo {author} {\bibfnamefont {Y.}~\bibnamefont {Hu}}, \bibinfo {author} {\bibfnamefont {A.}~\bibnamefont {Stroppa}},\ and\ \bibinfo {author} {\bibfnamefont {Q.}~\bibnamefont {Liu}},\ }\bibfield  {title} {\bibinfo {title} {{Ferroelectric Switchable Altermagnetism}},\ }\href {https://doi.org/10.1103/PhysRevLett.134.106802} {\bibfield  {journal} {\bibinfo  {journal} {Physical Review Letters}\ }\textbf {\bibinfo {volume} {134}},\ \bibinfo {pages} {106802} (\bibinfo {year} {2025})}\BibitemShut {NoStop}%
\bibitem [{\citenamefont {Duan}\ \emph {et~al.}(2025)\citenamefont {Duan}, \citenamefont {Zhang}, \citenamefont {Zhu}, \citenamefont {Liu}, \citenamefont {Zhang}, \citenamefont {\ifmmode \check{Z}\else \v{Z}\fi{}uti\ifmmode~\acute{c}\else \'{c}\fi{}},\ and\ \citenamefont {Zhou}}]{Duan2025}%
  \BibitemOpen
  \bibfield  {author} {\bibinfo {author} {\bibfnamefont {X.}~\bibnamefont {Duan}}, \bibinfo {author} {\bibfnamefont {J.}~\bibnamefont {Zhang}}, \bibinfo {author} {\bibfnamefont {Z.}~\bibnamefont {Zhu}}, \bibinfo {author} {\bibfnamefont {Y.}~\bibnamefont {Liu}}, \bibinfo {author} {\bibfnamefont {Z.}~\bibnamefont {Zhang}}, \bibinfo {author} {\bibfnamefont {I.}~\bibnamefont {\ifmmode \check{Z}\else \v{Z}\fi{}uti\ifmmode~\acute{c}\else \'{c}\fi{}}},\ and\ \bibinfo {author} {\bibfnamefont {T.}~\bibnamefont {Zhou}},\ }\bibfield  {title} {\bibinfo {title} {{Antiferroelectric Altermagnets: Antiferroelectricity Alters Magnets}},\ }\href {https://doi.org/10.1103/PhysRevLett.134.106801} {\bibfield  {journal} {\bibinfo  {journal} {Physical Review Letters}\ }\textbf {\bibinfo {volume} {134}},\ \bibinfo {pages} {106801} (\bibinfo {year} {2025})}\BibitemShut {NoStop}%
\bibitem [{\citenamefont {Mazin}\ \emph {et~al.}(2021)\citenamefont {Mazin}, \citenamefont {Koepernik}, \citenamefont {Johannes}, \citenamefont {Gonz{\'a}lez-Hern{\'a}ndez},\ and\ \citenamefont {{\v{S}}mejkal}}]{Mazin2021}%
  \BibitemOpen
  \bibfield  {author} {\bibinfo {author} {\bibfnamefont {I.~I.}\ \bibnamefont {Mazin}}, \bibinfo {author} {\bibfnamefont {K.}~\bibnamefont {Koepernik}}, \bibinfo {author} {\bibfnamefont {M.~D.}\ \bibnamefont {Johannes}}, \bibinfo {author} {\bibfnamefont {R.}~\bibnamefont {Gonz{\'a}lez-Hern{\'a}ndez}},\ and\ \bibinfo {author} {\bibfnamefont {L.}~\bibnamefont {{\v{S}}mejkal}},\ }\bibfield  {title} {\bibinfo {title} {{Prediction of unconventional magnetism in doped $\text{FeSb}_2$}},\ }\href {https://doi.org/10.1073/pnas.2108924118} {\bibfield  {journal} {\bibinfo  {journal} {Proceedings of the National Academy of Sciences}\ }\textbf {\bibinfo {volume} {118}},\ \bibinfo {pages} {e2108924118} (\bibinfo {year} {2021})}\BibitemShut {NoStop}%
\bibitem [{\citenamefont {Guo}\ \emph {et~al.}(2023)\citenamefont {Guo}, \citenamefont {Liu}, \citenamefont {Janson}, \citenamefont {Fulga}, \citenamefont {{van den Brink}},\ and\ \citenamefont {Facio}}]{Facio2023}%
  \BibitemOpen
  \bibfield  {author} {\bibinfo {author} {\bibfnamefont {Y.}~\bibnamefont {Guo}}, \bibinfo {author} {\bibfnamefont {H.}~\bibnamefont {Liu}}, \bibinfo {author} {\bibfnamefont {O.}~\bibnamefont {Janson}}, \bibinfo {author} {\bibfnamefont {I.~C.}\ \bibnamefont {Fulga}}, \bibinfo {author} {\bibfnamefont {J.}~\bibnamefont {{van den Brink}}},\ and\ \bibinfo {author} {\bibfnamefont {J.~I.}\ \bibnamefont {Facio}},\ }\bibfield  {title} {\bibinfo {title} {{Spin-split collinear antiferromagnets: A large-scale ab-initio study}},\ }\href {https://doi.org/https://doi.org/10.1016/j.mtphys.2023.100991} {\bibfield  {journal} {\bibinfo  {journal} {Mater. Today Phys.}\ }\textbf {\bibinfo {volume} {32}},\ \bibinfo {pages} {100991} (\bibinfo {year} {2023})}\BibitemShut {NoStop}%
\bibitem [{\citenamefont {Mazin}(2023)}]{Mazin2023}%
  \BibitemOpen
  \bibfield  {author} {\bibinfo {author} {\bibfnamefont {I.}~\bibnamefont {Mazin}},\ }\bibfield  {title} {\bibinfo {title} {{Altermagnetism in $\text{MnTe}$: Origin, predicted manifestations, and routes to detwinning}},\ }\href {https://doi.org/https://doi.org/10.1103/PhysRevB.107.L100418} {\bibfield  {journal} {\bibinfo  {journal} {Physical Review B}\ }\textbf {\bibinfo {volume} {107}},\ \bibinfo {pages} {L100418} (\bibinfo {year} {2023})}\BibitemShut {NoStop}%
\bibitem [{\citenamefont {Gao}\ \emph {et~al.}(2023)\citenamefont {Gao}, \citenamefont {Qu}, \citenamefont {Zeng}, \citenamefont {Wen}, \citenamefont {Sun}, \citenamefont {Guo},\ and\ \citenamefont {Lu}}]{Gao2023}%
  \BibitemOpen
  \bibfield  {author} {\bibinfo {author} {\bibfnamefont {Z.-F.}\ \bibnamefont {Gao}}, \bibinfo {author} {\bibfnamefont {S.}~\bibnamefont {Qu}}, \bibinfo {author} {\bibfnamefont {B.}~\bibnamefont {Zeng}}, \bibinfo {author} {\bibfnamefont {J.-R.}\ \bibnamefont {Wen}}, \bibinfo {author} {\bibfnamefont {H.}~\bibnamefont {Sun}}, \bibinfo {author} {\bibfnamefont {P.}~\bibnamefont {Guo}},\ and\ \bibinfo {author} {\bibfnamefont {Z.-Y.}\ \bibnamefont {Lu}},\ }\bibfield  {title} {\bibinfo {title} {{{AI}-accelerated Discovery of Altermagnetic Materials}},\ }\href {https://arxiv.org/pdf/2311.04418} {\bibfield  {journal} {\bibinfo  {journal} {arXiv:2311.04418}\ } (\bibinfo {year} {2023})}\BibitemShut {NoStop}%
\bibitem [{\citenamefont {Jaeschke-Ubiergo}\ \emph {et~al.}(2024)\citenamefont {Jaeschke-Ubiergo}, \citenamefont {Bharadwaj}, \citenamefont {Jungwirth}, \citenamefont {\ifmmode~\check{S}\else \v{S}\fi{}mejkal},\ and\ \citenamefont {Sinova}}]{Jaeschke2024}%
  \BibitemOpen
  \bibfield  {author} {\bibinfo {author} {\bibfnamefont {R.}~\bibnamefont {Jaeschke-Ubiergo}}, \bibinfo {author} {\bibfnamefont {V.~K.}\ \bibnamefont {Bharadwaj}}, \bibinfo {author} {\bibfnamefont {T.}~\bibnamefont {Jungwirth}}, \bibinfo {author} {\bibfnamefont {L.}~\bibnamefont {\ifmmode~\check{S}\else \v{S}\fi{}mejkal}},\ and\ \bibinfo {author} {\bibfnamefont {J.}~\bibnamefont {Sinova}},\ }\bibfield  {title} {\bibinfo {title} {{Supercell altermagnets}},\ }\href {https://doi.org/10.1103/PhysRevB.109.094425} {\bibfield  {journal} {\bibinfo  {journal} {Physical Review B}\ }\textbf {\bibinfo {volume} {109}},\ \bibinfo {pages} {094425} (\bibinfo {year} {2024})}\BibitemShut {NoStop}%
\bibitem [{\citenamefont {Wan}\ \emph {et~al.}(2024)\citenamefont {Wan}, \citenamefont {Mandal}, \citenamefont {Guo},\ and\ \citenamefont {Haule}}]{Haule2024}%
  \BibitemOpen
  \bibfield  {author} {\bibinfo {author} {\bibfnamefont {X.}~\bibnamefont {Wan}}, \bibinfo {author} {\bibfnamefont {S.}~\bibnamefont {Mandal}}, \bibinfo {author} {\bibfnamefont {Y.}~\bibnamefont {Guo}},\ and\ \bibinfo {author} {\bibfnamefont {K.}~\bibnamefont {Haule}},\ }\bibfield  {title} {\bibinfo {title} {{High-throughput Search for Metallic Altermagnets by Embedded Dynamical Mean Field Theory}},\ }\href {https://arxiv.org/pdf/2412.10356} {\bibfield  {journal} {\bibinfo  {journal} {arXiv:2412.10356}\ } (\bibinfo {year} {2024})}\BibitemShut {NoStop}%
\bibitem [{\citenamefont {S{\o}dequist}\ and\ \citenamefont {Olsen}(2024)}]{Sodequist2024}%
  \BibitemOpen
  \bibfield  {author} {\bibinfo {author} {\bibfnamefont {J.}~\bibnamefont {S{\o}dequist}}\ and\ \bibinfo {author} {\bibfnamefont {T.}~\bibnamefont {Olsen}},\ }\bibfield  {title} {\bibinfo {title} {{Two-dimensional altermagnets from high throughput computational screening: Symmetry requirements, chiral magnons, and spin-orbit effects}},\ }\href {https://pubs.aip.org/aip/apl/article/124/18/182409/3288014/Two-dimensional-altermagnets-from-high-throughput} {\bibfield  {journal} {\bibinfo  {journal} {Applied Physics Letters}\ }\textbf {\bibinfo {volume} {124}} (\bibinfo {year} {2024})}\BibitemShut {NoStop}%
\bibitem [{\citenamefont {Li}\ \emph {et~al.}(2024{\natexlab{a}})\citenamefont {Li}, \citenamefont {Fan}, \citenamefont {Wang}, \citenamefont {Zhang},\ and\ \citenamefont {Li}}]{Li2024strain}%
  \BibitemOpen
  \bibfield  {author} {\bibinfo {author} {\bibfnamefont {J.-Y.}\ \bibnamefont {Li}}, \bibinfo {author} {\bibfnamefont {A.-D.}\ \bibnamefont {Fan}}, \bibinfo {author} {\bibfnamefont {Y.-K.}\ \bibnamefont {Wang}}, \bibinfo {author} {\bibfnamefont {Y.}~\bibnamefont {Zhang}},\ and\ \bibinfo {author} {\bibfnamefont {S.}~\bibnamefont {Li}},\ }\bibfield  {title} {\bibinfo {title} {{Strain-induced valley polarization, topological states, and piezomagnetism in two-dimensional altermagnetic V$_2$Te$_2$O, V$_2$STeO, V$_2$SSeO, and V$_2$S$_2$O}},\ }\href {https://pubs.aip.org/aip/apl/article/125/22/222404/3322459/Strain-induced-valley-polarization-topological} {\bibfield  {journal} {\bibinfo  {journal} {Applied Physics Letters}\ }\textbf {\bibinfo {volume} {125}} (\bibinfo {year} {2024}{\natexlab{a}})}\BibitemShut {NoStop}%
\bibitem [{\citenamefont {Jiang}\ \emph {et~al.}(2024{\natexlab{b}})\citenamefont {Jiang}, \citenamefont {Zhang}, \citenamefont {Bai}, \citenamefont {Tian}, \citenamefont {Gong},\ and\ \citenamefont {Kong}}]{Jiang2024monolayer}%
  \BibitemOpen
  \bibfield  {author} {\bibinfo {author} {\bibfnamefont {Y.}~\bibnamefont {Jiang}}, \bibinfo {author} {\bibfnamefont {X.}~\bibnamefont {Zhang}}, \bibinfo {author} {\bibfnamefont {H.}~\bibnamefont {Bai}}, \bibinfo {author} {\bibfnamefont {Y.}~\bibnamefont {Tian}}, \bibinfo {author} {\bibfnamefont {W.-J.}\ \bibnamefont {Gong}},\ and\ \bibinfo {author} {\bibfnamefont {X.}~\bibnamefont {Kong}},\ }\bibfield  {title} {\bibinfo {title} {{Strain-engineering spin-valley locking effect in altermagnetic monolayer with multipiezo properties}},\ }\href {https://arxiv.org/pdf/2412.05597} {\bibfield  {journal} {\bibinfo  {journal} {arXiv:2412.05597}\ } (\bibinfo {year} {2024}{\natexlab{b}})}\BibitemShut {NoStop}%
\bibitem [{\citenamefont {Wei}\ \emph {et~al.}(2025)\citenamefont {Wei}, \citenamefont {Li}, \citenamefont {Hatt}, \citenamefont {Huai}, \citenamefont {Liu}, \citenamefont {Singh}, \citenamefont {Kim}, \citenamefont {Fernandes}, \citenamefont {Cardon}, \citenamefont {Zhao}, \citenamefont {Tran}, \citenamefont {Frandsen}, \citenamefont {Burch}, \citenamefont {Liu},\ and\ \citenamefont {Ji}}]{Ji2025}%
  \BibitemOpen
  \bibfield  {author} {\bibinfo {author} {\bibfnamefont {C.-C.}\ \bibnamefont {Wei}}, \bibinfo {author} {\bibfnamefont {X.}~\bibnamefont {Li}}, \bibinfo {author} {\bibfnamefont {S.}~\bibnamefont {Hatt}}, \bibinfo {author} {\bibfnamefont {X.}~\bibnamefont {Huai}}, \bibinfo {author} {\bibfnamefont {J.}~\bibnamefont {Liu}}, \bibinfo {author} {\bibfnamefont {B.}~\bibnamefont {Singh}}, \bibinfo {author} {\bibfnamefont {K.-M.}\ \bibnamefont {Kim}}, \bibinfo {author} {\bibfnamefont {R.~M.}\ \bibnamefont {Fernandes}}, \bibinfo {author} {\bibfnamefont {P.}~\bibnamefont {Cardon}}, \bibinfo {author} {\bibfnamefont {L.}~\bibnamefont {Zhao}}, \bibinfo {author} {\bibfnamefont {T.~T.}\ \bibnamefont {Tran}}, \bibinfo {author} {\bibfnamefont {B.~A.}\ \bibnamefont {Frandsen}}, \bibinfo {author} {\bibfnamefont {K.~S.}\ \bibnamefont {Burch}}, \bibinfo {author} {\bibfnamefont {F.}~\bibnamefont {Liu}},\ and\ \bibinfo {author} {\bibfnamefont {H.}~\bibnamefont {Ji}},\ }\bibfield  {title} {\bibinfo {title}
  {{${\mathrm{La}}_{2}{\mathrm{O}}_{3}{\mathrm{Mn}}_{2}{\mathrm{Se}}_{2}$: A correlated insulating layered d-wave altermagnet}},\ }\href {https://doi.org/10.1103/PhysRevMaterials.9.024402} {\bibfield  {journal} {\bibinfo  {journal} {Physical Review Materials}\ }\textbf {\bibinfo {volume} {9}},\ \bibinfo {pages} {024402} (\bibinfo {year} {2025})}\BibitemShut {NoStop}%
\bibitem [{\citenamefont {Krempask{\'y}}\ \emph {et~al.}(2024)\citenamefont {Krempask{\'y}}, \citenamefont {{\v{S}}mejkal}, \citenamefont {D'Souza}, \citenamefont {Hajlaoui}, \citenamefont {Springholz}, \citenamefont {Uhl{\'\i}{\v{r}}ov{\'a}}, \citenamefont {Alarab}, \citenamefont {Constantinou}, \citenamefont {Strocov}, \citenamefont {Usanov} \emph {et~al.}}]{Krempasky2024}%
  \BibitemOpen
  \bibfield  {author} {\bibinfo {author} {\bibfnamefont {J.}~\bibnamefont {Krempask{\'y}}}, \bibinfo {author} {\bibfnamefont {L.}~\bibnamefont {{\v{S}}mejkal}}, \bibinfo {author} {\bibfnamefont {S.}~\bibnamefont {D'Souza}}, \bibinfo {author} {\bibfnamefont {M.}~\bibnamefont {Hajlaoui}}, \bibinfo {author} {\bibfnamefont {G.}~\bibnamefont {Springholz}}, \bibinfo {author} {\bibfnamefont {K.}~\bibnamefont {Uhl{\'\i}{\v{r}}ov{\'a}}}, \bibinfo {author} {\bibfnamefont {F.}~\bibnamefont {Alarab}}, \bibinfo {author} {\bibfnamefont {P.}~\bibnamefont {Constantinou}}, \bibinfo {author} {\bibfnamefont {V.}~\bibnamefont {Strocov}}, \bibinfo {author} {\bibfnamefont {D.}~\bibnamefont {Usanov}}, \emph {et~al.},\ }\bibfield  {title} {\bibinfo {title} {{Altermagnetic lifting of Kramers spin degeneracy}},\ }\href {https://doi.org/https://doi.org/10.1038/s41586-023-06907-7} {\bibfield  {journal} {\bibinfo  {journal} {Nature}\ }\textbf {\bibinfo {volume} {626}},\ \bibinfo {pages} {517} (\bibinfo {year} {2024})}\BibitemShut
  {NoStop}%
\bibitem [{\citenamefont {Amin}\ \emph {et~al.}(2024)\citenamefont {Amin}, \citenamefont {Dal~Din}, \citenamefont {Golias}, \citenamefont {Niu}, \citenamefont {Zakharov}, \citenamefont {Fromage}, \citenamefont {Fields}, \citenamefont {Heywood}, \citenamefont {Cousins}, \citenamefont {Maccherozzi} \emph {et~al.}}]{Amin2024}%
  \BibitemOpen
  \bibfield  {author} {\bibinfo {author} {\bibfnamefont {O.}~\bibnamefont {Amin}}, \bibinfo {author} {\bibfnamefont {A.}~\bibnamefont {Dal~Din}}, \bibinfo {author} {\bibfnamefont {E.}~\bibnamefont {Golias}}, \bibinfo {author} {\bibfnamefont {Y.}~\bibnamefont {Niu}}, \bibinfo {author} {\bibfnamefont {A.}~\bibnamefont {Zakharov}}, \bibinfo {author} {\bibfnamefont {S.}~\bibnamefont {Fromage}}, \bibinfo {author} {\bibfnamefont {C.}~\bibnamefont {Fields}}, \bibinfo {author} {\bibfnamefont {S.}~\bibnamefont {Heywood}}, \bibinfo {author} {\bibfnamefont {R.}~\bibnamefont {Cousins}}, \bibinfo {author} {\bibfnamefont {F.}~\bibnamefont {Maccherozzi}}, \emph {et~al.},\ }\bibfield  {title} {\bibinfo {title} {{Nanoscale imaging and control of altermagnetism in MnTe}},\ }\href {https://www.nature.com/articles/s41586-024-08234-x} {\bibfield  {journal} {\bibinfo  {journal} {Nature}\ }\textbf {\bibinfo {volume} {636}},\ \bibinfo {pages} {348} (\bibinfo {year} {2024})}\BibitemShut {NoStop}%
\bibitem [{\citenamefont {Lee}\ \emph {et~al.}(2024)\citenamefont {Lee}, \citenamefont {Lee}, \citenamefont {Jung}, \citenamefont {Jung}, \citenamefont {Kim}, \citenamefont {Lee}, \citenamefont {Seok}, \citenamefont {Kim}, \citenamefont {Park}, \citenamefont {{\v{S}}mejkal}, \citenamefont {Kang},\ and\ \citenamefont {Kim}}]{Lee2023}%
  \BibitemOpen
  \bibfield  {author} {\bibinfo {author} {\bibfnamefont {S.}~\bibnamefont {Lee}}, \bibinfo {author} {\bibfnamefont {S.}~\bibnamefont {Lee}}, \bibinfo {author} {\bibfnamefont {S.}~\bibnamefont {Jung}}, \bibinfo {author} {\bibfnamefont {J.}~\bibnamefont {Jung}}, \bibinfo {author} {\bibfnamefont {D.}~\bibnamefont {Kim}}, \bibinfo {author} {\bibfnamefont {Y.}~\bibnamefont {Lee}}, \bibinfo {author} {\bibfnamefont {B.}~\bibnamefont {Seok}}, \bibinfo {author} {\bibfnamefont {J.}~\bibnamefont {Kim}}, \bibinfo {author} {\bibfnamefont {B.~G.}\ \bibnamefont {Park}}, \bibinfo {author} {\bibfnamefont {L.}~\bibnamefont {{\v{S}}mejkal}}, \bibinfo {author} {\bibfnamefont {C.-J.}\ \bibnamefont {Kang}},\ and\ \bibinfo {author} {\bibfnamefont {C.}~\bibnamefont {Kim}},\ }\bibfield  {title} {\bibinfo {title} {{Broken {Kramers'} degeneracy in altermagnetic {MnTe}}},\ }\href {https://doi.org/10.1103/PhysRevLett.132.036702} {\bibfield  {journal} {\bibinfo  {journal} {Physical Review Letters}\ }\textbf {\bibinfo {volume} {132}},\
  \bibinfo {pages} {036702} (\bibinfo {year} {2024})}\BibitemShut {NoStop}%
\bibitem [{\citenamefont {Osumi}\ \emph {et~al.}(2024)\citenamefont {Osumi}, \citenamefont {Souma}, \citenamefont {Aoyama}, \citenamefont {Yamauchi}, \citenamefont {Honma}, \citenamefont {Nakayama}, \citenamefont {Takahashi}, \citenamefont {Ohgushi},\ and\ \citenamefont {Sato}}]{Osumi2023}%
  \BibitemOpen
  \bibfield  {author} {\bibinfo {author} {\bibfnamefont {T.}~\bibnamefont {Osumi}}, \bibinfo {author} {\bibfnamefont {S.}~\bibnamefont {Souma}}, \bibinfo {author} {\bibfnamefont {T.}~\bibnamefont {Aoyama}}, \bibinfo {author} {\bibfnamefont {K.}~\bibnamefont {Yamauchi}}, \bibinfo {author} {\bibfnamefont {A.}~\bibnamefont {Honma}}, \bibinfo {author} {\bibfnamefont {K.}~\bibnamefont {Nakayama}}, \bibinfo {author} {\bibfnamefont {T.}~\bibnamefont {Takahashi}}, \bibinfo {author} {\bibfnamefont {K.}~\bibnamefont {Ohgushi}},\ and\ \bibinfo {author} {\bibfnamefont {T.}~\bibnamefont {Sato}},\ }\bibfield  {title} {\bibinfo {title} {{Observation of a giant band splitting in altermagnetic {MnTe}}},\ }\href {https://doi.org/10.1103/PhysRevB.109.115102} {\bibfield  {journal} {\bibinfo  {journal} {Physical Review B}\ }\textbf {\bibinfo {volume} {109}},\ \bibinfo {pages} {115102} (\bibinfo {year} {2024})}\BibitemShut {NoStop}%
\bibitem [{\citenamefont {Reimers}\ \emph {et~al.}(2024)\citenamefont {Reimers}, \citenamefont {Odenbreit}, \citenamefont {{\v S}mejkal}, \citenamefont {Strocov}, \citenamefont {Constantinou}, \citenamefont {Hellenes}, \citenamefont {Ubiergo}, \citenamefont {Campos}, \citenamefont {Bharadwaj}, \citenamefont {Chakraborty}, \citenamefont {Denneulin}, \citenamefont {Shi}, \citenamefont {Dunin-Borkowski}, \citenamefont {Das}, \citenamefont {Kl\"{a}ui}, \citenamefont {Sinova},\ and\ \citenamefont {Jourdan}}]{Reimers2023}%
  \BibitemOpen
  \bibfield  {author} {\bibinfo {author} {\bibfnamefont {S.}~\bibnamefont {Reimers}}, \bibinfo {author} {\bibfnamefont {L.}~\bibnamefont {Odenbreit}}, \bibinfo {author} {\bibfnamefont {L.}~\bibnamefont {{\v S}mejkal}}, \bibinfo {author} {\bibfnamefont {V.~N.}\ \bibnamefont {Strocov}}, \bibinfo {author} {\bibfnamefont {P.}~\bibnamefont {Constantinou}}, \bibinfo {author} {\bibfnamefont {A.~B.}\ \bibnamefont {Hellenes}}, \bibinfo {author} {\bibfnamefont {R.~J.}\ \bibnamefont {Ubiergo}}, \bibinfo {author} {\bibfnamefont {W.~H.}\ \bibnamefont {Campos}}, \bibinfo {author} {\bibfnamefont {V.~K.}\ \bibnamefont {Bharadwaj}}, \bibinfo {author} {\bibfnamefont {A.}~\bibnamefont {Chakraborty}}, \bibinfo {author} {\bibfnamefont {T.}~\bibnamefont {Denneulin}}, \bibinfo {author} {\bibfnamefont {W.}~\bibnamefont {Shi}}, \bibinfo {author} {\bibfnamefont {R.~E.}\ \bibnamefont {Dunin-Borkowski}}, \bibinfo {author} {\bibfnamefont {S.}~\bibnamefont {Das}}, \bibinfo {author} {\bibfnamefont {M.}~\bibnamefont {Kl\"{a}ui}}, \bibinfo
  {author} {\bibfnamefont {J.}~\bibnamefont {Sinova}},\ and\ \bibinfo {author} {\bibfnamefont {M.}~\bibnamefont {Jourdan}},\ }\bibfield  {title} {\bibinfo {title} {{Direct observation of altermagnetic band splitting in {CrSb} thin films}},\ }\href {https://doi.org/10.1038/s41467-024-46476-5} {\bibfield  {journal} {\bibinfo  {journal} {Nat. Commun.}\ }\textbf {\bibinfo {volume} {15}},\ \bibinfo {pages} {2116} (\bibinfo {year} {2024})}\BibitemShut {NoStop}%
\bibitem [{\citenamefont {Yang}\ \emph {et~al.}(2024)\citenamefont {Yang}, \citenamefont {Li}, \citenamefont {Yang}, \citenamefont {Li}, \citenamefont {Zheng}, \citenamefont {Zhu}, \citenamefont {Pan}, \citenamefont {Xu}, \citenamefont {Cao}, \citenamefont {Zhao}, \citenamefont {Jana}, \citenamefont {Zhang}, \citenamefont {Ye}, \citenamefont {Song}, \citenamefont {Hu}, \citenamefont {Yang}, \citenamefont {Fujii}, \citenamefont {Vobornik}, \citenamefont {Shi}, \citenamefont {Yuan}, \citenamefont {Zhang}, \citenamefont {Xu},\ and\ \citenamefont {Liu}}]{Yang2024}%
  \BibitemOpen
  \bibfield  {author} {\bibinfo {author} {\bibfnamefont {G.}~\bibnamefont {Yang}}, \bibinfo {author} {\bibfnamefont {Z.}~\bibnamefont {Li}}, \bibinfo {author} {\bibfnamefont {S.}~\bibnamefont {Yang}}, \bibinfo {author} {\bibfnamefont {J.}~\bibnamefont {Li}}, \bibinfo {author} {\bibfnamefont {H.}~\bibnamefont {Zheng}}, \bibinfo {author} {\bibfnamefont {W.}~\bibnamefont {Zhu}}, \bibinfo {author} {\bibfnamefont {Z.}~\bibnamefont {Pan}}, \bibinfo {author} {\bibfnamefont {Y.}~\bibnamefont {Xu}}, \bibinfo {author} {\bibfnamefont {S.}~\bibnamefont {Cao}}, \bibinfo {author} {\bibfnamefont {W.}~\bibnamefont {Zhao}}, \bibinfo {author} {\bibfnamefont {A.}~\bibnamefont {Jana}}, \bibinfo {author} {\bibfnamefont {J.}~\bibnamefont {Zhang}}, \bibinfo {author} {\bibfnamefont {M.}~\bibnamefont {Ye}}, \bibinfo {author} {\bibfnamefont {Y.}~\bibnamefont {Song}}, \bibinfo {author} {\bibfnamefont {L.-H.}\ \bibnamefont {Hu}}, \bibinfo {author} {\bibfnamefont {L.}~\bibnamefont {Yang}}, \bibinfo {author} {\bibfnamefont
  {J.}~\bibnamefont {Fujii}}, \bibinfo {author} {\bibfnamefont {I.}~\bibnamefont {Vobornik}}, \bibinfo {author} {\bibfnamefont {M.}~\bibnamefont {Shi}}, \bibinfo {author} {\bibfnamefont {H.}~\bibnamefont {Yuan}}, \bibinfo {author} {\bibfnamefont {Y.}~\bibnamefont {Zhang}}, \bibinfo {author} {\bibfnamefont {Y.}~\bibnamefont {Xu}},\ and\ \bibinfo {author} {\bibfnamefont {Y.}~\bibnamefont {Liu}},\ }\bibfield  {title} {\bibinfo {title} {{Three-dimensional mapping and electronic origin of large altermagnetic splitting near Fermi level in CrSb}},\ }\href {https://arxiv.org/pdf/2405.12575} {\bibfield  {journal} {\bibinfo  {journal} {arXiv:2405.12575}\ } (\bibinfo {year} {2024})}\BibitemShut {NoStop}%
\bibitem [{\citenamefont {Li}\ \emph {et~al.}(2024{\natexlab{b}})\citenamefont {Li}, \citenamefont {Hu}, \citenamefont {Li}, \citenamefont {Wang}, \citenamefont {Chen}, \citenamefont {Thiagarajan}, \citenamefont {Leandersson}, \citenamefont {Polley}, \citenamefont {Kim}, \citenamefont {Liu}, \citenamefont {Fulga}, \citenamefont {Vergniory}, \citenamefont {Janson}, \citenamefont {Tjernberg},\ and\ \citenamefont {van~den Brink}}]{Li2024}%
  \BibitemOpen
  \bibfield  {author} {\bibinfo {author} {\bibfnamefont {C.}~\bibnamefont {Li}}, \bibinfo {author} {\bibfnamefont {M.}~\bibnamefont {Hu}}, \bibinfo {author} {\bibfnamefont {Z.}~\bibnamefont {Li}}, \bibinfo {author} {\bibfnamefont {Y.}~\bibnamefont {Wang}}, \bibinfo {author} {\bibfnamefont {W.}~\bibnamefont {Chen}}, \bibinfo {author} {\bibfnamefont {B.}~\bibnamefont {Thiagarajan}}, \bibinfo {author} {\bibfnamefont {M.}~\bibnamefont {Leandersson}}, \bibinfo {author} {\bibfnamefont {C.}~\bibnamefont {Polley}}, \bibinfo {author} {\bibfnamefont {T.}~\bibnamefont {Kim}}, \bibinfo {author} {\bibfnamefont {H.}~\bibnamefont {Liu}}, \bibinfo {author} {\bibfnamefont {C.}~\bibnamefont {Fulga}}, \bibinfo {author} {\bibfnamefont {M.~G.}\ \bibnamefont {Vergniory}}, \bibinfo {author} {\bibfnamefont {O.}~\bibnamefont {Janson}}, \bibinfo {author} {\bibfnamefont {O.}~\bibnamefont {Tjernberg}},\ and\ \bibinfo {author} {\bibfnamefont {J.}~\bibnamefont {van~den Brink}},\ }\bibfield  {title} {\bibinfo {title} {{Topological Weyl
  Altermagnetism in CrSb}},\ }\href {https://arxiv.org/pdf/2405.14777} {\bibfield  {journal} {\bibinfo  {journal} {arXiv:2405.14777}\ } (\bibinfo {year} {2024}{\natexlab{b}})}\BibitemShut {NoStop}%
\bibitem [{\citenamefont {Ding}\ \emph {et~al.}(2024)\citenamefont {Ding}, \citenamefont {Jiang}, \citenamefont {Chen}, \citenamefont {Tao}, \citenamefont {Liu}, \citenamefont {Li}, \citenamefont {Liu}, \citenamefont {Sun}, \citenamefont {Cheng}, \citenamefont {Liu}, \citenamefont {Yang}, \citenamefont {Zhang}, \citenamefont {Deng}, \citenamefont {Jing}, \citenamefont {Huang}, \citenamefont {Shi}, \citenamefont {Ye}, \citenamefont {Qiao}, \citenamefont {Wang}, \citenamefont {Guo}, \citenamefont {Feng},\ and\ \citenamefont {Shen}}]{Ding2024}%
  \BibitemOpen
  \bibfield  {author} {\bibinfo {author} {\bibfnamefont {J.}~\bibnamefont {Ding}}, \bibinfo {author} {\bibfnamefont {Z.}~\bibnamefont {Jiang}}, \bibinfo {author} {\bibfnamefont {X.}~\bibnamefont {Chen}}, \bibinfo {author} {\bibfnamefont {Z.}~\bibnamefont {Tao}}, \bibinfo {author} {\bibfnamefont {Z.}~\bibnamefont {Liu}}, \bibinfo {author} {\bibfnamefont {T.}~\bibnamefont {Li}}, \bibinfo {author} {\bibfnamefont {J.}~\bibnamefont {Liu}}, \bibinfo {author} {\bibfnamefont {J.}~\bibnamefont {Sun}}, \bibinfo {author} {\bibfnamefont {J.}~\bibnamefont {Cheng}}, \bibinfo {author} {\bibfnamefont {J.}~\bibnamefont {Liu}}, \bibinfo {author} {\bibfnamefont {Y.}~\bibnamefont {Yang}}, \bibinfo {author} {\bibfnamefont {R.}~\bibnamefont {Zhang}}, \bibinfo {author} {\bibfnamefont {L.}~\bibnamefont {Deng}}, \bibinfo {author} {\bibfnamefont {W.}~\bibnamefont {Jing}}, \bibinfo {author} {\bibfnamefont {Y.}~\bibnamefont {Huang}}, \bibinfo {author} {\bibfnamefont {Y.}~\bibnamefont {Shi}}, \bibinfo {author} {\bibfnamefont
  {M.}~\bibnamefont {Ye}}, \bibinfo {author} {\bibfnamefont {S.}~\bibnamefont {Qiao}}, \bibinfo {author} {\bibfnamefont {Y.}~\bibnamefont {Wang}}, \bibinfo {author} {\bibfnamefont {Y.}~\bibnamefont {Guo}}, \bibinfo {author} {\bibfnamefont {D.}~\bibnamefont {Feng}},\ and\ \bibinfo {author} {\bibfnamefont {D.}~\bibnamefont {Shen}},\ }\bibfield  {title} {\bibinfo {title} {{Large Band Splitting in $g$-Wave Altermagnet CrSb}},\ }\href {https://doi.org/10.1103/PhysRevLett.133.206401} {\bibfield  {journal} {\bibinfo  {journal} {Physical Review Letters}\ }\textbf {\bibinfo {volume} {133}},\ \bibinfo {pages} {206401} (\bibinfo {year} {2024})}\BibitemShut {NoStop}%
\bibitem [{\citenamefont {Babu~Regmi}\ \emph {et~al.}(2024)\citenamefont {Babu~Regmi}, \citenamefont {Bhandari}, \citenamefont {Thapa}, \citenamefont {Hao}, \citenamefont {Sharma}, \citenamefont {McKenzie}, \citenamefont {Chen}, \citenamefont {Nayak}, \citenamefont {El~Gazzah}, \citenamefont {G{\'a}bor~M{\'a}rkus} \emph {et~al.}}]{Babu2024}%
  \BibitemOpen
  \bibfield  {author} {\bibinfo {author} {\bibfnamefont {R.}~\bibnamefont {Babu~Regmi}}, \bibinfo {author} {\bibfnamefont {H.}~\bibnamefont {Bhandari}}, \bibinfo {author} {\bibfnamefont {B.}~\bibnamefont {Thapa}}, \bibinfo {author} {\bibfnamefont {Y.}~\bibnamefont {Hao}}, \bibinfo {author} {\bibfnamefont {N.}~\bibnamefont {Sharma}}, \bibinfo {author} {\bibfnamefont {J.}~\bibnamefont {McKenzie}}, \bibinfo {author} {\bibfnamefont {X.}~\bibnamefont {Chen}}, \bibinfo {author} {\bibfnamefont {A.}~\bibnamefont {Nayak}}, \bibinfo {author} {\bibfnamefont {M.}~\bibnamefont {El~Gazzah}}, \bibinfo {author} {\bibfnamefont {B.}~\bibnamefont {G{\'a}bor~M{\'a}rkus}}, \emph {et~al.},\ }\bibfield  {title} {\bibinfo {title} {{Altermagnetism in the layered intercalated transition metal dichalcogenide CoNb$_4$Se$_8$}},\ }\href {https://arxiv.org/pdf/2408.08835} {\bibfield  {journal} {\bibinfo  {journal} {arXiv:2408.08835}\ } (\bibinfo {year} {2024})}\BibitemShut {NoStop}%
\bibitem [{\citenamefont {De~Vita}\ \emph {et~al.}(2025)\citenamefont {De~Vita}, \citenamefont {Bigi}, \citenamefont {Romanin}, \citenamefont {Watson}, \citenamefont {Polewczyk}, \citenamefont {Zonno}, \citenamefont {Bertran}, \citenamefont {Petersen}, \citenamefont {Motti}, \citenamefont {Vinai} \emph {et~al.}}]{Vita2025}%
  \BibitemOpen
  \bibfield  {author} {\bibinfo {author} {\bibfnamefont {A.}~\bibnamefont {De~Vita}}, \bibinfo {author} {\bibfnamefont {C.}~\bibnamefont {Bigi}}, \bibinfo {author} {\bibfnamefont {D.}~\bibnamefont {Romanin}}, \bibinfo {author} {\bibfnamefont {M.~D.}\ \bibnamefont {Watson}}, \bibinfo {author} {\bibfnamefont {V.}~\bibnamefont {Polewczyk}}, \bibinfo {author} {\bibfnamefont {M.}~\bibnamefont {Zonno}}, \bibinfo {author} {\bibfnamefont {F.}~\bibnamefont {Bertran}}, \bibinfo {author} {\bibfnamefont {M.~B.}\ \bibnamefont {Petersen}}, \bibinfo {author} {\bibfnamefont {F.}~\bibnamefont {Motti}}, \bibinfo {author} {\bibfnamefont {G.}~\bibnamefont {Vinai}}, \emph {et~al.},\ }\bibfield  {title} {\bibinfo {title} {{Optical switching in a layered altermagnet}},\ }\href {https://arxiv.org/pdf/2502.20010} {\bibfield  {journal} {\bibinfo  {journal} {arXiv:2502.20010}\ } (\bibinfo {year} {2025})}\BibitemShut {NoStop}%
\bibitem [{\citenamefont {Graham}\ \emph {et~al.}(2025)\citenamefont {Graham}, \citenamefont {Hicken}, \citenamefont {Regmi}, \citenamefont {Janoschek}, \citenamefont {Mazin}, \citenamefont {Luetkens}, \citenamefont {Ghimire},\ and\ \citenamefont {Guguchia}}]{Graham2025}%
  \BibitemOpen
  \bibfield  {author} {\bibinfo {author} {\bibfnamefont {J.}~\bibnamefont {Graham}}, \bibinfo {author} {\bibfnamefont {T.}~\bibnamefont {Hicken}}, \bibinfo {author} {\bibfnamefont {R.}~\bibnamefont {Regmi}}, \bibinfo {author} {\bibfnamefont {M.}~\bibnamefont {Janoschek}}, \bibinfo {author} {\bibfnamefont {I.}~\bibnamefont {Mazin}}, \bibinfo {author} {\bibfnamefont {H.}~\bibnamefont {Luetkens}}, \bibinfo {author} {\bibfnamefont {N.}~\bibnamefont {Ghimire}},\ and\ \bibinfo {author} {\bibfnamefont {Z.}~\bibnamefont {Guguchia}},\ }\bibfield  {title} {\bibinfo {title} {{Local probe evidence supporting altermagnetism in Co$_{1/4}$NbSe$_2$}},\ }\href {https://arxiv.org/pdf/2503.09193} {\bibfield  {journal} {\bibinfo  {journal} {arXiv:2503.09193}\ } (\bibinfo {year} {2025})}\BibitemShut {NoStop}%
\bibitem [{\citenamefont {Jiang}\ \emph {et~al.}(2024{\natexlab{c}})\citenamefont {Jiang}, \citenamefont {Hu}, \citenamefont {Bai}, \citenamefont {Song}, \citenamefont {Mu}, \citenamefont {Qu}, \citenamefont {Li}, \citenamefont {Zhu}, \citenamefont {Pi}, \citenamefont {Wei} \emph {et~al.}}]{Jiang2024discovery}%
  \BibitemOpen
  \bibfield  {author} {\bibinfo {author} {\bibfnamefont {B.}~\bibnamefont {Jiang}}, \bibinfo {author} {\bibfnamefont {M.}~\bibnamefont {Hu}}, \bibinfo {author} {\bibfnamefont {J.}~\bibnamefont {Bai}}, \bibinfo {author} {\bibfnamefont {Z.}~\bibnamefont {Song}}, \bibinfo {author} {\bibfnamefont {C.}~\bibnamefont {Mu}}, \bibinfo {author} {\bibfnamefont {G.}~\bibnamefont {Qu}}, \bibinfo {author} {\bibfnamefont {W.}~\bibnamefont {Li}}, \bibinfo {author} {\bibfnamefont {W.}~\bibnamefont {Zhu}}, \bibinfo {author} {\bibfnamefont {H.}~\bibnamefont {Pi}}, \bibinfo {author} {\bibfnamefont {Z.}~\bibnamefont {Wei}}, \emph {et~al.},\ }\bibfield  {title} {\bibinfo {title} {{Discovery of a metallic room-temperature d-wave altermagnet KV$_2$Se$_2$O}},\ }\href {https://arxiv.org/pdf/2408.00320} {\bibfield  {journal} {\bibinfo  {journal} {arXiv:2408.00320}\ } (\bibinfo {year} {2024}{\natexlab{c}})}\BibitemShut {NoStop}%
\bibitem [{\citenamefont {Feng}\ \emph {et~al.}(2022)\citenamefont {Feng}, \citenamefont {Zhou}, \citenamefont {{\v{S}}mejkal}, \citenamefont {Wu}, \citenamefont {Zhu}, \citenamefont {Guo}, \citenamefont {Gonz{\'a}lez-Hern{\'a}ndez}, \citenamefont {Wang}, \citenamefont {Yan}, \citenamefont {Qin} \emph {et~al.}}]{Feng2022}%
  \BibitemOpen
  \bibfield  {author} {\bibinfo {author} {\bibfnamefont {Z.}~\bibnamefont {Feng}}, \bibinfo {author} {\bibfnamefont {X.}~\bibnamefont {Zhou}}, \bibinfo {author} {\bibfnamefont {L.}~\bibnamefont {{\v{S}}mejkal}}, \bibinfo {author} {\bibfnamefont {L.}~\bibnamefont {Wu}}, \bibinfo {author} {\bibfnamefont {Z.}~\bibnamefont {Zhu}}, \bibinfo {author} {\bibfnamefont {H.}~\bibnamefont {Guo}}, \bibinfo {author} {\bibfnamefont {R.}~\bibnamefont {Gonz{\'a}lez-Hern{\'a}ndez}}, \bibinfo {author} {\bibfnamefont {X.}~\bibnamefont {Wang}}, \bibinfo {author} {\bibfnamefont {H.}~\bibnamefont {Yan}}, \bibinfo {author} {\bibfnamefont {P.}~\bibnamefont {Qin}}, \emph {et~al.},\ }\bibfield  {title} {\bibinfo {title} {{An anomalous Hall effect in altermagnetic ruthenium dioxide}},\ }\href {https://doi.org/https://doi.org/10.1038/s41928-022-00866-z} {\bibfield  {journal} {\bibinfo  {journal} {Nature Electronics}\ }\textbf {\bibinfo {volume} {5}},\ \bibinfo {pages} {735} (\bibinfo {year} {2022})}\BibitemShut {NoStop}%
\bibitem [{\citenamefont {Gonzalez~Betancourt}\ \emph {et~al.}(2023)\citenamefont {Gonzalez~Betancourt}, \citenamefont {Zub\'a\ifmmode~\check{c}\else \v{c}\fi{}}, \citenamefont {Gonzalez-Hernandez}, \citenamefont {Geishendorf}, \citenamefont {\ifmmode \check{S}\else \v{S}\fi{}ob\'a\ifmmode~\check{n}\else \v{n}\fi{}}, \citenamefont {Springholz}, \citenamefont {Olejn\'{\i}k}, \citenamefont {\ifmmode~\check{S}\else \v{S}\fi{}mejkal}, \citenamefont {Sinova}, \citenamefont {Jungwirth}, \citenamefont {Goennenwein}, \citenamefont {Thomas}, \citenamefont {Reichlov\'a}, \citenamefont {\ifmmode~\check{Z}\else \v{Z}\fi{}elezn\'y},\ and\ \citenamefont {Kriegner}}]{Betancourt2023}%
  \BibitemOpen
  \bibfield  {author} {\bibinfo {author} {\bibfnamefont {R.~D.}\ \bibnamefont {Gonzalez~Betancourt}}, \bibinfo {author} {\bibfnamefont {J.}~\bibnamefont {Zub\'a\ifmmode~\check{c}\else \v{c}\fi{}}}, \bibinfo {author} {\bibfnamefont {R.}~\bibnamefont {Gonzalez-Hernandez}}, \bibinfo {author} {\bibfnamefont {K.}~\bibnamefont {Geishendorf}}, \bibinfo {author} {\bibfnamefont {Z.}~\bibnamefont {\ifmmode \check{S}\else \v{S}\fi{}ob\'a\ifmmode~\check{n}\else \v{n}\fi{}}}, \bibinfo {author} {\bibfnamefont {G.}~\bibnamefont {Springholz}}, \bibinfo {author} {\bibfnamefont {K.}~\bibnamefont {Olejn\'{\i}k}}, \bibinfo {author} {\bibfnamefont {L.}~\bibnamefont {\ifmmode~\check{S}\else \v{S}\fi{}mejkal}}, \bibinfo {author} {\bibfnamefont {J.}~\bibnamefont {Sinova}}, \bibinfo {author} {\bibfnamefont {T.}~\bibnamefont {Jungwirth}}, \bibinfo {author} {\bibfnamefont {S.~T.~B.}\ \bibnamefont {Goennenwein}}, \bibinfo {author} {\bibfnamefont {A.}~\bibnamefont {Thomas}}, \bibinfo {author} {\bibfnamefont {H.}~\bibnamefont
  {Reichlov\'a}}, \bibinfo {author} {\bibfnamefont {J.}~\bibnamefont {\ifmmode~\check{Z}\else \v{Z}\fi{}elezn\'y}},\ and\ \bibinfo {author} {\bibfnamefont {D.}~\bibnamefont {Kriegner}},\ }\bibfield  {title} {\bibinfo {title} {{Spontaneous Anomalous Hall Effect Arising from an Unconventional Compensated Magnetic Phase in a Semiconductor}},\ }\href {https://doi.org/https://doi.org/10.1103/PhysRevLett.130.036702} {\bibfield  {journal} {\bibinfo  {journal} {Physical Review Letters}\ }\textbf {\bibinfo {volume} {130}},\ \bibinfo {pages} {036702} (\bibinfo {year} {2023})}\BibitemShut {NoStop}%
\bibitem [{\citenamefont {Fedchenko}\ \emph {et~al.}(2024)\citenamefont {Fedchenko}, \citenamefont {Min{\'a}r}, \citenamefont {Akashdeep}, \citenamefont {D'Souza}, \citenamefont {Vasilyev}, \citenamefont {Tkach}, \citenamefont {Odenbreit}, \citenamefont {Nguyen}, \citenamefont {Kutnyakhov}, \citenamefont {Wind}, \citenamefont {Wenthaus}, \citenamefont {Scholz}, \citenamefont {Rossnagel}, \citenamefont {Hoesch}, \citenamefont {Aeschlimann}, \citenamefont {Stadtm{\"u}ller}, \citenamefont {Kl{\"a}ui}, \citenamefont {Sch{\"o}nhense}, \citenamefont {Jungwirth}, \citenamefont {Hellenes}, \citenamefont {Jakob}, \citenamefont {{\v S}mejkal}, \citenamefont {Sinova},\ and\ \citenamefont {Elmers}}]{Fedchenko2023}%
  \BibitemOpen
  \bibfield  {author} {\bibinfo {author} {\bibfnamefont {O.}~\bibnamefont {Fedchenko}}, \bibinfo {author} {\bibfnamefont {J.}~\bibnamefont {Min{\'a}r}}, \bibinfo {author} {\bibfnamefont {A.}~\bibnamefont {Akashdeep}}, \bibinfo {author} {\bibfnamefont {S.}~\bibnamefont {D'Souza}}, \bibinfo {author} {\bibfnamefont {D.}~\bibnamefont {Vasilyev}}, \bibinfo {author} {\bibfnamefont {O.}~\bibnamefont {Tkach}}, \bibinfo {author} {\bibfnamefont {L.}~\bibnamefont {Odenbreit}}, \bibinfo {author} {\bibfnamefont {Q.~L.}\ \bibnamefont {Nguyen}}, \bibinfo {author} {\bibfnamefont {D.}~\bibnamefont {Kutnyakhov}}, \bibinfo {author} {\bibfnamefont {N.}~\bibnamefont {Wind}}, \bibinfo {author} {\bibfnamefont {L.}~\bibnamefont {Wenthaus}}, \bibinfo {author} {\bibfnamefont {M.}~\bibnamefont {Scholz}}, \bibinfo {author} {\bibfnamefont {K.}~\bibnamefont {Rossnagel}}, \bibinfo {author} {\bibfnamefont {M.}~\bibnamefont {Hoesch}}, \bibinfo {author} {\bibfnamefont {M.}~\bibnamefont {Aeschlimann}}, \bibinfo {author} {\bibfnamefont
  {B.}~\bibnamefont {Stadtm{\"u}ller}}, \bibinfo {author} {\bibfnamefont {M.}~\bibnamefont {Kl{\"a}ui}}, \bibinfo {author} {\bibfnamefont {G.}~\bibnamefont {Sch{\"o}nhense}}, \bibinfo {author} {\bibfnamefont {T.}~\bibnamefont {Jungwirth}}, \bibinfo {author} {\bibfnamefont {A.~B.}\ \bibnamefont {Hellenes}}, \bibinfo {author} {\bibfnamefont {G.}~\bibnamefont {Jakob}}, \bibinfo {author} {\bibfnamefont {L.}~\bibnamefont {{\v S}mejkal}}, \bibinfo {author} {\bibfnamefont {J.}~\bibnamefont {Sinova}},\ and\ \bibinfo {author} {\bibfnamefont {H.-J.}\ \bibnamefont {Elmers}},\ }\bibfield  {title} {\bibinfo {title} {{Observation of time-reversal symmetry breaking in the band structure of altermagnetic {RuO$_2$}}},\ }\href {https://doi.org/10.1126/sciadv.adj4883} {\bibfield  {journal} {\bibinfo  {journal} {Sci. Adv.}\ }\textbf {\bibinfo {volume} {10}},\ \bibinfo {pages} {eadj4883} (\bibinfo {year} {2024})}\BibitemShut {NoStop}%
\bibitem [{\citenamefont {Bai}\ \emph {et~al.}(2023)\citenamefont {Bai}, \citenamefont {Zhang}, \citenamefont {Zhou}, \citenamefont {Chen}, \citenamefont {Wan}, \citenamefont {Han}, \citenamefont {Zhu}, \citenamefont {Liang}, \citenamefont {Su}, \citenamefont {Han}, \citenamefont {Pan},\ and\ \citenamefont {Song}}]{Bai2023}%
  \BibitemOpen
  \bibfield  {author} {\bibinfo {author} {\bibfnamefont {H.}~\bibnamefont {Bai}}, \bibinfo {author} {\bibfnamefont {Y.~C.}\ \bibnamefont {Zhang}}, \bibinfo {author} {\bibfnamefont {Y.~J.}\ \bibnamefont {Zhou}}, \bibinfo {author} {\bibfnamefont {P.}~\bibnamefont {Chen}}, \bibinfo {author} {\bibfnamefont {C.~H.}\ \bibnamefont {Wan}}, \bibinfo {author} {\bibfnamefont {L.}~\bibnamefont {Han}}, \bibinfo {author} {\bibfnamefont {W.~X.}\ \bibnamefont {Zhu}}, \bibinfo {author} {\bibfnamefont {S.~X.}\ \bibnamefont {Liang}}, \bibinfo {author} {\bibfnamefont {Y.~C.}\ \bibnamefont {Su}}, \bibinfo {author} {\bibfnamefont {X.~F.}\ \bibnamefont {Han}}, \bibinfo {author} {\bibfnamefont {F.}~\bibnamefont {Pan}},\ and\ \bibinfo {author} {\bibfnamefont {C.}~\bibnamefont {Song}},\ }\bibfield  {title} {\bibinfo {title} {{Efficient Spin-to-Charge Conversion via Altermagnetic Spin Splitting Effect in Antiferromagnet ${\mathrm{RuO}}_{2}$}},\ }\href {https://doi.org/https://doi.org/10.1103/PhysRevLett.130.216701} {\bibfield
  {journal} {\bibinfo  {journal} {Physical Review Letters}\ }\textbf {\bibinfo {volume} {130}},\ \bibinfo {pages} {216701} (\bibinfo {year} {2023})}\BibitemShut {NoStop}%
\bibitem [{\citenamefont {Lin}\ \emph {et~al.}(2024)\citenamefont {Lin}, \citenamefont {Chen}, \citenamefont {Lu}, \citenamefont {Liang}, \citenamefont {Feng}, \citenamefont {Yamagami}, \citenamefont {Osiecki}, \citenamefont {Leandersson}, \citenamefont {Thiagarajan}, \citenamefont {Liu}, \citenamefont {Felser},\ and\ \citenamefont {Ma}}]{Lin2024observation}%
  \BibitemOpen
  \bibfield  {author} {\bibinfo {author} {\bibfnamefont {Z.}~\bibnamefont {Lin}}, \bibinfo {author} {\bibfnamefont {D.}~\bibnamefont {Chen}}, \bibinfo {author} {\bibfnamefont {W.}~\bibnamefont {Lu}}, \bibinfo {author} {\bibfnamefont {X.}~\bibnamefont {Liang}}, \bibinfo {author} {\bibfnamefont {S.}~\bibnamefont {Feng}}, \bibinfo {author} {\bibfnamefont {K.}~\bibnamefont {Yamagami}}, \bibinfo {author} {\bibfnamefont {J.}~\bibnamefont {Osiecki}}, \bibinfo {author} {\bibfnamefont {M.}~\bibnamefont {Leandersson}}, \bibinfo {author} {\bibfnamefont {B.}~\bibnamefont {Thiagarajan}}, \bibinfo {author} {\bibfnamefont {J.}~\bibnamefont {Liu}}, \bibinfo {author} {\bibfnamefont {C.}~\bibnamefont {Felser}},\ and\ \bibinfo {author} {\bibfnamefont {J.}~\bibnamefont {Ma}},\ }\bibfield  {title} {\bibinfo {title} {{Observation of Giant Spin Splitting and d-wave Spin Texture in Room Temperature Altermagnet {RuO$_2$}}},\ }\href {https://arxiv.org/pdf/2402.04995} {\bibfield  {journal} {\bibinfo  {journal} {arXiv:2402.04995}\ }
  (\bibinfo {year} {2024})}\BibitemShut {NoStop}%
\bibitem [{\citenamefont {Smolyanyuk}\ \emph {et~al.}(2024)\citenamefont {Smolyanyuk}, \citenamefont {Mazin}, \citenamefont {Garcia-Gassull},\ and\ \citenamefont {Valent{\'\i}}}]{Smolyanyuk2023}%
  \BibitemOpen
  \bibfield  {author} {\bibinfo {author} {\bibfnamefont {A.}~\bibnamefont {Smolyanyuk}}, \bibinfo {author} {\bibfnamefont {I.~I.}\ \bibnamefont {Mazin}}, \bibinfo {author} {\bibfnamefont {L.}~\bibnamefont {Garcia-Gassull}},\ and\ \bibinfo {author} {\bibfnamefont {R.}~\bibnamefont {Valent{\'\i}}},\ }\bibfield  {title} {\bibinfo {title} {{Fragility of the magnetic order in the prototypical altermagnet {RuO$_2$}}},\ }\href {https://doi.org/10.1103/PhysRevB.109.134424} {\bibfield  {journal} {\bibinfo  {journal} {Physical Review B}\ }\textbf {\bibinfo {volume} {109}},\ \bibinfo {pages} {134424} (\bibinfo {year} {2024})}\BibitemShut {NoStop}%
\bibitem [{\citenamefont {Ke{\ss}ler}\ \emph {et~al.}(2024)\citenamefont {Ke{\ss}ler}, \citenamefont {Garcia-Gassull}, \citenamefont {Suter}, \citenamefont {Prokscha}, \citenamefont {Salman}, \citenamefont {Khalyavin}, \citenamefont {Manuel}, \citenamefont {Orlandi}, \citenamefont {Mazin}, \citenamefont {Valent{\'\i}} \emph {et~al.}}]{Kessler2024}%
  \BibitemOpen
  \bibfield  {author} {\bibinfo {author} {\bibfnamefont {P.}~\bibnamefont {Ke{\ss}ler}}, \bibinfo {author} {\bibfnamefont {L.}~\bibnamefont {Garcia-Gassull}}, \bibinfo {author} {\bibfnamefont {A.}~\bibnamefont {Suter}}, \bibinfo {author} {\bibfnamefont {T.}~\bibnamefont {Prokscha}}, \bibinfo {author} {\bibfnamefont {Z.}~\bibnamefont {Salman}}, \bibinfo {author} {\bibfnamefont {D.}~\bibnamefont {Khalyavin}}, \bibinfo {author} {\bibfnamefont {P.}~\bibnamefont {Manuel}}, \bibinfo {author} {\bibfnamefont {F.}~\bibnamefont {Orlandi}}, \bibinfo {author} {\bibfnamefont {I.~I.}\ \bibnamefont {Mazin}}, \bibinfo {author} {\bibfnamefont {R.}~\bibnamefont {Valent{\'\i}}}, \emph {et~al.},\ }\bibfield  {title} {\bibinfo {title} {{Absence of magnetic order in RuO$_2$: insights from $\mu$ SR spectroscopy and neutron diffraction}},\ }\href {https://www.nature.com/articles/s44306-024-00055-y} {\bibfield  {journal} {\bibinfo  {journal} {npj Spintronics}\ }\textbf {\bibinfo {volume} {2}},\ \bibinfo {pages} {50} (\bibinfo {year}
  {2024})}\BibitemShut {NoStop}%
\bibitem [{\citenamefont {Hiraishi}\ \emph {et~al.}(2024)\citenamefont {Hiraishi}, \citenamefont {Okabe}, \citenamefont {Koda}, \citenamefont {Kadono}, \citenamefont {Muroi}, \citenamefont {Hirai},\ and\ \citenamefont {Hiroi}}]{Hiraishi2024}%
  \BibitemOpen
  \bibfield  {author} {\bibinfo {author} {\bibfnamefont {M.}~\bibnamefont {Hiraishi}}, \bibinfo {author} {\bibfnamefont {H.}~\bibnamefont {Okabe}}, \bibinfo {author} {\bibfnamefont {A.}~\bibnamefont {Koda}}, \bibinfo {author} {\bibfnamefont {R.}~\bibnamefont {Kadono}}, \bibinfo {author} {\bibfnamefont {T.}~\bibnamefont {Muroi}}, \bibinfo {author} {\bibfnamefont {D.}~\bibnamefont {Hirai}},\ and\ \bibinfo {author} {\bibfnamefont {Z.}~\bibnamefont {Hiroi}},\ }\bibfield  {title} {\bibinfo {title} {{Nonmagnetic Ground State in ${\mathrm{RuO}}_{2}$ Revealed by Muon Spin Rotation}},\ }\href {https://doi.org/10.1103/PhysRevLett.132.166702} {\bibfield  {journal} {\bibinfo  {journal} {Physical Review Letters}\ }\textbf {\bibinfo {volume} {132}},\ \bibinfo {pages} {166702} (\bibinfo {year} {2024})}\BibitemShut {NoStop}%
\bibitem [{\citenamefont {Jeong}\ \emph {et~al.}(2024)\citenamefont {Jeong}, \citenamefont {Choi}, \citenamefont {Nair}, \citenamefont {Buiarelli}, \citenamefont {Pourbahari}, \citenamefont {Oh}, \citenamefont {Bassim}, \citenamefont {Seo}, \citenamefont {Choi}, \citenamefont {Fernandes} \emph {et~al.}}]{Jeong2024}%
  \BibitemOpen
  \bibfield  {author} {\bibinfo {author} {\bibfnamefont {S.~G.}\ \bibnamefont {Jeong}}, \bibinfo {author} {\bibfnamefont {I.~H.}\ \bibnamefont {Choi}}, \bibinfo {author} {\bibfnamefont {S.}~\bibnamefont {Nair}}, \bibinfo {author} {\bibfnamefont {L.}~\bibnamefont {Buiarelli}}, \bibinfo {author} {\bibfnamefont {B.}~\bibnamefont {Pourbahari}}, \bibinfo {author} {\bibfnamefont {J.~Y.}\ \bibnamefont {Oh}}, \bibinfo {author} {\bibfnamefont {N.}~\bibnamefont {Bassim}}, \bibinfo {author} {\bibfnamefont {A.}~\bibnamefont {Seo}}, \bibinfo {author} {\bibfnamefont {W.~S.}\ \bibnamefont {Choi}}, \bibinfo {author} {\bibfnamefont {R.~M.}\ \bibnamefont {Fernandes}}, \emph {et~al.},\ }\bibfield  {title} {\bibinfo {title} {{Altermagnetic polar metallic phase in ultra-thin epitaxially-strained RuO$_2$ films}},\ }\href {https://arxiv.org/pdf/2405.05838} {\bibfield  {journal} {\bibinfo  {journal} {arXiv:2405.05838}\ } (\bibinfo {year} {2024})}\BibitemShut {NoStop}%
\bibitem [{\citenamefont {Naka}\ \emph {et~al.}(2019)\citenamefont {Naka}, \citenamefont {Hayami}, \citenamefont {Kusunose}, \citenamefont {Yanagi}, \citenamefont {Motome},\ and\ \citenamefont {Seo}}]{Naka2019}%
  \BibitemOpen
  \bibfield  {author} {\bibinfo {author} {\bibfnamefont {M.}~\bibnamefont {Naka}}, \bibinfo {author} {\bibfnamefont {S.}~\bibnamefont {Hayami}}, \bibinfo {author} {\bibfnamefont {H.}~\bibnamefont {Kusunose}}, \bibinfo {author} {\bibfnamefont {Y.}~\bibnamefont {Yanagi}}, \bibinfo {author} {\bibfnamefont {Y.}~\bibnamefont {Motome}},\ and\ \bibinfo {author} {\bibfnamefont {H.}~\bibnamefont {Seo}},\ }\bibfield  {title} {\bibinfo {title} {{Spin current generation in organic antiferromagnets}},\ }\href {https://doi.org/10.1038/s41467-019-12229-y} {\bibfield  {journal} {\bibinfo  {journal} {Nat. Commun.}\ }\textbf {\bibinfo {volume} {10}},\ \bibinfo {pages} {4305} (\bibinfo {year} {2019})}\BibitemShut {NoStop}%
\bibitem [{\citenamefont {Hayami}\ \emph {et~al.}(2019)\citenamefont {Hayami}, \citenamefont {Yanagi},\ and\ \citenamefont {Kusunose}}]{Hayami2019}%
  \BibitemOpen
  \bibfield  {author} {\bibinfo {author} {\bibfnamefont {S.}~\bibnamefont {Hayami}}, \bibinfo {author} {\bibfnamefont {Y.}~\bibnamefont {Yanagi}},\ and\ \bibinfo {author} {\bibfnamefont {H.}~\bibnamefont {Kusunose}},\ }\bibfield  {title} {\bibinfo {title} {{Momentum-Dependent Spin Splitting by Collinear Antiferromagnetic Ordering}},\ }\href {https://doi.org/10.7566/JPSJ.88.123702} {\bibfield  {journal} {\bibinfo  {journal} {J. Phys. Soc. Jpn.}\ }\textbf {\bibinfo {volume} {88}},\ \bibinfo {pages} {123702} (\bibinfo {year} {2019})}\BibitemShut {NoStop}%
\bibitem [{\citenamefont {Yuan}\ \emph {et~al.}(2020)\citenamefont {Yuan}, \citenamefont {Wang}, \citenamefont {Luo}, \citenamefont {Rashba},\ and\ \citenamefont {Zunger}}]{Zunger2020}%
  \BibitemOpen
  \bibfield  {author} {\bibinfo {author} {\bibfnamefont {L.-D.}\ \bibnamefont {Yuan}}, \bibinfo {author} {\bibfnamefont {Z.}~\bibnamefont {Wang}}, \bibinfo {author} {\bibfnamefont {J.-W.}\ \bibnamefont {Luo}}, \bibinfo {author} {\bibfnamefont {E.~I.}\ \bibnamefont {Rashba}},\ and\ \bibinfo {author} {\bibfnamefont {A.}~\bibnamefont {Zunger}},\ }\bibfield  {title} {\bibinfo {title} {{Giant momentum-dependent spin splitting in centrosymmetric low-$Z$ antiferromagnets}},\ }\href {https://doi.org/10.1103/PhysRevB.102.014422} {\bibfield  {journal} {\bibinfo  {journal} {Physical Review B}\ }\textbf {\bibinfo {volume} {102}},\ \bibinfo {pages} {014422} (\bibinfo {year} {2020})}\BibitemShut {NoStop}%
\bibitem [{\citenamefont {Hayami}\ \emph {et~al.}(2020)\citenamefont {Hayami}, \citenamefont {Yanagi},\ and\ \citenamefont {Kusunose}}]{Kusunose2020}%
  \BibitemOpen
  \bibfield  {author} {\bibinfo {author} {\bibfnamefont {S.}~\bibnamefont {Hayami}}, \bibinfo {author} {\bibfnamefont {Y.}~\bibnamefont {Yanagi}},\ and\ \bibinfo {author} {\bibfnamefont {H.}~\bibnamefont {Kusunose}},\ }\bibfield  {title} {\bibinfo {title} {{Bottom-up design of spin-split and reshaped electronic band structures in antiferromagnets without spin-orbit coupling: Procedure on the basis of augmented multipoles}},\ }\href {https://doi.org/10.1103/PhysRevB.102.144441} {\bibfield  {journal} {\bibinfo  {journal} {Physical Review B}\ }\textbf {\bibinfo {volume} {102}},\ \bibinfo {pages} {144441} (\bibinfo {year} {2020})}\BibitemShut {NoStop}%
\bibitem [{\citenamefont {Jaeschke-Ubiergo}\ \emph {et~al.}(2025)\citenamefont {Jaeschke-Ubiergo}, \citenamefont {Bharadwaj}, \citenamefont {Campos}, \citenamefont {Zarzuela}, \citenamefont {Biniskos}, \citenamefont {Fernandes}, \citenamefont {Jungwirth}, \citenamefont {Sinova},\ and\ \citenamefont {{\v{S}}mejkal}}]{Jaeschke2025}%
  \BibitemOpen
  \bibfield  {author} {\bibinfo {author} {\bibfnamefont {R.}~\bibnamefont {Jaeschke-Ubiergo}}, \bibinfo {author} {\bibfnamefont {V.-K.}\ \bibnamefont {Bharadwaj}}, \bibinfo {author} {\bibfnamefont {W.}~\bibnamefont {Campos}}, \bibinfo {author} {\bibfnamefont {R.}~\bibnamefont {Zarzuela}}, \bibinfo {author} {\bibfnamefont {N.}~\bibnamefont {Biniskos}}, \bibinfo {author} {\bibfnamefont {R.~M.}\ \bibnamefont {Fernandes}}, \bibinfo {author} {\bibfnamefont {T.}~\bibnamefont {Jungwirth}}, \bibinfo {author} {\bibfnamefont {J.}~\bibnamefont {Sinova}},\ and\ \bibinfo {author} {\bibfnamefont {L.}~\bibnamefont {{\v{S}}mejkal}},\ }\bibfield  {title} {\bibinfo {title} {Atomic altermagnetism},\ }\href {https://arxiv.org/pdf/2503.10797} {\bibfield  {journal} {\bibinfo  {journal} {arXiv:2503.10797}\ } (\bibinfo {year} {2025})}\BibitemShut {NoStop}%
\bibitem [{\citenamefont {Santini}\ \emph {et~al.}(2009)\citenamefont {Santini}, \citenamefont {Carretta}, \citenamefont {Amoretti}, \citenamefont {Caciuffo}, \citenamefont {Magnani},\ and\ \citenamefont {Lander}}]{Santini2009}%
  \BibitemOpen
  \bibfield  {author} {\bibinfo {author} {\bibfnamefont {P.}~\bibnamefont {Santini}}, \bibinfo {author} {\bibfnamefont {S.}~\bibnamefont {Carretta}}, \bibinfo {author} {\bibfnamefont {G.}~\bibnamefont {Amoretti}}, \bibinfo {author} {\bibfnamefont {R.}~\bibnamefont {Caciuffo}}, \bibinfo {author} {\bibfnamefont {N.}~\bibnamefont {Magnani}},\ and\ \bibinfo {author} {\bibfnamefont {G.~H.}\ \bibnamefont {Lander}},\ }\bibfield  {title} {\bibinfo {title} {{Multipolar interactions in $f$-electron systems: The paradigm of actinide dioxides}},\ }\href {https://doi.org/10.1103/RevModPhys.81.807} {\bibfield  {journal} {\bibinfo  {journal} {Rev. Mod. Phys.}\ }\textbf {\bibinfo {volume} {81}},\ \bibinfo {pages} {807} (\bibinfo {year} {2009})}\BibitemShut {NoStop}%
\bibitem [{\citenamefont {Voleti}\ \emph {et~al.}(2020)\citenamefont {Voleti}, \citenamefont {Maharaj}, \citenamefont {Gaulin}, \citenamefont {Luke},\ and\ \citenamefont {Paramekanti}}]{Voleti2020}%
  \BibitemOpen
  \bibfield  {author} {\bibinfo {author} {\bibfnamefont {S.}~\bibnamefont {Voleti}}, \bibinfo {author} {\bibfnamefont {D.~D.}\ \bibnamefont {Maharaj}}, \bibinfo {author} {\bibfnamefont {B.~D.}\ \bibnamefont {Gaulin}}, \bibinfo {author} {\bibfnamefont {G.}~\bibnamefont {Luke}},\ and\ \bibinfo {author} {\bibfnamefont {A.}~\bibnamefont {Paramekanti}},\ }\bibfield  {title} {\bibinfo {title} {{Multipolar magnetism in d-orbital systems: Crystal field levels, octupolar order, and orbital loop currents}},\ }\href {https://doi.org/https://doi.org/10.1103/PhysRevB.101.155118} {\bibfield  {journal} {\bibinfo  {journal} {Physical Review B}\ }\textbf {\bibinfo {volume} {101}},\ \bibinfo {pages} {155118} (\bibinfo {year} {2020})}\BibitemShut {NoStop}%
\bibitem [{\citenamefont {Fiore~Mosca}\ \emph {et~al.}(2022)\citenamefont {Fiore~Mosca}, \citenamefont {Pourovskii},\ and\ \citenamefont {Franchini}}]{Fiore2022}%
  \BibitemOpen
  \bibfield  {author} {\bibinfo {author} {\bibfnamefont {D.}~\bibnamefont {Fiore~Mosca}}, \bibinfo {author} {\bibfnamefont {L.~V.}\ \bibnamefont {Pourovskii}},\ and\ \bibinfo {author} {\bibfnamefont {C.}~\bibnamefont {Franchini}},\ }\bibfield  {title} {\bibinfo {title} {{Modeling magnetic multipolar phases in density functional theory}},\ }\href {https://doi.org/https://doi.org/10.1103/PhysRevB.106.035127} {\bibfield  {journal} {\bibinfo  {journal} {Physical Review B}\ }\textbf {\bibinfo {volume} {106}},\ \bibinfo {pages} {035127} (\bibinfo {year} {2022})}\BibitemShut {NoStop}%
\bibitem [{\citenamefont {Winkler}\ and\ \citenamefont {Z{\"u}licke}(2023)}]{winkler2023theory}%
  \BibitemOpen
  \bibfield  {author} {\bibinfo {author} {\bibfnamefont {R.}~\bibnamefont {Winkler}}\ and\ \bibinfo {author} {\bibfnamefont {U.}~\bibnamefont {Z{\"u}licke}},\ }\bibfield  {title} {\bibinfo {title} {{Theory of electric, magnetic, and toroidal polarizations in crystalline solids with applications to hexagonal lonsdaleite and cubic diamond}},\ }\href {https://doi.org/https://doi.org/10.1103/PhysRevB.107.155201} {\bibfield  {journal} {\bibinfo  {journal} {Physical Review B}\ }\textbf {\bibinfo {volume} {107}},\ \bibinfo {pages} {155201} (\bibinfo {year} {2023})}\BibitemShut {NoStop}%
\bibitem [{\citenamefont {{Pomeranchuk, I Ia and others}}(1958)}]{Pomeranchuk1958}%
  \BibitemOpen
  \bibfield  {author} {\bibinfo {author} {\bibnamefont {{Pomeranchuk, I Ia and others}}},\ }\bibfield  {title} {\bibinfo {title} {{On the stability of a Fermi liquid}},\ }\href {http://jetp.ras.ru/cgi-bin/dn/e_008_02_0361.pdf} {\bibfield  {journal} {\bibinfo  {journal} {Sov. Phys. JETP}\ }\textbf {\bibinfo {volume} {8}},\ \bibinfo {pages} {361} (\bibinfo {year} {1958})}\BibitemShut {NoStop}%
\bibitem [{\citenamefont {Wu}\ \emph {et~al.}(2007)\citenamefont {Wu}, \citenamefont {Sun}, \citenamefont {Fradkin},\ and\ \citenamefont {Zhang}}]{Wu2007}%
  \BibitemOpen
  \bibfield  {author} {\bibinfo {author} {\bibfnamefont {C.}~\bibnamefont {Wu}}, \bibinfo {author} {\bibfnamefont {K.}~\bibnamefont {Sun}}, \bibinfo {author} {\bibfnamefont {E.}~\bibnamefont {Fradkin}},\ and\ \bibinfo {author} {\bibfnamefont {S.-C.}\ \bibnamefont {Zhang}},\ }\bibfield  {title} {\bibinfo {title} {{Fermi liquid instabilities in the spin channel}},\ }\href {https://doi.org/10.1103/PhysRevB.75.115103} {\bibfield  {journal} {\bibinfo  {journal} {Physical Review B}\ }\textbf {\bibinfo {volume} {75}},\ \bibinfo {pages} {115103} (\bibinfo {year} {2007})}\BibitemShut {NoStop}%
\bibitem [{\citenamefont {Ahn}\ \emph {et~al.}(2019)\citenamefont {Ahn}, \citenamefont {Hariki}, \citenamefont {Lee},\ and\ \citenamefont {Kune\ifmmode~\check{s}\else \v{s}\fi{}}}]{Kunes2019}%
  \BibitemOpen
  \bibfield  {author} {\bibinfo {author} {\bibfnamefont {K.-H.}\ \bibnamefont {Ahn}}, \bibinfo {author} {\bibfnamefont {A.}~\bibnamefont {Hariki}}, \bibinfo {author} {\bibfnamefont {K.-W.}\ \bibnamefont {Lee}},\ and\ \bibinfo {author} {\bibfnamefont {J.}~\bibnamefont {Kune\ifmmode~\check{s}\else \v{s}\fi{}}},\ }\bibfield  {title} {\bibinfo {title} {{Antiferromagnetism in {RuO}$_{2}$ as $d$-wave {Pomeranchuk} instability}},\ }\href {https://doi.org/10.1103/PhysRevB.99.184432} {\bibfield  {journal} {\bibinfo  {journal} {Physical Review B}\ }\textbf {\bibinfo {volume} {99}},\ \bibinfo {pages} {184432} (\bibinfo {year} {2019})}\BibitemShut {NoStop}%
\bibitem [{\citenamefont {Jungwirth}\ \emph {et~al.}(2024{\natexlab{b}})\citenamefont {Jungwirth}, \citenamefont {Fernandes}, \citenamefont {Fradkin}, \citenamefont {MacDonald}, \citenamefont {Sinova},\ and\ \citenamefont {Smejkal}}]{Jungwirth2024supefluid}%
  \BibitemOpen
  \bibfield  {author} {\bibinfo {author} {\bibfnamefont {T.}~\bibnamefont {Jungwirth}}, \bibinfo {author} {\bibfnamefont {R.}~\bibnamefont {Fernandes}}, \bibinfo {author} {\bibfnamefont {E.}~\bibnamefont {Fradkin}}, \bibinfo {author} {\bibfnamefont {A.}~\bibnamefont {MacDonald}}, \bibinfo {author} {\bibfnamefont {J.}~\bibnamefont {Sinova}},\ and\ \bibinfo {author} {\bibfnamefont {L.}~\bibnamefont {Smejkal}},\ }\bibfield  {title} {\bibinfo {title} {{From supefluid $^3$He to altermagnets}},\ }\href {https://arxiv.org/pdf/2411.00717} {\bibfield  {journal} {\bibinfo  {journal} {arXiv:2411.00717}\ } (\bibinfo {year} {2024}{\natexlab{b}})}\BibitemShut {NoStop}%
\bibitem [{\citenamefont {Aoyama}\ and\ \citenamefont {Ohgushi}(2024)}]{Aoyama2024_piezomagnetism}%
  \BibitemOpen
  \bibfield  {author} {\bibinfo {author} {\bibfnamefont {T.}~\bibnamefont {Aoyama}}\ and\ \bibinfo {author} {\bibfnamefont {K.}~\bibnamefont {Ohgushi}},\ }\bibfield  {title} {\bibinfo {title} {{Piezomagnetic properties in altermagnetic MnTe}},\ }\href {https://journals.aps.org/prmaterials/pdf/10.1103/PhysRevMaterials.8.L041402} {\bibfield  {journal} {\bibinfo  {journal} {Physical Review Materials}\ }\textbf {\bibinfo {volume} {8}},\ \bibinfo {pages} {L041402} (\bibinfo {year} {2024})}\BibitemShut {NoStop}%
\bibitem [{\citenamefont {Yershov}\ \emph {et~al.}(2024)\citenamefont {Yershov}, \citenamefont {Kravchuk}, \citenamefont {Daghofer},\ and\ \citenamefont {van~den Brink}}]{vandenBrink2024}%
  \BibitemOpen
  \bibfield  {author} {\bibinfo {author} {\bibfnamefont {K.~V.}\ \bibnamefont {Yershov}}, \bibinfo {author} {\bibfnamefont {V.~P.}\ \bibnamefont {Kravchuk}}, \bibinfo {author} {\bibfnamefont {M.}~\bibnamefont {Daghofer}},\ and\ \bibinfo {author} {\bibfnamefont {J.}~\bibnamefont {van~den Brink}},\ }\bibfield  {title} {\bibinfo {title} {{Fluctuation-induced piezomagnetism in local moment altermagnets}},\ }\href {https://doi.org/10.1103/PhysRevB.110.144421} {\bibfield  {journal} {\bibinfo  {journal} {Physical Review B}\ }\textbf {\bibinfo {volume} {110}},\ \bibinfo {pages} {144421} (\bibinfo {year} {2024})}\BibitemShut {NoStop}%
\bibitem [{\citenamefont {Dzialoshinskii}(1958)}]{Dzialoshinskii1958_piezomagnetism}%
  \BibitemOpen
  \bibfield  {author} {\bibinfo {author} {\bibfnamefont {I.}~\bibnamefont {Dzialoshinskii}},\ }\bibfield  {title} {\bibinfo {title} {{The problem of piezomagnetism}},\ }\href {http://www.jetp.ras.ru/cgi-bin/dn/e_006_03_0621.pdf} {\bibfield  {journal} {\bibinfo  {journal} {Sov. Phys. JETP}\ }\textbf {\bibinfo {volume} {6}},\ \bibinfo {pages} {621} (\bibinfo {year} {1958})}\BibitemShut {NoStop}%
\bibitem [{\citenamefont {Lee}(1955)}]{Lee1955_magnetostriction}%
  \BibitemOpen
  \bibfield  {author} {\bibinfo {author} {\bibfnamefont {E.~W.}\ \bibnamefont {Lee}},\ }\bibfield  {title} {\bibinfo {title} {{Magnetostriction and magnetomechanical effects}},\ }\href {https://iopscience.iop.org/article/10.1088/0034-4885/18/1/305/pdf} {\bibfield  {journal} {\bibinfo  {journal} {Reports on progress in physics}\ }\textbf {\bibinfo {volume} {18}},\ \bibinfo {pages} {184} (\bibinfo {year} {1955})}\BibitemShut {NoStop}%
\bibitem [{\citenamefont {Callen}(1968)}]{Callen1968_magnetostriction}%
  \BibitemOpen
  \bibfield  {author} {\bibinfo {author} {\bibfnamefont {E.}~\bibnamefont {Callen}},\ }\bibfield  {title} {\bibinfo {title} {{Magnetostriction}},\ }\href {https://pubs.aip.org/aip/jap/article/39/2/519/4802/Magnetostriction} {\bibfield  {journal} {\bibinfo  {journal} {Journal of Applied Physics}\ }\textbf {\bibinfo {volume} {39}},\ \bibinfo {pages} {519} (\bibinfo {year} {1968})}\BibitemShut {NoStop}%
\bibitem [{\citenamefont {Ma}\ \emph {et~al.}(1996)\citenamefont {Ma}, \citenamefont {Wang}, \citenamefont {Zhu},\ and\ \citenamefont {Sheng}}]{Ma1996_RF}%
  \BibitemOpen
  \bibfield  {author} {\bibinfo {author} {\bibfnamefont {Y.-Q.}\ \bibnamefont {Ma}}, \bibinfo {author} {\bibfnamefont {Z.}~\bibnamefont {Wang}}, \bibinfo {author} {\bibfnamefont {J.-X.}\ \bibnamefont {Zhu}},\ and\ \bibinfo {author} {\bibfnamefont {L.}~\bibnamefont {Sheng}},\ }\bibfield  {title} {\bibinfo {title} {{Gaussian random field p-spin-interaction ising model in a transverse field}},\ }\href {https://link.springer.com/article/10.1007/s002570050124} {\bibfield  {journal} {\bibinfo  {journal} {Zeitschrift f{\"u}r Physik B Condensed Matter}\ }\textbf {\bibinfo {volume} {100}},\ \bibinfo {pages} {295} (\bibinfo {year} {1996})}\BibitemShut {NoStop}%
\bibitem [{\citenamefont {Schneider}\ and\ \citenamefont {Pytte}(1977)}]{schneider1977_RF}%
  \BibitemOpen
  \bibfield  {author} {\bibinfo {author} {\bibfnamefont {T.}~\bibnamefont {Schneider}}\ and\ \bibinfo {author} {\bibfnamefont {E.}~\bibnamefont {Pytte}},\ }\bibfield  {title} {\bibinfo {title} {{Random-field instability of the ferromagnetic state}},\ }\href {https://journals.aps.org/prb/pdf/10.1103/PhysRevB.15.1519} {\bibfield  {journal} {\bibinfo  {journal} {Physical Review B}\ }\textbf {\bibinfo {volume} {15}},\ \bibinfo {pages} {1519} (\bibinfo {year} {1977})}\BibitemShut {NoStop}%
\bibitem [{\citenamefont {Binder}(1983)}]{Binder1983}%
  \BibitemOpen
  \bibfield  {author} {\bibinfo {author} {\bibfnamefont {K.}~\bibnamefont {Binder}},\ }\bibfield  {title} {\bibinfo {title} {{Random-field induced interface widths in Ising systems}},\ }\href {https://link.springer.com/article/10.1007/BF01470045} {\bibfield  {journal} {\bibinfo  {journal} {Zeitschrift f{\"u}r Physik B Condensed Matter}\ }\textbf {\bibinfo {volume} {50}},\ \bibinfo {pages} {343} (\bibinfo {year} {1983})}\BibitemShut {NoStop}%
\bibitem [{\citenamefont {Nattermann}(1988)}]{Nattermann1988_RF}%
  \BibitemOpen
  \bibfield  {author} {\bibinfo {author} {\bibfnamefont {T.}~\bibnamefont {Nattermann}},\ }\bibfield  {title} {\bibinfo {title} {{Dipolar interaction in random-field systems}},\ }\href {https://iopscience.iop.org/article/10.1088/0305-4470/21/12/005/pdf} {\bibfield  {journal} {\bibinfo  {journal} {Journal of Physics A: Mathematical and General}\ }\textbf {\bibinfo {volume} {21}},\ \bibinfo {pages} {L645} (\bibinfo {year} {1988})}\BibitemShut {NoStop}%
\bibitem [{\citenamefont {Toh}(1992)}]{Toh1992_RF}%
  \BibitemOpen
  \bibfield  {author} {\bibinfo {author} {\bibfnamefont {H.}~\bibnamefont {Toh}},\ }\bibfield  {title} {\bibinfo {title} {{Structural phase transitions with random strains}},\ }\href {https://iopscience.iop.org/article/10.1088/0305-4470/25/18/012} {\bibfield  {journal} {\bibinfo  {journal} {Journal of Physics A: Mathematical and General}\ }\textbf {\bibinfo {volume} {25}},\ \bibinfo {pages} {4767} (\bibinfo {year} {1992})}\BibitemShut {NoStop}%
\bibitem [{\citenamefont {Fishman}\ and\ \citenamefont {Aharony}(1979)}]{Fishman1979}%
  \BibitemOpen
  \bibfield  {author} {\bibinfo {author} {\bibfnamefont {S.}~\bibnamefont {Fishman}}\ and\ \bibinfo {author} {\bibfnamefont {A.}~\bibnamefont {Aharony}},\ }\bibfield  {title} {\bibinfo {title} {{Random field effects in disordered anisotropic antiferromagnets}},\ }\href {https://iopscience.iop.org/article/10.1088/0022-3719/12/18/006/pdf} {\bibfield  {journal} {\bibinfo  {journal} {Journal of Physics C: Solid State Physics}\ }\textbf {\bibinfo {volume} {12}},\ \bibinfo {pages} {L729} (\bibinfo {year} {1979})}\BibitemShut {NoStop}%
\bibitem [{\citenamefont {Wen}\ \emph {et~al.}(2010)\citenamefont {Wen}, \citenamefont {Subedi}, \citenamefont {Bo}, \citenamefont {Yeshurun}, \citenamefont {Sarachik}, \citenamefont {Kent}, \citenamefont {Millis}, \citenamefont {Lampropoulos},\ and\ \citenamefont {Christou}}]{Christou2010}%
  \BibitemOpen
  \bibfield  {author} {\bibinfo {author} {\bibfnamefont {B.}~\bibnamefont {Wen}}, \bibinfo {author} {\bibfnamefont {P.}~\bibnamefont {Subedi}}, \bibinfo {author} {\bibfnamefont {L.}~\bibnamefont {Bo}}, \bibinfo {author} {\bibfnamefont {Y.}~\bibnamefont {Yeshurun}}, \bibinfo {author} {\bibfnamefont {M.~P.}\ \bibnamefont {Sarachik}}, \bibinfo {author} {\bibfnamefont {A.~D.}\ \bibnamefont {Kent}}, \bibinfo {author} {\bibfnamefont {A.~J.}\ \bibnamefont {Millis}}, \bibinfo {author} {\bibfnamefont {C.}~\bibnamefont {Lampropoulos}},\ and\ \bibinfo {author} {\bibfnamefont {G.}~\bibnamefont {Christou}},\ }\bibfield  {title} {\bibinfo {title} {{Realization of random-field Ising ferromagnetism in a molecular magnet}},\ }\href {https://doi.org/10.1103/PhysRevB.82.014406} {\bibfield  {journal} {\bibinfo  {journal} {Physical Review B}\ }\textbf {\bibinfo {volume} {82}},\ \bibinfo {pages} {014406} (\bibinfo {year} {2010})}\BibitemShut {NoStop}%
\bibitem [{\citenamefont {Carlson}\ \emph {et~al.}(2006)\citenamefont {Carlson}, \citenamefont {Dahmen}, \citenamefont {Fradkin},\ and\ \citenamefont {Kivelson}}]{Carlson2006}%
  \BibitemOpen
  \bibfield  {author} {\bibinfo {author} {\bibfnamefont {E.~W.}\ \bibnamefont {Carlson}}, \bibinfo {author} {\bibfnamefont {K.~A.}\ \bibnamefont {Dahmen}}, \bibinfo {author} {\bibfnamefont {E.}~\bibnamefont {Fradkin}},\ and\ \bibinfo {author} {\bibfnamefont {S.~A.}\ \bibnamefont {Kivelson}},\ }\bibfield  {title} {\bibinfo {title} {{Hysteresis and Noise from Electronic Nematicity in High-Temperature Superconductors}},\ }\href {https://doi.org/10.1103/PhysRevLett.96.097003} {\bibfield  {journal} {\bibinfo  {journal} {Physical Review Letters}\ }\textbf {\bibinfo {volume} {96}},\ \bibinfo {pages} {097003} (\bibinfo {year} {2006})}\BibitemShut {NoStop}%
\bibitem [{\citenamefont {Silevitch}\ \emph {et~al.}(2007)\citenamefont {Silevitch}, \citenamefont {Bitko}, \citenamefont {Brooke}, \citenamefont {Ghosh}, \citenamefont {Aeppli},\ and\ \citenamefont {Rosenbaum}}]{Silevitch2007}%
  \BibitemOpen
  \bibfield  {author} {\bibinfo {author} {\bibfnamefont {D.}~\bibnamefont {Silevitch}}, \bibinfo {author} {\bibfnamefont {D.}~\bibnamefont {Bitko}}, \bibinfo {author} {\bibfnamefont {J.}~\bibnamefont {Brooke}}, \bibinfo {author} {\bibfnamefont {S.}~\bibnamefont {Ghosh}}, \bibinfo {author} {\bibfnamefont {G.}~\bibnamefont {Aeppli}},\ and\ \bibinfo {author} {\bibfnamefont {T.}~\bibnamefont {Rosenbaum}},\ }\bibfield  {title} {\bibinfo {title} {{A ferromagnet in a continuously tunable random field}},\ }\href {https://www.nature.com/articles/nature06050} {\bibfield  {journal} {\bibinfo  {journal} {Nature}\ }\textbf {\bibinfo {volume} {448}},\ \bibinfo {pages} {567} (\bibinfo {year} {2007})}\BibitemShut {NoStop}%
\bibitem [{\citenamefont {Senthil}(1998)}]{Senthil1998_RF}%
  \BibitemOpen
  \bibfield  {author} {\bibinfo {author} {\bibfnamefont {T.}~\bibnamefont {Senthil}},\ }\bibfield  {title} {\bibinfo {title} {{Properties of the random-field Ising model in a transverse magnetic field}},\ }\href {https://journals.aps.org/prb/pdf/10.1103/PhysRevB.57.8375} {\bibfield  {journal} {\bibinfo  {journal} {Physical Review B}\ }\textbf {\bibinfo {volume} {57}},\ \bibinfo {pages} {8375} (\bibinfo {year} {1998})}\BibitemShut {NoStop}%
\bibitem [{\citenamefont {Zhang}\ \emph {et~al.}(2024{\natexlab{b}})\citenamefont {Zhang}, \citenamefont {Cheng}, \citenamefont {Yin}, \citenamefont {Liu}, \citenamefont {Deng}, \citenamefont {Qiao}, \citenamefont {Shi}, \citenamefont {Zhang}, \citenamefont {Lin}, \citenamefont {Liu} \emph {et~al.}}]{Zhang2024crystal}%
  \BibitemOpen
  \bibfield  {author} {\bibinfo {author} {\bibfnamefont {F.}~\bibnamefont {Zhang}}, \bibinfo {author} {\bibfnamefont {X.}~\bibnamefont {Cheng}}, \bibinfo {author} {\bibfnamefont {Z.}~\bibnamefont {Yin}}, \bibinfo {author} {\bibfnamefont {C.}~\bibnamefont {Liu}}, \bibinfo {author} {\bibfnamefont {L.}~\bibnamefont {Deng}}, \bibinfo {author} {\bibfnamefont {Y.}~\bibnamefont {Qiao}}, \bibinfo {author} {\bibfnamefont {Z.}~\bibnamefont {Shi}}, \bibinfo {author} {\bibfnamefont {S.}~\bibnamefont {Zhang}}, \bibinfo {author} {\bibfnamefont {J.}~\bibnamefont {Lin}}, \bibinfo {author} {\bibfnamefont {Z.}~\bibnamefont {Liu}}, \emph {et~al.},\ }\bibfield  {title} {\bibinfo {title} {{Crystal-symmetry-paired spin-valley locking in a layered room-temperature antiferromagnet}},\ }\href {https://arxiv.org/pdf/2407.19555} {\bibfield  {journal} {\bibinfo  {journal} {arXiv:2407.19555}\ } (\bibinfo {year} {2024}{\natexlab{b}})}\BibitemShut {NoStop}%
\bibitem [{\citenamefont {Chou}\ and\ \citenamefont {Nelson}(1996)}]{Chou1996}%
  \BibitemOpen
  \bibfield  {author} {\bibinfo {author} {\bibfnamefont {T.}~\bibnamefont {Chou}}\ and\ \bibinfo {author} {\bibfnamefont {D.~R.}\ \bibnamefont {Nelson}},\ }\bibfield  {title} {\bibinfo {title} {{Dislocation-mediated melting near isostructural critical points}},\ }\href {https://journals.aps.org/pre/pdf/10.1103/PhysRevE.53.2560} {\bibfield  {journal} {\bibinfo  {journal} {Physical Review E}\ }\textbf {\bibinfo {volume} {53}},\ \bibinfo {pages} {2560} (\bibinfo {year} {1996})}\BibitemShut {NoStop}%
\bibitem [{\citenamefont {Paul}\ and\ \citenamefont {Garst}(2017)}]{Paul2017}%
  \BibitemOpen
  \bibfield  {author} {\bibinfo {author} {\bibfnamefont {I.}~\bibnamefont {Paul}}\ and\ \bibinfo {author} {\bibfnamefont {M.}~\bibnamefont {Garst}},\ }\bibfield  {title} {\bibinfo {title} {{Lattice Effects on Nematic Quantum Criticality in Metals}},\ }\href {https://doi.org/10.1103/PhysRevLett.118.227601} {\bibfield  {journal} {\bibinfo  {journal} {Physical Review Letters}\ }\textbf {\bibinfo {volume} {118}},\ \bibinfo {pages} {227601} (\bibinfo {year} {2017})}\BibitemShut {NoStop}%
\bibitem [{\citenamefont {Dutta}\ \emph {et~al.}(1996)\citenamefont {Dutta}, \citenamefont {Chakrabarti},\ and\ \citenamefont {Stinchcombe}}]{Dutta1996_RF}%
  \BibitemOpen
  \bibfield  {author} {\bibinfo {author} {\bibfnamefont {A.}~\bibnamefont {Dutta}}, \bibinfo {author} {\bibfnamefont {B.}~\bibnamefont {Chakrabarti}},\ and\ \bibinfo {author} {\bibfnamefont {R.}~\bibnamefont {Stinchcombe}},\ }\bibfield  {title} {\bibinfo {title} {{Phase transitions in the random field Ising model in the presence of a transverse field}},\ }\href {https://iopscience.iop.org/article/10.1088/0305-4470/29/17/007/pdf} {\bibfield  {journal} {\bibinfo  {journal} {Journal of Physics A: Mathematical and General}\ }\textbf {\bibinfo {volume} {29}},\ \bibinfo {pages} {5285} (\bibinfo {year} {1996})}\BibitemShut {NoStop}%
\bibitem [{\citenamefont {Yokota}\ and\ \citenamefont {Sugiyama}(1988)}]{yokota1988reentrant}%
  \BibitemOpen
  \bibfield  {author} {\bibinfo {author} {\bibfnamefont {T.}~\bibnamefont {Yokota}}\ and\ \bibinfo {author} {\bibfnamefont {Y.}~\bibnamefont {Sugiyama}},\ }\bibfield  {title} {\bibinfo {title} {{Reentrant phase transitions in a quantum spin system with random fields}},\ }\href {https://journals.aps.org/prb/pdf/10.1103/PhysRevB.37.5657} {\bibfield  {journal} {\bibinfo  {journal} {Physical Review B}\ }\textbf {\bibinfo {volume} {37}},\ \bibinfo {pages} {5657} (\bibinfo {year} {1988})}\BibitemShut {NoStop}%
\bibitem [{\citenamefont {Steward}\ \emph {et~al.}(2025)\citenamefont {Steward}, \citenamefont {Palle}, \citenamefont {Garst}, \citenamefont {Schmalian},\ and\ \citenamefont {Jang}}]{Steward2025}%
  \BibitemOpen
  \bibfield  {author} {\bibinfo {author} {\bibfnamefont {C.~R.}\ \bibnamefont {Steward}}, \bibinfo {author} {\bibfnamefont {G.}~\bibnamefont {Palle}}, \bibinfo {author} {\bibfnamefont {M.}~\bibnamefont {Garst}}, \bibinfo {author} {\bibfnamefont {J.}~\bibnamefont {Schmalian}},\ and\ \bibinfo {author} {\bibfnamefont {I.}~\bibnamefont {Jang}},\ }\bibfield  {title} {\bibinfo {title} {{Elastic Quantum Criticality in Nematics and Altermagnets via the Elasto-Caloric Effect}},\ }\href {https://arxiv.org/pdf/2502.14033} {\bibfield  {journal} {\bibinfo  {journal} {arXiv:2502.14033}\ } (\bibinfo {year} {2025})}\BibitemShut {NoStop}%
\bibitem [{\citenamefont {Fernandes}\ \emph {et~al.}(2010)\citenamefont {Fernandes}, \citenamefont {VanBebber}, \citenamefont {Bhattacharya}, \citenamefont {Chandra}, \citenamefont {Keppens}, \citenamefont {Mandrus}, \citenamefont {McGuire}, \citenamefont {Sales}, \citenamefont {Sefat},\ and\ \citenamefont {Schmalian}}]{Fernandes2010}%
  \BibitemOpen
  \bibfield  {author} {\bibinfo {author} {\bibfnamefont {R.~M.}\ \bibnamefont {Fernandes}}, \bibinfo {author} {\bibfnamefont {L.~H.}\ \bibnamefont {VanBebber}}, \bibinfo {author} {\bibfnamefont {S.}~\bibnamefont {Bhattacharya}}, \bibinfo {author} {\bibfnamefont {P.}~\bibnamefont {Chandra}}, \bibinfo {author} {\bibfnamefont {V.}~\bibnamefont {Keppens}}, \bibinfo {author} {\bibfnamefont {D.}~\bibnamefont {Mandrus}}, \bibinfo {author} {\bibfnamefont {M.~A.}\ \bibnamefont {McGuire}}, \bibinfo {author} {\bibfnamefont {B.~C.}\ \bibnamefont {Sales}}, \bibinfo {author} {\bibfnamefont {A.~S.}\ \bibnamefont {Sefat}},\ and\ \bibinfo {author} {\bibfnamefont {J.}~\bibnamefont {Schmalian}},\ }\bibfield  {title} {\bibinfo {title} {{Effects of Nematic Fluctuations on the Elastic Properties of Iron Arsenide Superconductors}},\ }\href {https://doi.org/10.1103/PhysRevLett.105.157003} {\bibfield  {journal} {\bibinfo  {journal} {Physical Review Letters}\ }\textbf {\bibinfo {volume} {105}},\ \bibinfo {pages} {157003} (\bibinfo
  {year} {2010})}\BibitemShut {NoStop}%
\bibitem [{\citenamefont {Fradkin}\ \emph {et~al.}(2010)\citenamefont {Fradkin}, \citenamefont {Kivelson}, \citenamefont {Lawler}, \citenamefont {Eisenstein},\ and\ \citenamefont {Mackenzie}}]{Fradkin2010}%
  \BibitemOpen
  \bibfield  {author} {\bibinfo {author} {\bibfnamefont {E.}~\bibnamefont {Fradkin}}, \bibinfo {author} {\bibfnamefont {S.~A.}\ \bibnamefont {Kivelson}}, \bibinfo {author} {\bibfnamefont {M.~J.}\ \bibnamefont {Lawler}}, \bibinfo {author} {\bibfnamefont {J.~P.}\ \bibnamefont {Eisenstein}},\ and\ \bibinfo {author} {\bibfnamefont {A.~P.}\ \bibnamefont {Mackenzie}},\ }\bibfield  {title} {\bibinfo {title} {{Nematic Fermi fluids in condensed matter physics}},\ }\href {https://doi.org/10.1146/annurev-conmatphys-070909-103925} {\bibfield  {journal} {\bibinfo  {journal} {Annu. Rev. Condens. Matter Phys.}\ }\textbf {\bibinfo {volume} {1}},\ \bibinfo {pages} {153} (\bibinfo {year} {2010})}\BibitemShut {NoStop}%
\bibitem [{\citenamefont {Fernandes}\ \emph {et~al.}(2014)\citenamefont {Fernandes}, \citenamefont {Chubukov},\ and\ \citenamefont {Schmalian}}]{Fernandes2014}%
  \BibitemOpen
  \bibfield  {author} {\bibinfo {author} {\bibfnamefont {R.}~\bibnamefont {Fernandes}}, \bibinfo {author} {\bibfnamefont {A.}~\bibnamefont {Chubukov}},\ and\ \bibinfo {author} {\bibfnamefont {J.}~\bibnamefont {Schmalian}},\ }\bibfield  {title} {\bibinfo {title} {{What drives nematic order in iron-based superconductors?}},\ }\href {https://doi.org/10.1038/NPHYS2877} {\bibfield  {journal} {\bibinfo  {journal} {Nature Physics}\ }\textbf {\bibinfo {volume} {10}},\ \bibinfo {pages} {97} (\bibinfo {year} {2014})}\BibitemShut {NoStop}%
\bibitem [{\citenamefont {Ikeda}\ \emph {et~al.}(2019)\citenamefont {Ikeda}, \citenamefont {Straquadine}, \citenamefont {Hristov}, \citenamefont {Worasaran}, \citenamefont {Palmstrom}, \citenamefont {Sorensen}, \citenamefont {Walmsley},\ and\ \citenamefont {Fisher}}]{ikeda2019ac}%
  \BibitemOpen
  \bibfield  {author} {\bibinfo {author} {\bibfnamefont {M.~S.}\ \bibnamefont {Ikeda}}, \bibinfo {author} {\bibfnamefont {J.~A.}\ \bibnamefont {Straquadine}}, \bibinfo {author} {\bibfnamefont {A.~T.}\ \bibnamefont {Hristov}}, \bibinfo {author} {\bibfnamefont {T.}~\bibnamefont {Worasaran}}, \bibinfo {author} {\bibfnamefont {J.~C.}\ \bibnamefont {Palmstrom}}, \bibinfo {author} {\bibfnamefont {M.}~\bibnamefont {Sorensen}}, \bibinfo {author} {\bibfnamefont {P.}~\bibnamefont {Walmsley}},\ and\ \bibinfo {author} {\bibfnamefont {I.~R.}\ \bibnamefont {Fisher}},\ }\bibfield  {title} {\bibinfo {title} {{AC elastocaloric effect as a probe for thermodynamic signatures of continuous phase transitions}},\ }\href {https://pubs.aip.org/aip/rsi/article/90/8/083902/360017/AC-elastocaloric-effect-as-a-probe-for} {\bibfield  {journal} {\bibinfo  {journal} {Review of Scientific Instruments}\ }\textbf {\bibinfo {volume} {90}} (\bibinfo {year} {2019})}\BibitemShut {NoStop}%
\bibitem [{\citenamefont {Ikeda}\ \emph {et~al.}(2021)\citenamefont {Ikeda}, \citenamefont {Worasaran}, \citenamefont {Rosenberg}, \citenamefont {Palmstrom}, \citenamefont {Kivelson},\ and\ \citenamefont {Fisher}}]{Ikeda2021_ECE}%
  \BibitemOpen
  \bibfield  {author} {\bibinfo {author} {\bibfnamefont {M.~S.}\ \bibnamefont {Ikeda}}, \bibinfo {author} {\bibfnamefont {T.}~\bibnamefont {Worasaran}}, \bibinfo {author} {\bibfnamefont {E.~W.}\ \bibnamefont {Rosenberg}}, \bibinfo {author} {\bibfnamefont {J.~C.}\ \bibnamefont {Palmstrom}}, \bibinfo {author} {\bibfnamefont {S.~A.}\ \bibnamefont {Kivelson}},\ and\ \bibinfo {author} {\bibfnamefont {I.~R.}\ \bibnamefont {Fisher}},\ }\bibfield  {title} {\bibinfo {title} {{Elastocaloric signature of nematic fluctuations}},\ }\href {https://www.pnas.org/doi/10.1073/pnas.2105911118} {\bibfield  {journal} {\bibinfo  {journal} {Proceedings of the National Academy of Sciences}\ }\textbf {\bibinfo {volume} {118}},\ \bibinfo {pages} {e2105911118} (\bibinfo {year} {2021})}\BibitemShut {NoStop}%
\bibitem [{\citenamefont {Rosenberg}\ \emph {et~al.}(2024)\citenamefont {Rosenberg}, \citenamefont {Ikeda},\ and\ \citenamefont {Fisher}}]{rosenberg2024nematic}%
  \BibitemOpen
  \bibfield  {author} {\bibinfo {author} {\bibfnamefont {E.~W.}\ \bibnamefont {Rosenberg}}, \bibinfo {author} {\bibfnamefont {M.}~\bibnamefont {Ikeda}},\ and\ \bibinfo {author} {\bibfnamefont {I.~R.}\ \bibnamefont {Fisher}},\ }\bibfield  {title} {\bibinfo {title} {{The nematic susceptibility of the ferroquadrupolar metal TmAg$_2$ measured via the elastocaloric effect}},\ }\href {https://www.nature.com/articles/s41535-024-00658-y} {\bibfield  {journal} {\bibinfo  {journal} {npj Quantum Materials}\ }\textbf {\bibinfo {volume} {9}},\ \bibinfo {pages} {46} (\bibinfo {year} {2024})}\BibitemShut {NoStop}%
\bibitem [{\citenamefont {Palle}\ \emph {et~al.}(2023)\citenamefont {Palle}, \citenamefont {Hicks}, \citenamefont {Valent{\'\i}}, \citenamefont {Hu}, \citenamefont {Li}, \citenamefont {Rost}, \citenamefont {Nicklas}, \citenamefont {Mackenzie},\ and\ \citenamefont {Schmalian}}]{palle2023constraints}%
  \BibitemOpen
  \bibfield  {author} {\bibinfo {author} {\bibfnamefont {G.}~\bibnamefont {Palle}}, \bibinfo {author} {\bibfnamefont {C.}~\bibnamefont {Hicks}}, \bibinfo {author} {\bibfnamefont {R.}~\bibnamefont {Valent{\'\i}}}, \bibinfo {author} {\bibfnamefont {Z.}~\bibnamefont {Hu}}, \bibinfo {author} {\bibfnamefont {Y.-S.}\ \bibnamefont {Li}}, \bibinfo {author} {\bibfnamefont {A.}~\bibnamefont {Rost}}, \bibinfo {author} {\bibfnamefont {M.}~\bibnamefont {Nicklas}}, \bibinfo {author} {\bibfnamefont {A.~P.}\ \bibnamefont {Mackenzie}},\ and\ \bibinfo {author} {\bibfnamefont {J.}~\bibnamefont {Schmalian}},\ }\bibfield  {title} {\bibinfo {title} {{Constraints on the superconducting state of Sr$_2$RuO$_4$ from elastocaloric measurements}},\ }\href {https://journals.aps.org/prb/pdf/10.1103/PhysRevB.108.094516} {\bibfield  {journal} {\bibinfo  {journal} {Physical Review B}\ }\textbf {\bibinfo {volume} {108}},\ \bibinfo {pages} {094516} (\bibinfo {year} {2023})}\BibitemShut {NoStop}%
\bibitem [{\citenamefont {Ghosh}\ \emph {et~al.}(2024)\citenamefont {Ghosh}, \citenamefont {Ikeda}, \citenamefont {Chakraborty}, \citenamefont {Worasaran}, \citenamefont {Theuss}, \citenamefont {Peralta}, \citenamefont {Lozano}, \citenamefont {Kim}, \citenamefont {Ryan}, \citenamefont {Ye} \emph {et~al.}}]{ghosh2024elastocaloric}%
  \BibitemOpen
  \bibfield  {author} {\bibinfo {author} {\bibfnamefont {S.}~\bibnamefont {Ghosh}}, \bibinfo {author} {\bibfnamefont {M.~S.}\ \bibnamefont {Ikeda}}, \bibinfo {author} {\bibfnamefont {A.~R.}\ \bibnamefont {Chakraborty}}, \bibinfo {author} {\bibfnamefont {T.}~\bibnamefont {Worasaran}}, \bibinfo {author} {\bibfnamefont {F.}~\bibnamefont {Theuss}}, \bibinfo {author} {\bibfnamefont {L.~B.}\ \bibnamefont {Peralta}}, \bibinfo {author} {\bibfnamefont {P.}~\bibnamefont {Lozano}}, \bibinfo {author} {\bibfnamefont {J.-W.}\ \bibnamefont {Kim}}, \bibinfo {author} {\bibfnamefont {P.~J.}\ \bibnamefont {Ryan}}, \bibinfo {author} {\bibfnamefont {L.}~\bibnamefont {Ye}}, \emph {et~al.},\ }\bibfield  {title} {\bibinfo {title} {{Elastocaloric evidence for a multicomponent superconductor stabilized within the nematic state in Ba(Fe$_{1-x}$Co$_x$)$_2$As$_2$}},\ }\href {https://arxiv.org/pdf/2402.17945} {\bibfield  {journal} {\bibinfo  {journal} {arXiv:2402.17945}\ } (\bibinfo {year} {2024})}\BibitemShut {NoStop}%
\bibitem [{\citenamefont {Li}\ \emph {et~al.}(2022)\citenamefont {Li}, \citenamefont {Garst}, \citenamefont {Schmalian}, \citenamefont {Ghosh}, \citenamefont {Kikugawa}, \citenamefont {Sokolov}, \citenamefont {Hicks}, \citenamefont {Jerzembeck}, \citenamefont {Ikeda}, \citenamefont {Hu} \emph {et~al.}}]{li2022elastocaloric}%
  \BibitemOpen
  \bibfield  {author} {\bibinfo {author} {\bibfnamefont {Y.-S.}\ \bibnamefont {Li}}, \bibinfo {author} {\bibfnamefont {M.}~\bibnamefont {Garst}}, \bibinfo {author} {\bibfnamefont {J.}~\bibnamefont {Schmalian}}, \bibinfo {author} {\bibfnamefont {S.}~\bibnamefont {Ghosh}}, \bibinfo {author} {\bibfnamefont {N.}~\bibnamefont {Kikugawa}}, \bibinfo {author} {\bibfnamefont {D.~A.}\ \bibnamefont {Sokolov}}, \bibinfo {author} {\bibfnamefont {C.~W.}\ \bibnamefont {Hicks}}, \bibinfo {author} {\bibfnamefont {F.}~\bibnamefont {Jerzembeck}}, \bibinfo {author} {\bibfnamefont {M.~S.}\ \bibnamefont {Ikeda}}, \bibinfo {author} {\bibfnamefont {Z.}~\bibnamefont {Hu}}, \emph {et~al.},\ }\bibfield  {title} {\bibinfo {title} {{Elastocaloric determination of the phase diagram of Sr$_2$RuO$_4$}},\ }\href {https://www.nature.com/articles/s41586-022-04820-z} {\bibfield  {journal} {\bibinfo  {journal} {Nature}\ }\textbf {\bibinfo {volume} {607}},\ \bibinfo {pages} {276} (\bibinfo {year} {2022})}\BibitemShut {NoStop}%
\bibitem [{\citenamefont {Ye}\ \emph {et~al.}(2023)\citenamefont {Ye}, \citenamefont {Sun}, \citenamefont {Sunko}, \citenamefont {Rodriguez-Nieva}, \citenamefont {Ikeda}, \citenamefont {Worasaran}, \citenamefont {Sorensen}, \citenamefont {Bachmann}, \citenamefont {Orenstein},\ and\ \citenamefont {Fisher}}]{ye2023elastocaloric}%
  \BibitemOpen
  \bibfield  {author} {\bibinfo {author} {\bibfnamefont {L.}~\bibnamefont {Ye}}, \bibinfo {author} {\bibfnamefont {Y.}~\bibnamefont {Sun}}, \bibinfo {author} {\bibfnamefont {V.}~\bibnamefont {Sunko}}, \bibinfo {author} {\bibfnamefont {J.~F.}\ \bibnamefont {Rodriguez-Nieva}}, \bibinfo {author} {\bibfnamefont {M.~S.}\ \bibnamefont {Ikeda}}, \bibinfo {author} {\bibfnamefont {T.}~\bibnamefont {Worasaran}}, \bibinfo {author} {\bibfnamefont {M.~E.}\ \bibnamefont {Sorensen}}, \bibinfo {author} {\bibfnamefont {M.~D.}\ \bibnamefont {Bachmann}}, \bibinfo {author} {\bibfnamefont {J.}~\bibnamefont {Orenstein}},\ and\ \bibinfo {author} {\bibfnamefont {I.~R.}\ \bibnamefont {Fisher}},\ }\bibfield  {title} {\bibinfo {title} {{Elastocaloric signatures of symmetric and antisymmetric strain-tuning of quadrupolar and magnetic phases in DyB$_2$C$_2$}},\ }\href {https://www.pnas.org/doi/10.1073/pnas.2302800120} {\bibfield  {journal} {\bibinfo  {journal} {Proceedings of the National Academy of Sciences}\ }\textbf {\bibinfo {volume}
  {120}},\ \bibinfo {pages} {e2302800120} (\bibinfo {year} {2023})}\BibitemShut {NoStop}%
\bibitem [{\citenamefont {Ye}\ \emph {et~al.}(2024)\citenamefont {Ye}, \citenamefont {Sorensen}, \citenamefont {Bachmann},\ and\ \citenamefont {Fisher}}]{Ye2024}%
  \BibitemOpen
  \bibfield  {author} {\bibinfo {author} {\bibfnamefont {L.}~\bibnamefont {Ye}}, \bibinfo {author} {\bibfnamefont {M.~E.}\ \bibnamefont {Sorensen}}, \bibinfo {author} {\bibfnamefont {M.~D.}\ \bibnamefont {Bachmann}},\ and\ \bibinfo {author} {\bibfnamefont {I.~R.}\ \bibnamefont {Fisher}},\ }\bibfield  {title} {\bibinfo {title} {{Measurement of the magnetic octupole susceptibility of $\text{PrV}_2\text{Al}_{20}$}},\ }\href {https://doi.org/10.1038/s41467-024-51269-x} {\bibfield  {journal} {\bibinfo  {journal} {Nature Communications}\ }\textbf {\bibinfo {volume} {15}},\ \bibinfo {pages} {7005} (\bibinfo {year} {2024})}\BibitemShut {NoStop}%
\bibitem [{\citenamefont {Chu}\ \emph {et~al.}(2012)\citenamefont {Chu}, \citenamefont {Kuo}, \citenamefont {Analytis},\ and\ \citenamefont {Fisher}}]{chu2012divergent}%
  \BibitemOpen
  \bibfield  {author} {\bibinfo {author} {\bibfnamefont {J.-H.}\ \bibnamefont {Chu}}, \bibinfo {author} {\bibfnamefont {H.-H.}\ \bibnamefont {Kuo}}, \bibinfo {author} {\bibfnamefont {J.~G.}\ \bibnamefont {Analytis}},\ and\ \bibinfo {author} {\bibfnamefont {I.~R.}\ \bibnamefont {Fisher}},\ }\bibfield  {title} {\bibinfo {title} {{Divergent nematic susceptibility in an iron arsenide superconductor}},\ }\href {https://www.science.org/doi/10.1126/science.1221713} {\bibfield  {journal} {\bibinfo  {journal} {Science}\ }\textbf {\bibinfo {volume} {337}},\ \bibinfo {pages} {710} (\bibinfo {year} {2012})}\BibitemShut {NoStop}%
\bibitem [{\citenamefont {Neto}\ and\ \citenamefont {Jones}(2005)}]{neto2005quantum}%
  \BibitemOpen
  \bibfield  {author} {\bibinfo {author} {\bibfnamefont {A.~C.}\ \bibnamefont {Neto}}\ and\ \bibinfo {author} {\bibfnamefont {B.}~\bibnamefont {Jones}},\ }\bibfield  {title} {\bibinfo {title} {{Quantum Griffiths effects in metallic systems}},\ }\href {https://journals.aps.org/prb/pdf/10.1103/PhysRevB.66.174433} {\bibfield  {journal} {\bibinfo  {journal} {Europhysics Letters}\ }\textbf {\bibinfo {volume} {71}},\ \bibinfo {pages} {790} (\bibinfo {year} {2005})}\BibitemShut {NoStop}%
\bibitem [{\citenamefont {Vojta}\ and\ \citenamefont {Lee}(2006)}]{vojta2006nonequilibrium}%
  \BibitemOpen
  \bibfield  {author} {\bibinfo {author} {\bibfnamefont {T.}~\bibnamefont {Vojta}}\ and\ \bibinfo {author} {\bibfnamefont {M.~Y.}\ \bibnamefont {Lee}},\ }\bibfield  {title} {\bibinfo {title} {{Nonequilibrium phase transition on a randomly diluted lattice}},\ }\href {https://journals.aps.org/prl/pdf/10.1103/PhysRevLett.96.035701} {\bibfield  {journal} {\bibinfo  {journal} {Physical Review Letters}\ }\textbf {\bibinfo {volume} {96}},\ \bibinfo {pages} {035701} (\bibinfo {year} {2006})}\BibitemShut {NoStop}%
\bibitem [{\citenamefont {Disa}\ \emph {et~al.}(2020)\citenamefont {Disa}, \citenamefont {Fechner}, \citenamefont {Nova}, \citenamefont {Liu}, \citenamefont {F{\"o}rst}, \citenamefont {Prabhakaran}, \citenamefont {Radaelli},\ and\ \citenamefont {Cavalleri}}]{Disa2020}%
  \BibitemOpen
  \bibfield  {author} {\bibinfo {author} {\bibfnamefont {A.~S.}\ \bibnamefont {Disa}}, \bibinfo {author} {\bibfnamefont {M.}~\bibnamefont {Fechner}}, \bibinfo {author} {\bibfnamefont {T.~F.}\ \bibnamefont {Nova}}, \bibinfo {author} {\bibfnamefont {B.}~\bibnamefont {Liu}}, \bibinfo {author} {\bibfnamefont {M.}~\bibnamefont {F{\"o}rst}}, \bibinfo {author} {\bibfnamefont {D.}~\bibnamefont {Prabhakaran}}, \bibinfo {author} {\bibfnamefont {P.~G.}\ \bibnamefont {Radaelli}},\ and\ \bibinfo {author} {\bibfnamefont {A.}~\bibnamefont {Cavalleri}},\ }\bibfield  {title} {\bibinfo {title} {{Polarizing an antiferromagnet by optical engineering of the crystal field}},\ }\href {https://www.nature.com/articles/s41567-020-0936-3} {\bibfield  {journal} {\bibinfo  {journal} {Nature Physics}\ }\textbf {\bibinfo {volume} {16}},\ \bibinfo {pages} {937} (\bibinfo {year} {2020})}\BibitemShut {NoStop}%
\bibitem [{\citenamefont {Borovik-Romanov}(1960)}]{Borovik1960}%
  \BibitemOpen
  \bibfield  {author} {\bibinfo {author} {\bibfnamefont {A.}~\bibnamefont {Borovik-Romanov}},\ }\bibfield  {title} {\bibinfo {title} {{Piezomagnetism in the antiferromagnetic fluorides of cobalt and manganese}},\ }\href {http://jetp.ras.ru/files/borovik1960_en.pdf} {\bibfield  {journal} {\bibinfo  {journal} {Soviet Phys JETP}\ }\textbf {\bibinfo {volume} {11}} (\bibinfo {year} {1960})}\BibitemShut {NoStop}%
\bibitem [{\citenamefont {Schleck}\ \emph {et~al.}(2010)\citenamefont {Schleck}, \citenamefont {Nahas}, \citenamefont {Lobo}, \citenamefont {Varignon}, \citenamefont {Lepetit}, \citenamefont {Nelson},\ and\ \citenamefont {Moreira}}]{Schleck2010}%
  \BibitemOpen
  \bibfield  {author} {\bibinfo {author} {\bibfnamefont {R.}~\bibnamefont {Schleck}}, \bibinfo {author} {\bibfnamefont {Y.}~\bibnamefont {Nahas}}, \bibinfo {author} {\bibfnamefont {R.~P. S.~M.}\ \bibnamefont {Lobo}}, \bibinfo {author} {\bibfnamefont {J.}~\bibnamefont {Varignon}}, \bibinfo {author} {\bibfnamefont {M.~B.}\ \bibnamefont {Lepetit}}, \bibinfo {author} {\bibfnamefont {C.~S.}\ \bibnamefont {Nelson}},\ and\ \bibinfo {author} {\bibfnamefont {R.~L.}\ \bibnamefont {Moreira}},\ }\bibfield  {title} {\bibinfo {title} {{Elastic and magnetic effects on the infrared phonon spectra of ${\text{MnF}}_{2}$}},\ }\href {https://doi.org/10.1103/PhysRevB.82.054412} {\bibfield  {journal} {\bibinfo  {journal} {Physical Review B}\ }\textbf {\bibinfo {volume} {82}},\ \bibinfo {pages} {054412} (\bibinfo {year} {2010})}\BibitemShut {NoStop}%
\bibitem [{\citenamefont {Bakshi}\ and\ \citenamefont {Hicks}(1982)}]{Bakshi1982}%
  \BibitemOpen
  \bibfield  {author} {\bibinfo {author} {\bibfnamefont {E.}~\bibnamefont {Bakshi}}\ and\ \bibinfo {author} {\bibfnamefont {T.}~\bibnamefont {Hicks}},\ }\bibfield  {title} {\bibinfo {title} {{The preparation and magnetic properties of MnF$_2$ and its mixtures with ZnF$_2$ and CoF$_2$}},\ }\href {https://iopscience.iop.org/article/10.1088/0022-3719/15/31/020/pdf} {\bibfield  {journal} {\bibinfo  {journal} {Journal of Physics C: Solid State Physics}\ }\textbf {\bibinfo {volume} {15}},\ \bibinfo {pages} {6449} (\bibinfo {year} {1982})}\BibitemShut {NoStop}%
\end{thebibliography}%

\appendix

\begin{widetext}
\section{Derivation of the elasto-caloric effect coefficient}
To obtain the ECE coefficient, we freeze out elastic modes and set $\tilde{J} = J$ in the infinite-order expression for $\bar{f}$ in Eq. (\ref{eq:free_energy_exact}). We let $b=\lambda H_z \varepsilon$ be an infinitesimal bias arising from an infinitesimal applied strain $\varepsilon$, and compute the specific entropy as
\begin{align}
 s=-\frac{\partial\bar{f}}{\partial T} & =\int_{-\infty}^{\infty}p(z-\Phi-b)\partial_{t}\Lambda(z)\mathrm{d}z\\
  & =\log2-\int_{-\infty}^{\infty}p(z-\Phi-b)\Big[\frac{\sqrt{z^{2}+\gamma^{2}}}{t}\tanh\frac{\sqrt{z^{2}+\gamma^{2}}}{t}-\log\cosh\frac{\sqrt{z^{2}+\gamma^{2}}}{t}\Big]\mathrm{d}z\label{eq:s_full}
 \end{align}
 where
 \begin{equation}
     p(z)\equiv \frac{1}{\sqrt{2\pi}w}e^{-z^2/2w^2}
 \end{equation}
 The term in the square brackets of the integrand in Eq. (\ref{eq:s_full})
 is a non-negative function bounded above by $\log2$, i.e., $s\le\log2$. Note that because $\tilde{J} = J$, the definitions for the scaled variables $t$, $\gamma$, and $w$ are modified slightly, namely, $t\equiv T$, $\gamma \equiv \Gamma$, and $w = 2hW$. Using this new definition, it follows that 
 \begin{align}
     \frac{\partial s}{\partial \varepsilon } &= \lambda H_z \Big{(}1+\frac{\partial \Phi}{\partial b}\Big{)}\int_{-\infty}^{\infty}p(z-\Phi-b)\frac{z}{T^2}\sech^2\frac{\sqrt{z^2+\Gamma^2}}{T}\mathrm{d}z  \label{eq:app1}\\
     \frac{\partial s}{\partial T} &= \int_{-\infty}^\infty p(z-\Phi-b) \Big(\frac{\partial \Phi}{\partial T} + \frac{\sqrt{z^2+\Gamma^2}}{T} \Big)\frac{\sqrt{z^2+\Gamma^2}}{T} \sech^2 \frac{\sqrt{z^2+\Gamma^2}}{T}\mathrm{d}z\label{eq:app2}
 \end{align}
 where $\Phi$ is the solution of $\partial \bar{f}/\partial \Phi=0$:
 \begin{equation}
     \Phi = \int_{-\infty}^\infty p(z-\Phi-b)\frac{z}{\sqrt{z^2+\Gamma^2}}\tanh\frac{\sqrt{z^2+\Gamma^2}}{T}\mathrm{d}z \label{eq:app3}
 \end{equation}
We solve Eq. (\ref{eq:app3}) numerically in the parameter space and combine Eq. (\ref{eq:app1}) and Eq. (\ref{eq:app2}) to obtain the ECE coefficient $\eta$ shown in Fig \ref{fig:ECE}.
\end{widetext}

\end{document}